\documentclass[aps,prd,11pt,a4paper,nofootinbib,oneside,superscriptaddress]{revtex4-1}
\pdfoutput = 1

\usepackage{amsmath,amssymb,amsfonts,color}
\usepackage{tensor,slashed,paralist,cases,mathrsfs}
\usepackage{float,cancel,xcolor}
\usepackage{graphicx}% Include figure files
\usepackage{dcolumn}% Align table columns on decimal point
\usepackage{bm}% bold math
\usepackage[colorlinks=true,urlcolor=blue,linkcolor=blue,citecolor=blue,linktocpage=true]{hyperref}

\newcommand{\beq}{\begin{eqnarray}}
\newcommand{\eeq}{\end{eqnarray}}
\newcommand{\bpmatrix}{\begin{pmatrix}}
\newcommand{\epmatrix}{\end{pmatrix}}
\newcommand{\ba}{\begin{array}}
\newcommand{\ea}{\end{array}}

\renewcommand{\eqref}[1]{Eq.~(\ref{#1})}

\newcommand{\bc}{\begin{center}}
\newcommand{\ec}{\end{center}}

\newcommand*{\defeq}{\stackrel{\small{\mathsf{def}}}{=}}

\begin{document}

\vspace*{1.5em}

%\title{Lepton-Specific Next-to-Minimal 2HDM Portal Dark Matter and the Galactic Center Gamma-Ray Excess}
%\title{Leptophilic Next-to-Minimal 2HDM Portal Vector Dark Matter for the Galactic Center Gamma-Ray Excess from the Two-Step Cascade Annihilation, 
%and Muon $g-2$}
%\title{Lepton-Specific N2HDM Portal Vector Dark Matter for the Galactic Center Gamma-Ray Excess from the Two-Step Cascade Annihilation, and Muon $g-2$}

\title{Hidden Higgs Portal Vector Dark Matter for the Galactic Center Gamma-Ray Excess from the Two-Step Cascade Annihilation, and Muon $g-2$}

\author{Kwei-Chou Yang}
\email{kcyang@cycu.edu.tw}

\affiliation{Department of Physics and Center for High Energy Physics, Chung Yuan Christian University, Taoyuan 320, Taiwan}

%\date{\today $\vphantom{\bigg|_{\bigg|}^|}$}

\begin{abstract}
We have built a lepton-specific next-to-minimal two-Higgs-doublet-portal vector dark matter model. The vector dark matter in the hidden sector does not directly couple to the visible sector, but instead annihilates into the hidden Higgs bosons which decay through a small coupling into the CP-odd Higgs bosons.  In this model,  the Galactic center gamma-ray excess is mainly due to the 2-step cascade annihilation with $\tau$'s in the  final state.
The obtained mass of the CP-odd Higgs $A$ in the Galactic center excess fit can explain the muon $g-2$ anomaly at the 2$\sigma$ level without violating the stringent constraints from the lepton universality and $\tau$ decays.
  We  show three different freeze-out types of the dark matter relic, called (i) the conventional WIMP dark matter, (ii) the unconventional WIMP dark matter and (iii) the cannibally  co-decaying dark matter, depending on the magnitudes of the mixing angles between the hidden Higgs and visible two-Higgs doublets.
  The dark matter in the hidden sector is secluded from detections in the direct searches or colliders, while the dark matter annihilation signals are not suppressed in a general hidden sector dark matter model. We discuss the  constraints from observations of  the dwarf spheroidal galaxies and the Fermi-LAT projected sensitivity. 

\end{abstract}
\maketitle
\newpage

\section{Introduction}

The evidence for the dark matter (DM) in the Universe has been well established from various astronomical observations and cosmological measurements.
The DM, which cannot be accounted for within the standard model (SM) scenario, indicates the existence of new physics.
The attractive candidates for the DM are the so-called weakly interacting massive particles (WIMPs) which, having a weak scale mass and annihilating into SM particles via weak scale couplings, can provide a correct thermal relic abundance, following the Boltzmann suppression before freeze-out.

 Several collaborations have reported an excess of GeV gamma-rays near the region of the Galactic center (GC) \cite{Goodenough:2009gk, Hooper:2010mq, Hooper:2011ti, Abazajian:2012pn, Gordon:2013vta, Huang:2013pda, Daylan:2014rsa, Calore:2014xka, Calore:2014nla,Karwin:2016tsw,TheFermi-LAT:2017vmf}, where the excess spectrum can be fitted using DM annihilation models.  Although the excess result might be explained by the millisecond pulsars or some other astrophysical sources  \cite{OLeary:2016cwz,Fermi-LAT:2017yoi,Ploeg:2017vai,Hooper:2015jlu,Cholis:2015dea,Petrovic:2014uda,Carlson:2014cwa}, the DM annihilation is a viable scenario from the particle physics point of view  \cite{Boehm:2014hva, Hektor:2014kga, Arina:2014yna, Hektor:2015zba, Lebedev:2011iq, Farzan:2012hh, Baek:2012se, Baek:2014goa, Ko:2014gha, Boucenna:2011hy, Escudero:2017yia,  Ko:2014loa, Abdullah:2014lla,  Martin:2014sxa, Berlin:2014pya, Kim:2016csm, Yang:2017zor}.
However, the DM models capable of explaining the GC gamma-ray excess are increasingly constrained by direct detection experiments and measurements at colliders. 
Some ideas have been proposed that can avoid overproduced signals in the latter two experiments;
for instance,  the DM annihilates into $b \bar{b}$ through a pseudoscalar mediator exchange \cite{Boehm:2014hva,Hektor:2014kga, Arina:2014yna,Yang:2017zor}.
The present work is motivated by the idea called ``secluded WIMP dark matter" in which the DM first annihilates to a pair of short-lived hidden mediators which subsequently decay into SM particles through very small couplings~\cite{Abdullah:2014lla,Ko:2014loa,Martin:2014sxa,Berlin:2014pya,Kim:2016csm,Escudero:2017yia,Yang:2017zor,Boehm:2014bia,Pospelov:2007mp,Hooper:2012cw,Mardon:2009rc,Profumo:2017obk}, so that it can easily evade the stringent constraints from current measurements, but provide observable gamma-ray signals in the indirect measurements.

In this paper, we will use the GC gamma-ray excess spectrum obtained by Calore, Cholis and Weniger (CCW) who analyzed the Fermi data in a consistent treatment of systematic uncertainties that came from the modeling of the Galactic diffuse emission background \cite{Calore:2014xka}.  
It is interesting to note that if the dark matter annihilates directly only into the $\tau^+ \tau^-$, the spectral fit to the GC gamma-ray GeV excess gives the best-fit result with dark matter mass $\sim 9.5$~GeV, and low-velocity annihilation cross section $\langle \sigma v \rangle\simeq 0.37\times 10^{-26}$~$\text{cm}^3\text{s}^{-1}$, nevertheless corresponding to a lower $p\text{-value} \sim 0.05$ for the goodness-of-fit test  \cite{Calore:2014xka}. 
The direct DM annihilation to $\tau^+ \tau^-$ produces a gamma-ray spectrum which peaks sharply at a little higher energies.  If the annihilation processes present some extra intermediate steps, the final state $\tau$'s  generated from the cascade decays are boosted, and therefore the resultant gamma-ray spectrum becomes broader and has a  better fit to the GC excess observation \cite{Elor:2015tva,Elor:2015bho}.  
We are interested in the two-step cascade annihilation process (see Fig.~\ref{fig:cascade-tau} for reference). 
The reason is that not only a much larger $p$-value $\sim0.22$ can be obtained, but also the fitted
DM mass and annihilation cross section are enlarged by a factor of $\sim 4$, compared with the direct annihilation to $\tau^+ \tau^-$. The resulting annihilation cross section required to explain the GC excess signals is thus in good agreement with that required by the correct relic abundance in the thermal WIMP scenario.

Here, we make an extension for the "secluded DM" idea. We will build a hidden sector dark matter model in which the produced gamma-ray spectrum is mostly generated by the  final state, $\tau$'s, resulted from the two-step cascade annihilation of the vector dark matter.  After spontaneous symmetry breaking, the hidden dark sector maintaining the dark discrete $Z_2$ symmetry  is composed of a singlet vector boson (the dark matter) and a real hidden Higgs (the mediator), where the latter will mix with the 2 neutral CP-even Higgs bosons of the lepton-specific two-Higgs-doublet model (2HDM) which is also called the type-X 2HDM. Moreover, in the two Higgs doublets the two extra neutral Higgs bosons, one CP-even and one CP-odd, couple to leptons are enhanced in large $\tan\beta$ limit, whereas their couplings to quarks are suppressed.  The main mechanism in this model for describing the GC excess is that the DM annihilation to the mediator pair is followed by the mediator decays to CP-odd Higgs bosons, $A$, which subsequently decay into taus.  We find that the resultant masses for DM and CP-odd Higgs are  $m_X\sim 25-50$~GeV and $m_A \sim 3.6 - 25$~GeV, respectively. 
Our result for $m_A$, in a good agreement with the allowed range given in the type-X 2HDM \cite{Wang:2014sda,Abe:2015oca,Chun:2016hzs}, can accommodate the muon $g-2$ anomaly at 2$\sigma$ level, under the constraints from the lepton universality and $\tau$ decays.

In the present case, the CP-odd boson $A$ is in chemical and thermal equilibrium with the SM thermal bath before the DM freeze out. However,
the dependence of the DM freeze-out temperature and corresponding thermally averaged annihilation cross section on the mixing angles of the hidden Higgs and 2HDM is subtle.   At a temperature below the mass  of the mediator $S$, the chemical equilibrium of $S$ with the thermal bath is maintained mainly through $S\leftrightarrow AA$. If  the mixing angles are not too small, the hidden Higgs mediator can be in thermal equilibrium with the bath, such that the DM particles behave like WIMPs, which exhibit the Boltzmann suppression until the freeze-out temperature.

 However, due to small mixing angles, resulting in that the coupling constant of the mediator to the $A$ boson is too small to ensure the required decay width of the $S$ boson to keep the dark sector in chemical equilibrium with the bath, the dark sector will decouple from the thermal background at the temperature below the mediator's mass. If so, the comoving number density of the dark sector will not  exponentially deplete until the occurrence of the mediator decaying to the $A$ pair. Such the mechanism was discussed by Dror, Kuflik, and Ng \cite{Dror:2016rxc} and by Farina, Pappadopulo, Ruderman, and Trevisan \cite{Farina:2016llk}, where the former used a degenerate hidden sector to illustrate the idea for the DM, called ``co-decaying dark matter". Our case is more relevant to the non-degenerate one, for which the hidden mediator undergoes cannibalism first \cite{Farina:2016llk,Pappadopulo:2016pkp} and, after that phase, the exponential suppression for comoving dark matter number density  could occur much earlier than that for the mediator due to a significantly suppressed up-scattering rate for the process. More detailed discussions will be presented in Sec.~\ref{sec:thetadelta-2}.

This paper is organized as follows. In Sec.~\ref{sec:LN2HDMVDM}, we describe the model. The relevant ingredients, including the Yukawa sectors, Higgs couplings, and  the decay widths of the mediator and CP-odd Higgs boson are presented. In Sec.~\ref{sec:ETconstraints}, we discuss the experimental and theoretical constraints on parameters related to the two-Higgs doublets and Yukawa sectors.
In Sec.~\ref{sec:CAconstraints}, we first outline the approach of determining the gamma-ray spectrum from a 2-step cascade annihilation to the final state $\tau$'s, and then describe the analysis and results concerning gamma-ray observations, compared with the relic abundance in the conventional WIMP dark matter scenario.  Sec.~\ref{sec:thetadelta} contains the analysis for the mixing angles between the hidden scalar and two neutron CP-even bosons in the two Higgs doublets. The discussions and conclusions are given in Sec.~\ref{sec:conclusion}.

\section{Lepton-Specific  Next-to-Minimal 2HDM Portal Vector Dark Matter}\label{sec:LN2HDMVDM}

\subsection{The Model}

We consider  a model with two Higgs doublets, $\Phi_1, \Phi_2$, and a complex scalar dark Higgs field $\Phi_S$.
%\subsection{The Higgs Sector}
 The CP-conserving potential for the Higgs sector is described by
\beq
V &=& m_{11}^2 |\Phi_1|^2 + m_{22}^2 |\Phi_2|^2 - m_{12}^2 (\Phi_1^\dagger
\Phi_2 + h.c.) + \frac{\lambda_1}{2} (\Phi_1^\dagger \Phi_1)^2 +
\frac{\lambda_2}{2} (\Phi_2^\dagger \Phi_2)^2 \nonumber \\
&& + \lambda_3
(\Phi_1^\dagger \Phi_1) (\Phi_2^\dagger \Phi_2) + \lambda_4
(\Phi_1^\dagger \Phi_2) (\Phi_2^\dagger \Phi_1) + \frac{\lambda_5}{2}
[(\Phi_1^\dagger \Phi_2)^2 + h.c.] \nonumber \\
&& + m_{33}^2 \Phi_S^\dagger \Phi_S + \frac{\lambda_6}{2} (\Phi_S^\dagger \Phi_S)^2 +
\lambda_7 (\Phi_1^\dagger \Phi_1) (\Phi_S^\dagger \Phi_S) +
\lambda_8 (\Phi_2^\dagger \Phi_2) (\Phi_S^\dagger \Phi_S) \;,
\label{eq:n2hdm-Vhiggs}
\eeq
where $\Phi_S$ is a singlet under the SM gauge fields. As usual, we have imposed a discrete $\mathbb{Z}_2$ symmetry to the Higgs potential, such that  $\Phi_1 \to \Phi_1$, $\Phi_2 \to -\Phi_2$, and $\Phi_S \to  \Phi_S$, under which the tree-level flavor changing neutral currents (FCNCs) are absent. The $\mathbb{Z}_2$ symmetry is softly broken by the term containing $m_{12}^2$.
On the other hand, we have considered that $\Phi_S$ is  charged in the dark $U_{\rm dm}(1)$ gauge group, while other Higgs fields and SM particles have no such quantum number. The $U_{\rm dm}(1)$ group contains an abelian  gauge boson, $X_\mu$. After spontaneous symmetry breaking, the vacuum expectation value (VEV) of $\Phi_S$ generates a mass for $X_\mu$, and a discrete $\mathbb{Z}_2^\prime$ symmetry: $X_\mu \to -X_\mu, \Phi_S \to \Phi_S^*$, is still maintained, such that $X_\mu$ is stable and can serve as a (vector) dark matter candidate. 

The relevant kinetic terms in the dark sector are given by
\begin{align}
{\cal L}_{\rm DM} =-\frac{1}{4} X_{\mu\nu} X^{\mu\nu} + (D_\mu\Phi_S)^\dagger (D^\mu \Phi_S) \,,
\end{align}
where $X_{\mu\nu} =\partial_\mu X_\nu -\partial_\nu X_\mu$, and the covariant derivative is defined as 
\begin{align}
D_\mu \Phi_S = (\partial_\mu + i g_X Q_{\Phi_S} X_\mu )\Phi_S,
\end{align}
with $Q_{\Phi_S}$ the  $U_{\rm dm}(1)$ charge of $\Phi_S$. After spontaneous symmetry breaking, we have
\begin{equation}
\Phi_S=\frac{1}{\sqrt{2}} (v_S + h_3),
\label{eq:vs}
\end{equation}
where the imaginary part of $\Phi_S$ is absorbed by the vector gauge boson (dark matter) due to the $\mathbb{Z}_2^\prime$ symmetry: $X_\mu \to -X_\mu$, and the vector gauge boson obtains a mass, $m_X=g_X Q_{\Phi_S} v_S$ (see also Refs.~\cite{Lebedev:2011iq, Farzan:2012hh, Baek:2012se, Baek:2014goa, Ko:2014gha,Karam:2016rsz,Karam:2015jta} for related discussions). In this paper, we will simply take $Q_{\Phi_S} =1$; in other words, $Q_{\Phi_S} $ and $g_X$ are lumped together.  The interacting terms  of the dark sector is given by
\begin{align}
{\cal L}_{\rm DM}^{\rm int} \supset
\frac{1}{2} g_X^2 X_\mu X^\mu h_3^2 +  g_X m_X X_\mu X^\mu h_3 \,,
\end{align}
where the hidden scalar field $h_3$ will mix with the neutral scalars in the two Higgs doublets through the interaction given by Eq.~(\ref{eq:n2hdm-Vhiggs}).

After spontaneous symmetry breaking, the version of the Higgs sector becomes the next-to-minimal two-Higgs-doublet model (N2HDM).  We decompose the Higgs doublet fields as
\begin{align}
\Phi_i = \left(\begin{array}{c}
h_i^+ \\
\frac{1}{\sqrt{2}}(v_i + h_i + i a_i )
\end{array}\right),\quad \text{with {\it i} = 1, 2}, 
\end{align}
where  the vacuum expectation values (VEVs) of the doublets approximately satisfy
$v_1^2+v_2^2 \approx v^2 =(246~\text{GeV})^2$, with $v$ being the SM VEV. 
The scalar fields in $\Phi_i$ and $\Phi_S$ can be expressed in terms of mass eigenstates of physical Higgs  states and Goldstone bosons as
\begin{align}
\left(\begin{array}{c}
h_1^\pm \\ h_2^\pm
\end{array}\right) 
&=
\left(\begin{array}{cc}
\cos\beta & -\sin\beta \\
\sin\beta & \cos\beta
\end{array}\right)
\left(\begin{array}{c}
G^\pm \\ H^\pm
\end{array}\right),  \\
\left(\begin{array}{c}
a_1 \\ a_2
\end{array}\right) 
&=
\left(\begin{array}{cc}
\cos\beta & -\sin\beta \\
\sin\beta & \cos\beta
\end{array}\right)
\left(\begin{array}{c}
G^0 \\ A
\end{array}\right),\\
\left(\begin{array}{c}
h_1 \\ h_2 \\h_3
\end{array}\right) 
&=
\left(\begin{array}{ccc}
\cos\alpha & -\sin\alpha & 0 \\  
\sin\alpha & \cos\alpha  & 0 \\ 
0 & 0 & 1 \\ 
\end{array}\right)
\left(\begin{array}{ccc}
1 & 0 & 0 \\  
0 & \cos\delta & -\sin\delta \\ 
0 & \sin\delta & \cos\delta \\ 
\end{array}\right)
\left(\begin{array}{ccc}
\cos\theta & 0 & -\sin\theta \\  
0 & 1 & 0 \\
\sin\theta & 0  & \cos\theta \\ 
\end{array}\right)
\left(\begin{array}{c}
H \\ h \\ S
\end{array}\right),
\end{align}
 where $(H,h)$ are the (heavy, light) Higgs CP-even scalars in the two Higgs doublets in the limit of $\delta, \theta \to 0$, $A$ the CP-odd scalar, $H^\pm$ the two charged Higgs bosons, and $(G^\pm,G^0)$ the Goldstone bosons corresponding to the longitudinal components of $W^\pm$ and $Z$, respectively. 
 Here $\beta$ is the mixing angle of the charged bosons, and $\alpha$ is  the mixing angle of neutral CP-even bosons $(h_1,h_2)$ in the limit of $\delta, \theta \to 0$, where the former is defined as $\tan\beta  = v_2/v_1$.

The theoretical requirements for the perturbativity, vacuum stability, and tree-level perturbative unitarity are given in Appendix~\ref{app:th-constraints}. From the results for the square of masses of $H^\pm$ and $A$,  square of mass matrix of $H, h$ and $S$, and the minimum conditions of the Higgs potential at the VEV, the quartic couplings $\lambda_i$, with $i\equiv 1,\dots, 8$, can be rewritten  in terms of the $m_h^2, m_H^2, m_A^2, m_{H^\pm}^2$ and $M^2 [\equiv m_{12}^2 / (s_\beta c_\beta)]$. We show the relations  in Appendix~\ref{app:quartic}.

\subsection{The Yukawa Sectors}

The type-X Yukawa interactions are imposed a $Z_2$ symmetry only to the right-handed quarks, $u_R \to -u_R$ and $d_R \to - d_R$. 
Thus, the Yukawa Lagrangian, describing the interactions of the Higgs doublets to the SM fermions, is given by 
\begin{align}
{\cal L}_{\rm Yukawa} =& - \overline{Q}_L y_u \tilde{\Phi}_2 u_R - \overline{Q}_L y_d \Phi_2 d_R -  \overline{L}_L y_\ell \Phi_1 \ell_R + h.c.,\label{eq:yukawa}
\end{align}
where $\tilde{\Phi}_2 = i\sigma_2 \Phi_2^*$, and $y_i$ is the $3\times 3$ Yukawa matrix.
In terms of the mass eigenstates of the scalar bosons, the Yukawa interaction terms can be rewritten by
\begin{align}
{\cal L}_{\rm Yukawa} \supset
 & -\sum_{f=u,d, \ell}\frac{m_f}{v}\left( \frac{\bar\xi_f}{c_\beta} h_f \bar ff  - i \, \text{sgn}(f) \, \xi_f \, A \bar f\gamma_5 f \right) \nonumber\\
 & + \left[ \sqrt{2}V_{ud} H^+ \bar u \left( \frac{m_u\xi_u}{v} P_L 
- \frac{m_d\xi_d}{v} P_R\right) d  -\frac{\sqrt{2} m_\ell \xi_\ell}{v} H^+ \bar \nu P_R \ell + h.c.\right], \label{eq:yukawa_int}
\end{align}
where $h_{u,d} \equiv h_2$, $h_\ell \equiv h_1$, $\bar\xi_{u,d}=\xi_{u,d}=\cot\beta$, $\bar\xi_\ell=1$, $\xi_\ell \equiv -\tan\beta$, 
$P_{R} =(1+ \gamma_5)/2$, $P_{L} =(1 - \gamma_5)/2$  and $V_{ud}$ is the Cabibbo-Kobayashi-Maskawa matrix element. 
Keeping small terms linear in $\sin\theta$ and $\sin\delta$, the Yukawa couplings of the neutral Higgs bosons in the type-X N2HDM,  normalized with respect to the SM Higgs, are given in Table~\ref{tab:effyukln2hdm}, where $g_{Aff} \equiv \text{sgn}(f) \,  \xi_f$.

For the type-X Yukawa interactions, the normalized lepton Yukawa coupling to the SM Higgs is given by
\begin{align}
g_{h\ell\ell}= -\frac{s_\alpha}{c_\beta} =s_{\beta -\alpha} - t_\beta c_{\beta-\alpha} \,.
\end{align}
Considering the LHC data but without muon $g-2$ constraint, the allowed parameters, consistent with the alignment limit of $s_{\beta-\alpha} \to 1$, lie in two different regions  \cite{Wang:2014sda}; in one region, the $g_{h\ell\ell}$ couplings ($\to 1$) have values near the SM ones,  while in the other region which is called the wrong-sign region, the $g_{h\ell\ell} \to -1 $ has  opposite sign to the SM Higgs couplings to $VV$, (normalized) $g_{hVV,hZZ} =s_{\beta-\alpha} \to 1$, and to the quark pair, $g_{hff}=c_\alpha/s_\beta =s_{\beta-\alpha} + c_{\beta-\alpha} /t_\beta \to 1$ for a large $\tan\beta$ satisfying $2 t_\beta c_{\beta-\alpha} \sim 2$. Only the wrong-sign region is favored by the muon $g-2$ measurement \cite{Wang:2014sda} (see also the following discussion in this work).

\begin{figure}[t!]
\begin{center}
\includegraphics[width=0.33\textwidth]{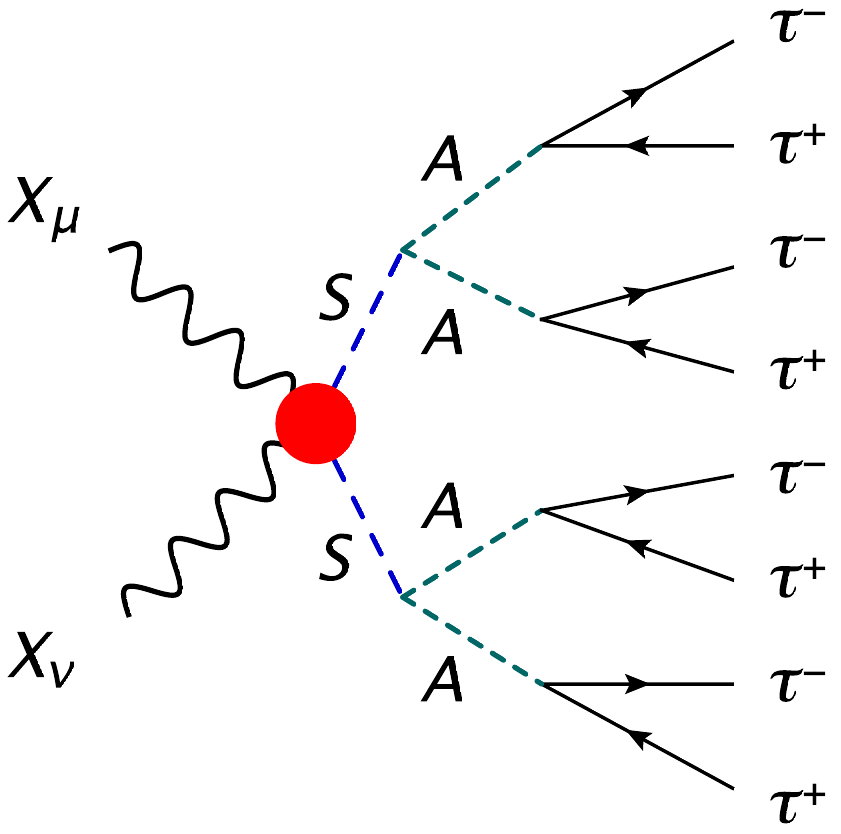}
\caption{The main DM annihilation process of two-step cascades, $X_\mu X_\nu \to S S \to 4 A\text{'s} \to 8 \tau\text{'s}$, relevant to the GC gamma-ray excess. The red shaded region contains interactions shown in Fig.~\ref{fig:vdm_ann}. }
\label{fig:cascade-tau}
\end{center}
\end{figure}

On the other hand, for the type-X Yukawa interactions, the couplings, $g_{A\ell\ell}$ and $g_{H\ell\ell} \propto \tan\beta$, are enhanced by a large $\tan\beta$, while $g_{Aqq}$ and $g_{Hqq} \propto 1/\tan\beta$ are suppressed, where $q\equiv$ quark. Therefore, as shown in Fig.~\ref{fig:cascade-tau}, the two-step cascade DM annihilation process via the on-shell pseudoscalar boson into the SM particles are dominated by $\tau$'s in the final states for a large $\tan\beta$.
Note that the DM annihilation into tau's cannot be through the heavier on-shell neutral Higgs, which is kinematically forbidden, because, as shown in this work, its mass $m_H\ \sim 300$~GeV is much larger than the DM mass. Note also that, in contrast with the type-X case, for the type-II Yukawa interactions, because both the down-type quark and lepton couplings of the heavier neutral Higgs boson are enhanced by $\tan\beta$, that model will be severely constrained by the extra Higgs search at the LHC and by the flavor physics \cite{Ferreira:2014naa}.

\begin{figure}[t!]
\begin{center}
\includegraphics[width=0.8\textwidth]{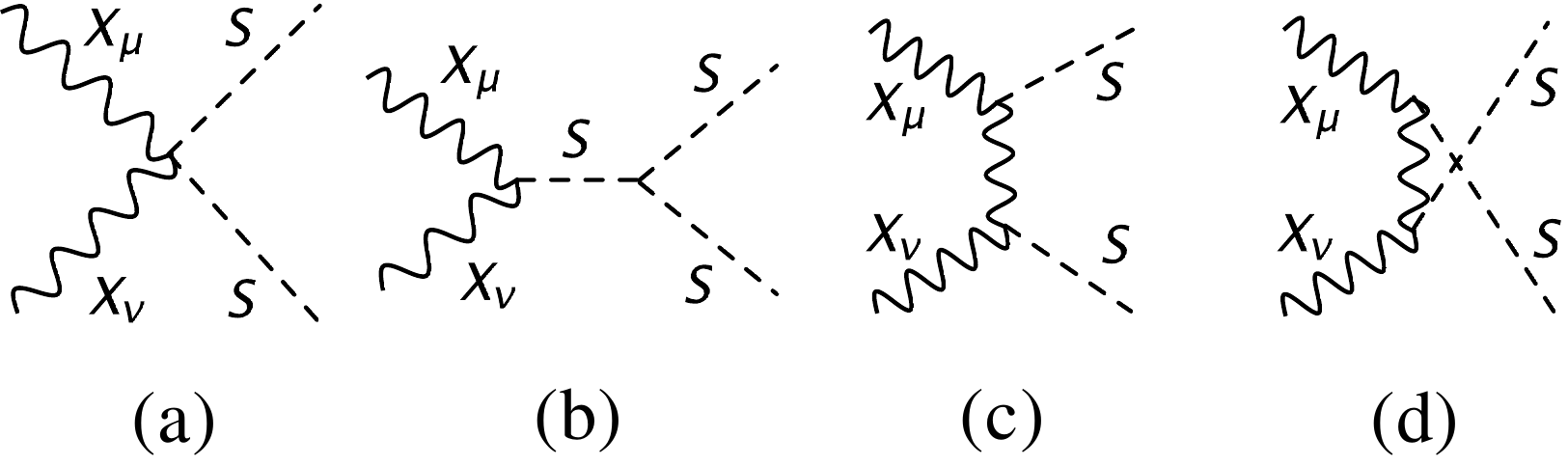}
\caption{Feynman diagrams that dominantly contribute to the DM annihilation cross section relevant to relic abundance and GC gamma-ray excess, where (a), (b), (c), and (d) are for the 4-vertex, $s$-, $t$-, $u$-channels, respectively. }
\label{fig:vdm_ann}
\end{center}
\end{figure}

In the present work, we study that the vector DM ($X$) first annihilates into the unstable hidden Higgs bosons ($S$), as shown in Fig.~\ref{fig:vdm_ann}, and then the $S$ dominantly decays  into the pseudoscalar pair.
In the following section, we will give the triple and quartic Higgs couplings, which are relevant to the $XX \to SS$ and $S\to AA$  processes. Moreover, these couplings are also relevant to the Boltzmann equations, which will be discussed in Sec.~\ref{sec:thetadelta-2}.

\begin{table}
\begin{center}
  \begin{tabular}{cccc}
% \multicolumn{4}{c}{Type X} \\
\hline\hline
 & $f=u,d$ &  & $f=\ell$ \\ 
$g_{hff}$ & $c_{\alpha} /s_\beta$ &      & $ -s_{\alpha} /c_\beta$ \\
$g_{Hff}$ & $s_{\alpha} /s_\beta$ &    & $c_{\alpha}/c_\beta$ \\
$g_{Sff}$ & $-(s_{\alpha} s_{\theta} + c_{\alpha} s_{\delta}  )/s_\beta$ &  & $(-c_{\alpha} s_{\theta} +
s_{\alpha} s_{\delta} ) /c_\beta$  \\ 
$g_{Aff}$ &  $\pm 1/t_\beta$ &  & $t_\beta$  \\ 
\hline\hline
\end{tabular}
\caption{ The tree level Yukawa couplings of the neutral type-X N2HDM Higgs bosons, keeping terms linear in $
\sin\theta$ and $\sin\delta$, with respect to that of the SM Higgs.
 \label{tab:effyukln2hdm} }
\end{center}
\end{table}

\subsection{The triple and quartic Higgs couplings}

We consider that the GC gamma-ray excess originates from the two-step cascade DM annihilation, so that the mixing angles, $\delta$ and $\theta$, are small, and only terms linear in $\sin\delta$ and $\sin\theta$ are kept in the effective couplings. However, because $\tan\beta$ needs to be larger ($\sim 35$) in the present case, we also keep the terms, which are quadratic in these two angles and involve $\tan\beta$. The constraints on $\delta$ and $\theta$ will be discussed in Sec.~\ref{sec:thetadelta}.

The Lagrangian, containing triple and quartic Higgs couplings, are relevant to the present study.
Using the result given in Appendix~\ref{app:quartic}, we can express these couplings in terms of the squares of physical Higgs masses and $M^2$ as 
\begin{align} 
{\cal L}_{\rm triple} & \supset 
\frac{1}{2} v \lambda_{hAA} hAA + \frac{1}{2} v \lambda_{HAA} HAA + \frac{1}{2} v \lambda_{SAA} SAA + \frac{1}{2} v \lambda_{hSS} hSS  + \frac{1}{2} v \lambda_{HSS} HSS \nonumber \\
& + \frac{1}{6}v \lambda_{SSS} SSS + \cdots \,,   \\
{\cal L}_{\rm quartic} &\supset 
 \frac{1}{4} \lambda_{SSAA} SSAA +  \frac{1}{24} \lambda_{SSSS} SSSS \cdots \,,  
\end{align}
where, neglecting all terms suppressed by $s_\delta^2, s_\theta^2$, and $s_\delta s_\theta$ except that enhanced by $t_\beta$, the couplings are given by
\begin{align}
\lambda_{hAA}  \simeq & \frac{1}{v^2} \Bigg[  \big(2M^2 -2 m_A^2 -m_h^2 \big) s_{\beta-\alpha} - (M^2 -m_h^2) \left(t_\beta - \frac{1}{t_\beta}\right) 
  c_{\beta-   \alpha} \Bigg] \,, \label{eq:triple-1-hAA} \\
 \lambda_{HAA}  \simeq & \frac{1}{v^2} \Bigg[  \big(2M^2 -2 m_A^2 -m_H^2 \big) c_{\beta-\alpha} + (M^2 -m_H^2) \left(t_\beta - \frac{1}{t_\beta}\right) 
  s_{\beta-   \alpha} \Bigg] \,,  \label{eq:triple-2-HAA} \\
\lambda_{SAA}  \simeq & \frac{1}{v^2} \Bigg[
   - M^2 \bigg( \frac{c_\alpha s_\theta - s_\alpha s_\delta}{c_\beta} + \frac{s_\alpha s_\theta + c_\alpha s_\delta}{s_\beta} \bigg)
  +m_S^2 \bigg( s_\beta^2 \frac{c_\alpha s_\theta - s_\alpha s_\delta}{c_\beta} + c_\beta^2 \frac{s_\alpha s_\theta + c_\alpha s_\delta}{s_\beta} \bigg)
  \nonumber\\
 &+  2 m_A^2 \Big( (c_\alpha s_\theta - s_\alpha s_\delta) c_\beta + (s_\alpha s_\theta + c_\alpha s_\delta) s_\beta\Big)
     \Bigg] \,,  \label{eq:triple-3-SAA} \\
\lambda_{hSS} \simeq &  -\frac{2m_S^2 +m_h^2 }{v v_S}
  \Big[ s_\delta - (c_\alpha s_\theta -s_\alpha s_\delta)^2 s_\alpha \frac{t_\beta}{s_\beta} \frac{v_s}{v} \Big]
  \nonumber\\
     & - \frac{3M^2}{v^2}  \frac{t_\beta (c_\alpha s_\theta - s_\alpha s_\delta )}{s_\beta}  
   \Big[  (s_\alpha s_\theta + c_\alpha s_\delta) c_\alpha -s_\delta s_\beta^2  \Big] , \\
   \lambda_{HSS} \simeq &  - \frac{2m_S^2 +m_h^2 }{v v_S}
  \Big[ s_\theta+ (c_\alpha s_\theta -s_\alpha s_\delta)^2 c_\alpha \frac{t_\beta}{s_\beta} \frac{v_s}{v} \Big]
  \nonumber\\
     & - \frac{3M^2}{v^2}  \frac{t_\beta (c_\alpha s_\theta - s_\alpha s_\delta )}{s_\beta}  
   \Big[  (s_\alpha s_\theta + c_\alpha s_\delta) s_\alpha -s_\theta s_\beta^2  \Big] , \\
\lambda_{SSS} \simeq & - \frac{3 m_S^2}{v v_S} \,, \\
\lambda_{SSAA} \simeq & \frac{1}{v v_S} 
 \Bigg[ s_\beta t_\beta 
    \Big( m_H^2 c_\alpha s_\theta -m_h^2 s_\alpha  s_\delta  - m_S^2 (c_\alpha s_\theta -s_\alpha s_\delta)\Big) 
    \nonumber\\
    & \qquad   + \frac{c_\beta}{ t_\beta} 
    \Big(  m_H^2 s_\alpha s_\theta + m_h^2 c_\alpha s_\delta  - m_S^2 (s_\alpha s_\theta + c_\alpha s_\delta)\Big) \Bigg] ,
    \\
\lambda_{SSSS} \simeq &  \frac{3 m_S^2}{v_S^2} \,.
\label{eq:triple-quartic-coupling}
\end{align}
Here and in the following, we adopt the abbreviations: $s_\sigma \equiv \sin\sigma, c_\sigma \equiv \cos\sigma$, and $t_\sigma\equiv \tan \sigma $.

\subsection{The decay widths for the $S$ and $A$ bosons}

The partial decay widths for $S$ and $A$ are relevant to the studies of the relic density and indirect detection searches.
 The  partial decay widths of the  $S$ boson are
\begin{align}
\Gamma(S \rightarrow \bar{f} f) 
 & = N_c^f  \frac{m_S}{8\pi} \left( \frac{m_f}{v}\right)^2 |g_{Sff}|^2  \left( 1- \frac{4 m_f^2}{m_S^2} \right)^{3/2} \theta(m_S-2m_f) \; ,  \label{eq:S-partial-width-1} \\
\Gamma(S \rightarrow g g) & = \frac{\alpha_s^2}{2  \pi^3 m_S}\left|\sum_{q\equiv\text{quarks}} \frac{m_q^2}{v} g_{Sqq} f_S \left( \frac{4 m_q^2}{m_S^2}\right)   \right|^2 \; , \label{eq:S-partial-width-2}  \\
\Gamma(S \rightarrow A A) & = \frac{\lambda_{SAA}^2 v^2 }{32 \pi m_S}  \left( 1- \frac{4 m_A^2}{m_S^2} \right)^{1/2} \theta(m_S-2m_A) \label{eq:S-partial-width-3}   \; ,
\end{align}
where $N_c^f \equiv 3\, (1)$ for quarks (leptons), and $f_S (\tau) =[1+ (1 -\tau) f( \tau)]$  with
\begin{equation}
f (\tau)  =  \left\{ \begin{array}{lr} \text{arcsin}^2 \sqrt{\tau^{-1}}
        \, , \ \ & \tau \geq 1 \\ -\frac{1}{4} \left[ \log \frac{1+\sqrt{1-\tau}}{1-\sqrt{1-\tau}} - i \pi\right]^2 \, , & \tau < 1 \end{array} \right. \; .
\label{eq:S-partial-width-f} 
\end{equation}

 The  partial decay widths of the CP-odd boson $A$ are given by
\begin{align}
\Gamma(A \rightarrow \bar{f} f) 
 & = N_c^f  \frac{m_A}{8\pi} \left( \frac{m_f}{v}\right)^2 |g_{Aff}|^2  \left( 1- \frac{4 m_f^2}{m_A^2} \right)^{1/2} \theta(m_A-2m_f) \; ,  \label{eq:A-partial-width-1} \\
\Gamma(A \rightarrow g g) & = \frac{\alpha_s^2}{2  \pi^3 m_A}\left|\sum_{q\equiv\text{quarks}} \frac{m_q^2}{v} g_{Aqq} f \left( \frac{4 m_q^2}{m_A^2}\right)   \right|^2 \,,  \label{eq:A-partial-width-2}  
\end{align}
with the couplings $g_{A\tau\tau} =g_{A\mu\mu}=\tan\beta$, $|g_{Aqq}|=1/\tan\beta$ in the consideration of the type-X Yukawa interactions. In the present case, because we take into account the large $\tan\beta (\sim 35)$ and $m_A =15 \sim 20$~GeV (see later discussions), we therefore have $\text{Br}(A\to \tau \tau) \simeq 1$, $\text{Br}(A\to \mu \mu) \simeq \text{Br}(A\to \tau \tau)\times (m_\mu /m_\tau)^2 \simeq 0.0035$, and neglect $A\to \bar{q} q$ and $A\to gg$ due to the $1/\tan^2\beta$ suppression in the decay rates.

\section{Use of the Parameters under the Experimental and Theoretical Constraints}\label{sec:ETconstraints}
% on the LN2HDM Parameter Space}
%\section{Theoretical and Experimental constraints on the LN2HDM parameter space}

For the CP-conserving LN2HDM, we adopt the observed Higgs resonance as one of the CP-even scalars: $h$ with mass $m_h = 125.09$~GeV \cite{Aad:2015zhl}, $v^2 \equiv (\sqrt{2} G_F)^{-1}\simeq (246~\text{GeV})^2$. Compared with the SM, the interactions contain 11 more independent parameters. We take the following remaining parameters as inputs:
\begin{eqnarray}
&&  g_X, \quad  \tan\beta, \quad \beta- \alpha  \,, \quad  \theta, \quad \delta \,, \nonumber\\
&& m_X, \quad m_S, \quad  m_H, \quad  m_A, \quad  m_{H^\pm}, \quad  M^2 \equiv m_{12}^2/(\sin\beta \cos\beta)  \,.
\end{eqnarray}
 In this parametrization, the tree-level couplings of the Higgs bosons to SM particles are functions of  $\tan\beta$, $\beta-\alpha$, $\theta$, and $\delta$.
In the following,  we will experimentally and theoretically constrain the parameters relevant to the two-Higgs doublets and Yukawa sectors.

\subsection{Experimental considerations}

In the present paper, we consider that the GC gamma excess is mostly due to the two-step  cascade annihilation of the dark matter into the final state $\tau$'s, for which the main process is schematically shown in  Fig.~\ref{fig:cascade-tau}, where the shaded region denotes the interactions, depicted in Fig.~\ref{fig:vdm_ann}, and is relevant to the DM annihilation cross section.  As shown in the following GC gamma-ray excess study that $m_A\sim 10-20~\text{GeV}<m_h/2$, we therefore need to consider the constraint on $\text{Br}(h\to AA)$, which is proportional to the square of $\lambda_{hAA}$. 
The magnitude of  $\lambda_{hAA}$ can be constrained from the measurement of $\text{Br}(h\to AA\to 4 \tau) \simeq \text{Br}(h\to AA)$, of which the current upper bound \cite{Khachatryan:2017mnf,Curtin:2013fra}  is about 0.2-0.4 for $8 \leq m_A \leq 30\, \text{GeV}$, resulting in $|\lambda_{hAA}|<1.95\times 10^{-2}$. Following Ref.~\cite{Abe:2015oca}, we will take $\lambda_{hAA}=0$,   i.e., 
\begin{align}
 \big(2M^2 -2 m_A^2 -m_h^2 \big) s_{\beta-\alpha} = (M^2 -m_h^2) \left(t_\beta - \frac{1}{t_\beta}\right) c_{\beta-\alpha}
\,,
\label{eq:hAAcondition}
\end{align}
from which, under the conditions of $ t_\beta \gg 1, m_A^2 / M^2 \ll 1$, $m_h^2 /M^2 \ll 1$ and $s_{\beta-\alpha}\to 1$, one can obtain the following approximations,
\begin{align}
\sin(\beta-\alpha)&\simeq 1-\frac{2}{\tan^2\beta}\left(1+\frac{m_h^2}{M^2}-\frac{2m_A^2}{M^2}  \right), 
\label{eq:s_b-a}  \\
\cos(\beta-\alpha)&\simeq \frac{2}{\tan\beta}\left(1+\frac{m_h^2}{2 M^2}-\frac{m_A^2}{M^2}  \right).  
\label{eq:c_b-a}
\end{align}
Using the results in Eqs.~(\ref{eq:s_b-a}) and (\ref{eq:c_b-a}), the normalized Yukawa coupling of the SM Higgs to the lepton pair can be expressed as 
\begin{align}\label{eq:ghll}
\frac{g_{h\ell \ell}}{g_{h\ell \ell}^{\text{SM}}} = -\frac{s_\alpha}{c_\beta} = s_{\beta-\alpha} -t _\beta c_{\beta-\alpha}
\simeq -1 - \frac{m_{h}^2}{M^2} + 2\frac{m_A^2}{M^2} 
 -\frac{2}{t_\beta^2} \left(1+ \frac{m_h^2}{M^2} -\frac{2 m_A^2}{M^2}\right)\,.
\end{align}
In this case, the alignment limit, $s_{\beta-\alpha} \to 1$, reproduces the wrong-sign SM coupling $g_{h\ell \ell} \to -1$. Taking the combination of the ATLAS and CMS $h\to \tau\tau$ measurements, the signal strength  reads $\mu_{\tau\tau} \equiv (\sigma_h \cdot \text{BR})_{\tau\tau}^\text{obs}/ ( \sigma \cdot \text{BR})_{\tau\tau}^\text{SM}= 1.11^{+0.24}_{-0.22}$ \cite{Khachatryan:2016vau}, which is defined as the observed product of the SM-like Higgs production cross section and the decay branching ratio $h\to \tau \tau$, normalized to the corresponding SM value. The corresponding requirement for $m_A \leq 20$~GeV is $|g_{h\tau\tau}| <1.26$ at 2$\sigma$ confidence level (C.L.), such that we have $M \gtrsim 245$~GeV.

The masses of Higgs bosons can be constrained by the electroweak precision measurements. Such new physics effects, which contribute the gauge vacuum polarization at the one-loop level, can be described by three oblique parameters, $S, U$ and $T$. We adopt the definition of these parameters, originally introduced by Peskin and Takeuchi~\cite{Peskin:1990zt,Peskin:1991sw}.  
Taking the limit $s_{\beta-\alpha} \to1$, $|m_{H^\pm} -m_H| \ll m_H$ and $m_A \ll m_H$, and keeping terms linear in $\sin\theta$ and $\sin\delta$, 
we obtain three oblique parameters from that given in Ref.~\cite{Grimus:2008nb}, where a general multi-Higgs-doublet model was studied. The results are collected in Appendix~\ref{ew:oblique}. In the limit that we take, the formulas are consistent with those in the two-Higgs doublet model, i.e. the correction due to the hidden Higgs boson $S$ is negligible,
 and the results approximately read
 \begin{align}
S &\approx -\frac{1}{24 \pi}  \left( \frac{5}{3} +\frac{4 (m_{H^\pm} -m_H) }{m_H } \right)
  \simeq -0.022 - 0.002 \times \frac{\text{300 GeV}}{m_H} \frac{m_{H^\pm} -m_H}{\text{10~GeV}}  ,\nonumber\\
T &\approx \frac{1}{32 \pi^2 \alpha_{\rm em} v^2}  m_H (m_{H^\pm} -m_H)
    \simeq  0.04\times \frac{m_H}{\text{300 GeV}} \frac{m_{H^\pm} -m_H}{\text{10~GeV}},\nonumber\\
U &\approx \frac{1}{12 \pi}  \left( \frac{m_{H^\pm} -m_H }{m_H } \right)
  \simeq 0.001 \times \frac{\text{300 GeV}}{m_H} \frac{m_{H^\pm} -m_H}{\text{10~GeV}}  ,
\end{align}
where the $T$ parameter is especially sensitive to the mass splitting, $m_{H^\pm} -m_H$. 
For the values $m_H\approx 300$~GeV and $|m_{H^\pm} -m_H| \sim {\cal O}(10)$~GeV, the theoretical prediction is consistent with that from the data fit which gives \cite{pdg}
\begin{align}
S=0.05\pm 0.10, \quad T=0.08\pm0.12, \quad U=0.02\pm0.10 .
\end{align}

\subsection{Theoretical considerations}

For this model, we need to have $m_H\sim m_{H^\pm} \sim M \gg m_A$.
To satisfy the perturbative bound, we impose the absolute values of all the quartic couplings to be less than $4\pi$. We can easily make the estimate on the mass bound for the heavy Higgs as follows. From Eq.~(\ref{app:lambda5}), we have $M^2 = \lambda_5 v^2 +m_A^2 < 4\pi v^2 +m_A^2$, so that $M^2 \lesssim (873~\text{GeV})^2$ for $m_A\lesssim 40$~GeV. From Eq.~(\ref{app:lambda4}), we get $m_{H^\pm}^2=(M^2 -m_A^2 -\lambda_4 v^2)/2 \lesssim (873^2 +4\pi \times 246^2)/2~\text{GeV}^2 $, i.e., $m_{H^\pm} \lesssim 873$~GeV. For small mixing angles, $\theta$ and $\delta$, which are relevant to the present work,  the tree-level perturbative unitarity, as the case of the type-X 2HDM, gives $m_{H^\pm} \lesssim 700$~GeV  \cite{Abe:2015oca}. On the other hand, the vacuum stability and perturbativity could be broken when we consider this model at higher scale, for which, again, in the limit of small $\theta$ and $\delta$,  the related bound is the same as the type-X 2HDM, and given by $m_{H^\pm} \lesssim (400)\ 310$~GeV for the cutoff scale $\Lambda \simeq  (10)\ 100$~TeV \cite{Abe:2015oca,Hektor:2015zba}. 

Neglecting the terms with power higher than that linear in $s_\theta$ and $s_\delta$, and taking the limit $ s_{\beta-\alpha} \to 1$ and $t_{\beta} \gg 1$, we have, from Eq. (\ref{app:lambda1}), that
\begin{align}
m_H^2 -M^2 & \cong \frac{\lambda_1 v^2}{ t_\beta^2}  + ( m_H^2 -m_h^2) c_{\beta-\alpha}^2
 -2(m_H^2 -m_h^2) \frac{s_{\beta-\alpha} c_{\beta-\alpha}}{t_\beta}
 -\frac{1}{t_\beta^2} (m_H^2 c_{\beta-\alpha}^2 + m_h^2 s_{\beta-\alpha}^2 ) \nonumber\\
 & \simeq \frac{\lambda_1 v^2}{ t_\beta^2}  - \frac{2 m_H^2}{t_\beta^2} \left( \frac{m_h^2}{M^2} - \frac{2 m_A^2}{M^2} \right)
  + \frac{m_h^2}{t_\beta^2} \left(1+ \frac{2 m_h^2}{M^2} - \frac{4 m_A^2}{M^2} \right) \,,
\end{align}
where the second line is obtained by using the relations given in Eqs.~(\ref{eq:s_b-a}) and (\ref{eq:c_b-a}). 
Considering the perturbativity and vacuum stability requirements: $0<\lambda_1 <4\pi$, we can get  $m_H- M \simeq \lambda_1 v^2/(2m_H t_\beta ^2) \lesssim 1 \times {300 \text{~GeV}\over m_H} \text{~GeV}$ for a large $\tan\beta \gtrsim 35$ and $m_H\gtrsim 250$~GeV.

\section{Cosmological and Astrophysical Constraints}\label{sec:CAconstraints}

\subsection{The gamma-ray spectrum originating from the two-step cascade dark matter annihilations: determining $m_X, m_S,$ and $m_A$}

The {\em differential gamma-ray flux}, arising from the two-step cascade annihilations of the vector DM, can be expressed by
\begin{eqnarray} 
\label{eq:gammaflux}
\frac{d \Phi_\gamma}{dE} = \frac{1}{8\pi m_X^2} 
\sum _f\langle \sigma v\rangle_f 
\Bigg(\frac{dN_\gamma^f}{dE}\Bigg)_X \,  \frac{1}{\Delta\Omega}
 \underbrace{ \int_{\Delta\Omega}  \int_{\rm l.o.s.} ds \rho^2(r(s,\psi))d\Omega }_{\text{J-factor}}  ,
\end{eqnarray}
where the J-factor is the integral of the DM density squared along the line of sight (l.o.s.) and over the solid angle $\Delta\Omega$ that covers the region of interest (ROI), and $\langle \sigma v\rangle_f$ and $(dN_\gamma^f / dE)_X$ are the low-velocity averaged annihilation cross section and the gamma-ray spectrum produced per annihilation with final state $f$, respectively. For illustration, the dominant process is depicted in Fig.~\ref{fig:cascade-tau}, where the final states are $\tau$'s, which mainly arise from the process, $\langle \sigma v \rangle_\tau \simeq \langle \sigma v\rangle_{XX\to SS} \times \text{Br}(S\to AA) \times \text{Br} (A\to \tau \tau)$, with  $\text{Br}(S\to AA)\simeq 1$ and $\text{Br} (A\to \tau \tau)\simeq 1$. Following the method given in Ref.~\cite{Elor:2015tva}, we can perform two-step Lorentz boosts to transform the gamma-ray spectrum given in the $A$ boson rest frame, $(dN_\gamma^\tau / dE)_A$, to the $X X$ center of mass (CM) frame; we first boost the spectrum to the $S$ rest frame and then to the  CM frame of the $X X$ pair.
For $(dN_\gamma^\tau / dE)_A$, we will use the PPPC4DMID result \cite{Cirelli:2010xx,Ciafaloni:2010ti}, which was generated by using PYTHIA 8.1 \cite{Sjostrand:2007gs}. Thus, $(dN_\gamma^\tau / dx_2)_X =m_X (dN_\gamma^\tau / dE)_X$ can be written as 
\begin{equation}
\bigg(\frac{d N_\gamma^\tau}{dx_2}  \bigg)_X= 
4 \int^{t_{\rm 2, max}}_{t_{\rm 2, min}} \frac{d x_1}{x_1 \sqrt{1-\epsilon_2^2}} 
\int^{t_{\rm 1, max}}_{t_{\rm 1, min}} \frac{d x_0}{x_0 \sqrt{1-\epsilon_1^2}} 
\bigg( \frac{d N_\gamma^\tau}{dx_0} \bigg)_A \,,
\end{equation}
where
 \begin{align}
 \epsilon_2 & =\frac{m_S}{m_X}, \quad \epsilon_1 = \frac{2m_A}{m_S}, \\
 0\leq x_2 & \leq \frac{1}{4} \Big(1+\sqrt{1-\epsilon_1^2} \Big) \Big(1+\sqrt{1-\epsilon_2^2}\Big), \quad x_2 =\frac{E}{m_X},  
 \quad  x_1 = \frac{2E_1}{m_S},  \quad x_0 = \frac{2E_0}{m_A}, 
\\
 t_{\rm 1, max} &= {\rm min} \bigg[1, \frac{2 x_1}{\epsilon_1^2} \Big(1+\sqrt{1-\epsilon_1^2} \Big) \bigg], \quad
  t_{\rm 1, min} = \frac{2 x_1}{\epsilon_1^2} \Big(1-\sqrt{1-\epsilon_1^2}\Big), \\
 t_{\rm 2, max} &= {\rm min} \bigg[\frac{1}{2} \Big(1+\sqrt{1-\epsilon_1^2}\Big), \frac{2 x_2}{\epsilon_2^2} \Big(1+\sqrt{1-\epsilon_2^2}\Big) \bigg], \quad
  t_{\rm 2, min} = \frac{2 x_2}{\epsilon_2^2} \Big(1-\sqrt{1-\epsilon_2^2}\Big),     \end{align}
with $E$, $E_1$, and $E_0$ being the photon energies in the $XX$ CM frame, $S$ rest frame, and $A$ rest frame, respectively.

\subsection{The Galactic Center gamma-ray excess}\label{subsec:GCE}

We use the GC gamma-ray excess spectrum obtained by CCW \cite{Calore:2014xka}, who have studied Fermi-LAT data covering the energy range 300 MeV$-$500 GeV in the inner Galaxy, where the ROI extended to a $40^\circ \times 40^\circ$ square region around the GC with the inner latitude less than $2^\circ$ masked out.  The systematic uncertainties of the background have been taken into account by CCW through a large number of  Galactic diffuse emission models.

 To see whether the present vector DM model can meet the observation, we perform a goodness-of-fit test, by calculating the $\chi^2$ test statistic,
\begin{equation}
\chi^2
=
\sum_{ij \in \text{bins}}
\left[\frac{d\Phi_\gamma}{dE_i}(m_X,\langle\sigma v\rangle ) -
\left(\frac{d\Phi_\gamma}{dE_i}\right)_{\text{obs}}\right]
\cdot \Sigma_{ij}^{-1} \cdot
\left[\frac{d\Phi_\gamma}{dE_j}(m_X,\langle\sigma v\rangle ) -
\left(\frac{d\Phi_\gamma}{dE_j}\right)_{\text{obs}}\right]\,,
\end{equation}
where 24 energy bins are adopted in the range 300 MeV$-$500 GeV,  $ d\Phi_\gamma/dE_i $ and $ (d\Phi_\gamma/dE_i )_\text{obs}$ are the model-predicted and observed flux in the $i{\rm th}$ bin, respectively. Here the covariance $\Sigma_{ij}$ contains  the uncorrelated statistical error, and correlated uncertainties, of which the latter is composed of the empirical model systematics and residual systematics.
Although CCW performed the analysis using  the older Fermi dataset, however, it was shown in Ref.~\cite{Linden:2016rcf} that the results have very little changes between Fermi Pass 7 and newer Pass 8 data. This appreciable difference at low energies might be due to the modeling for the point sources in various datasets \cite{TheFermi-LAT:2017vmf,Linden:2016rcf}. On the other hand, it is interesting to note that the central values of the low energy spectrum given by \cite{TheFermi-LAT:2017vmf} seem to be smaller than that obtained by CCW. If so, the best-fit DM mass will become larger compared with the present result.

Two physical parameters,  $\langle\sigma v\rangle$  and $m_X$, can thus be obtained from the fit. The value of  $\langle\sigma v\rangle$ is sensitive to the form of the Galactic DM density distribution, for which we use a generalized Navarro-Frenk-White (gNFW) halo profile  \cite{Navarro:1995iw,Navarro:1996gj},
 \begin{equation}
 \label{eq:gNFW}
 \rho(r)=\displaystyle \rho_{\odot} \left(\frac{r}{r_\odot}\right)^{-\gamma} \left(\frac{1+r/r_s}{1+r_\odot /r_s}\right)^{\gamma-3} \,,
 \end{equation}
 where  the scale radius $r_s=20$~kpc, $r$ is the distance to the GC, $-\gamma$ is the inner log slope of the halo density near the GC, and $\rho_\odot $ is the local DM density at $r_\odot=8.5$~kpc, which is the radial distance of the Sun from the GC.  We will take $\gamma=1.2$ and $\rho_\odot=0.357$~GeV as the canonical values.
However,  the uncertainties about the local dark matter density and the halo distribution near the GC remains large.  The resulting annihilation cross section in the fit due to  the variation of $\gamma \in [1.1,1.3]$ and $\rho_\odot \in [0.2, 0.6]$~GeV will be discussed later.

\subsection{The constraint from dwarf spheroidal observations}\label{subsec:dSph} 

In the present analysis, we will use the combined gamma-ray data of 28 confirmed and 17 candidate dwarf spheroidal galaxies (dSphs), recently reported by the Fermi-LAT and DES Collaborations \cite{Fermi-LAT:2016uux,FermiLatDesData}.  Compared with the earlier Fermi-Lat analysis \cite{Ackermann:2015zua}, where some point-like sources were modeled as extended ones, a consistent analysis across 45 targets were presented in Ref.~\cite{Fermi-LAT:2016uux}, and a limit weaker by a factor of $\sim 1.5$ were obtained in the low DM mass region ($\lesssim 70$~GeV).  Because there is no gamma-ray signal detected so far from this kind of objects, a bound on the DM annihilation can thus be set.  

We perform a combined likelihood analysis of 45 confirmed and candidate dSphs with 6 years of Fermi-LAT Pass 8 data in the energy range from 500~MeV to 500~GeV. The log-likelihood test statistic
 (TS) is given by
\begin{equation}
\text{TS}=-2 \sum_{k=1}^{N_{\rm dSph}} 
\ln \left[ \frac{ \mathcal{L}_{k} ( \langle \sigma v\rangle,  \hat{J}_k ; m_X | {\rm data}) }{ \mathcal{L}_{k} ( \overline{\langle \sigma v\rangle}, \bar{J}_k ; m_X | {\rm data}) }\right]  \,,
\end{equation}
with $N_{\rm dSph}=45$ and the profile likelihood of an individual target $k$,
\begin{eqnarray}
 \mathcal{L}_{k} ( \langle \sigma v\rangle,  J_k ; m_X | {\rm data}) 
 =
 \left(
 \sum_{i=1}^{N_{\rm bin}}  
 \mathcal{L}_{ki} ( \langle \sigma v\rangle,  J_k ; m_X | {\rm data}) 
 \right)
 \cdot  \mathcal{L}_{J_k}  \,,
 \end{eqnarray}
where $N_{\rm bin}=24$ are the numbers of bins, $ \mathcal{L}_{ki}$ is  the $i$-th binned likelihood of the target $k$~\cite{FermiLatDesData},  and
 the J-factor likelihood for a target $k$ is modeled by a normal distribution \cite{Lindholm:2015thesis},
\begin{equation}\label{eq:L_jfactor}
\begin{aligned}
  \mathcal{L}_{J_k}  = \frac{1}{\ln(10) J_{{\rm o},k}
    \sqrt{2 \pi} \sigma_k} e^{-\left(\log_{10} J_k- \log_{10} J_{{\rm o},k} \right)^2/(2\sigma_k^2)} \,.
\end{aligned}
\end{equation}
Here, $J_k$ is the expected  J-factor of a target $k$, while the nominal value $ J_{{\rm o},k}$ together with its error $\sigma_k$ is the spectroscopically determined value when possible, or the predicted one from the distance scaling relationship with an uncertainty of 0.6 dex, otherwise \cite{Fermi-LAT:2016uux}.
 For a given $m_X$, $\overline{\langle \sigma v\rangle}$ and $\bar{J}_k$ are the maximum likelihood estimators (MLEs), which maximize $\sum_{k=1}^{k= N_{\rm dSph}} \ln\mathcal{L}_{k}$. When $\langle \sigma v\rangle$ is fixed to a given value,  $\hat{J}_k$ are the conditional MLEs of the nuisance parameters. We can obtain the 95\% confidence level (C.L.) limit on low-velocity annihilation cross section $\langle \sigma v\rangle$ from the null measurement by increasing its value from $\overline{\langle \sigma v\rangle}$ until $\text{TS}=2.71$.

\begin{figure}[t!]
  \begin{center}
\includegraphics[width=0.5\textwidth]{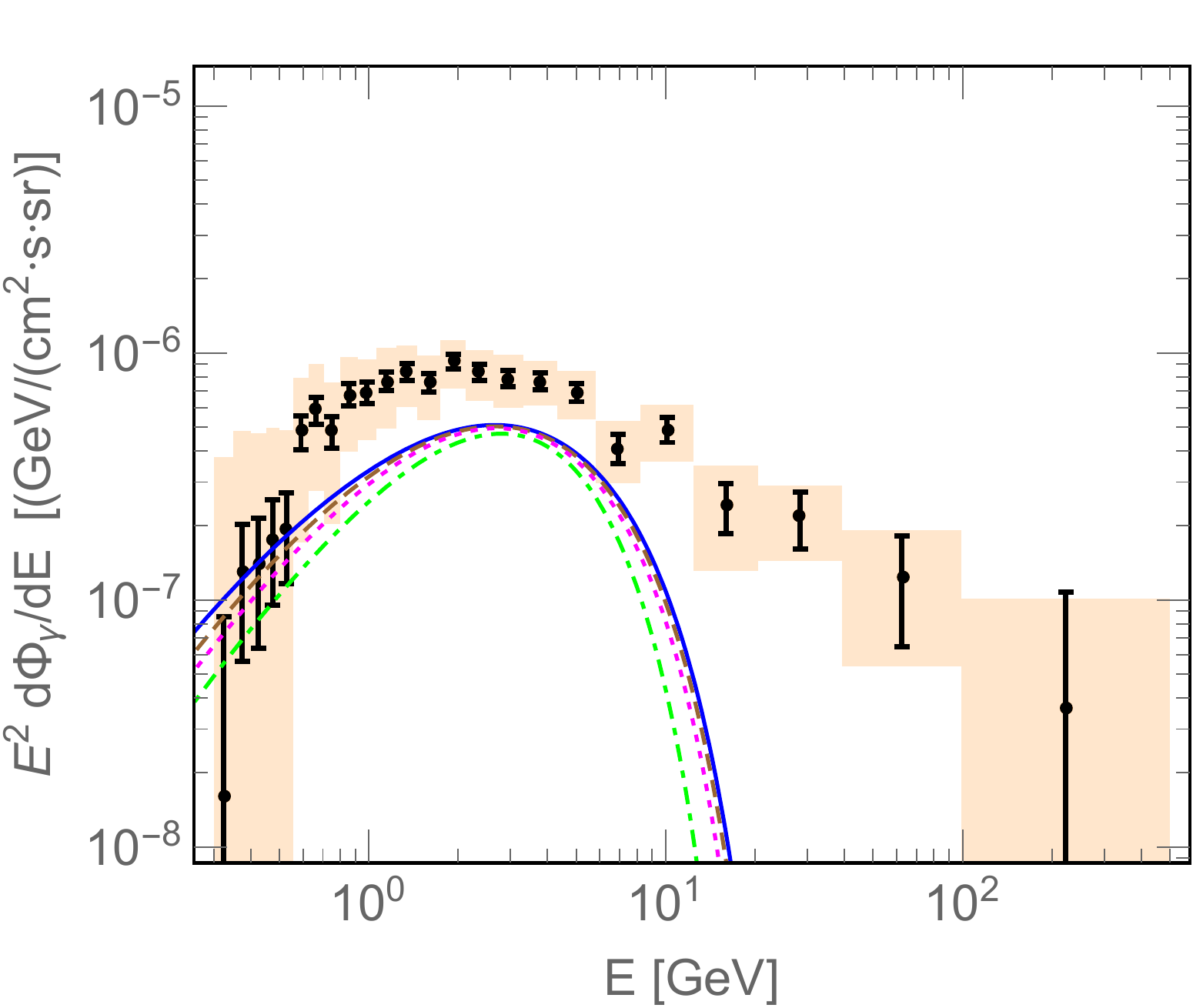}\caption{Comparison of the spectrum of the Galactic Center gamma-ray excess \cite{Calore:2014xka}  vs. the best-fit results for DM annihilating through a two-step cascade to the final state $\tau$'s, where, for the former, the statistical and systematical errors are shown by error bars and orange rectangles, respectively, while for the latter, the solid (blue),  dashed (brown),  dotted (magenta), and  dotdashed (green) curves are for the cases of $(m_S, m_A)=$  $(0.5 m_X, 0.2 m_X)$, $(0.7 m_X, 0.3 m_X)$, $(0.9 m_X, 0.36 m_X)$, and $(0.95 m_X, 0.45 m_X)$, respectively, with the corresponding $p$-values 0.22, 0.22, 0.23, 0.12. The corresponding good fit results, featuring by the $p$-values, are shown in the plane of $m_X$ and low-velocity annihilation cross section in Fig.~\ref{fig:gcmxsv}. 
}
\label{fig:gcfit}
\end{center}
\end{figure}

 \begin{figure}[t!]
  \begin{center}
\includegraphics[width=0.34\textwidth]{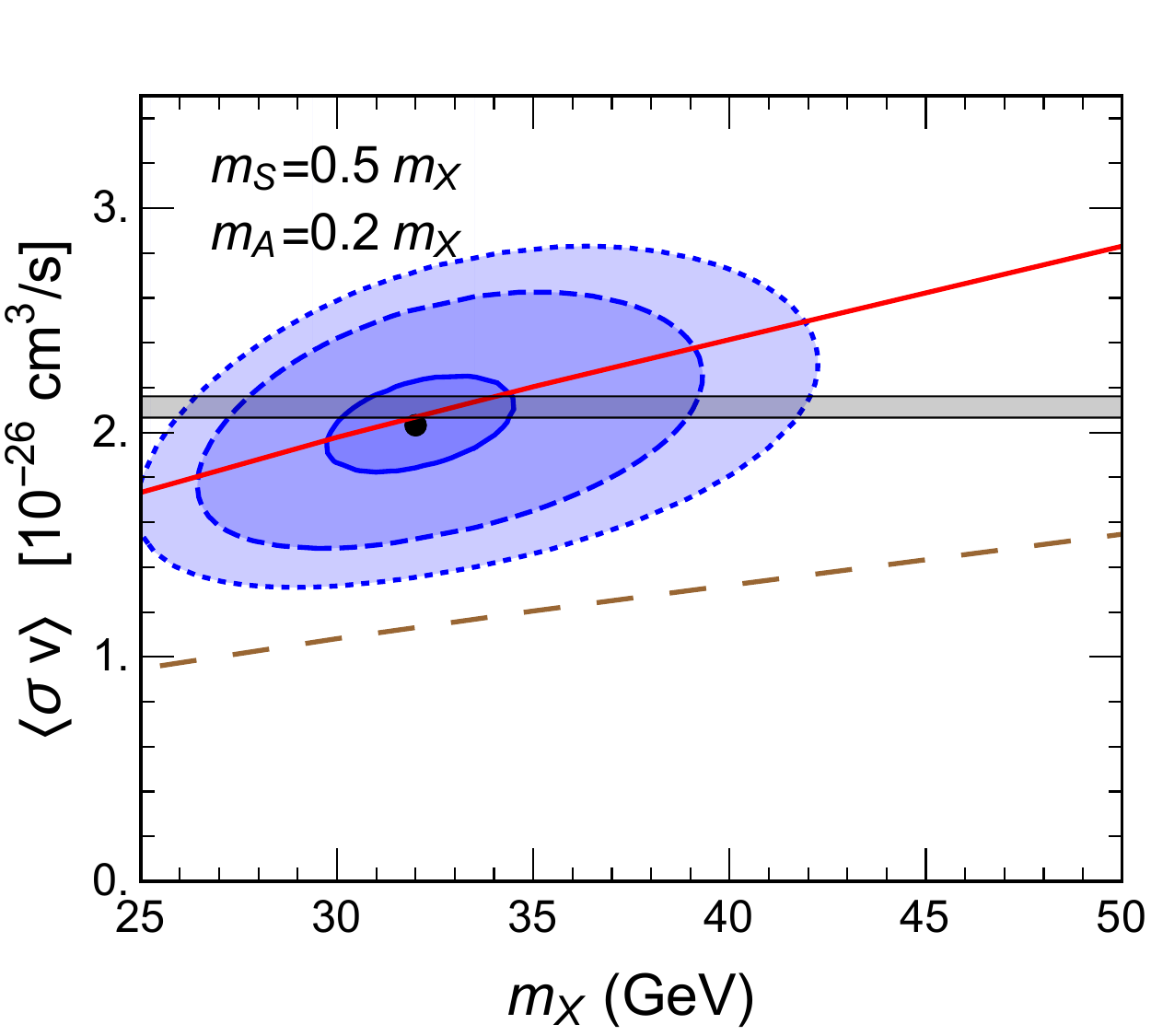}\hskip0.9cm
\includegraphics[width=0.34\textwidth]{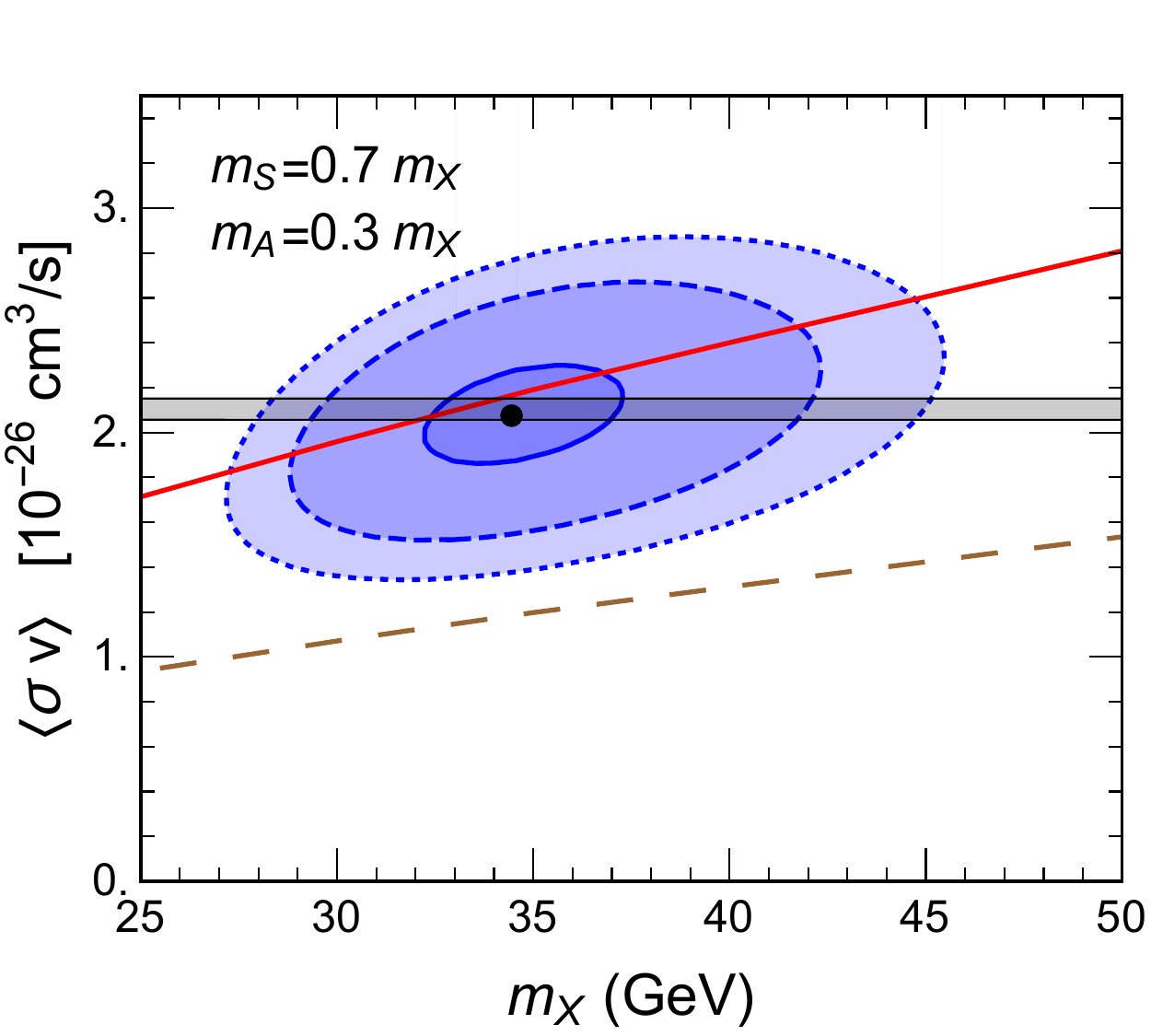} 
\\
\includegraphics[width=0.34\textwidth]{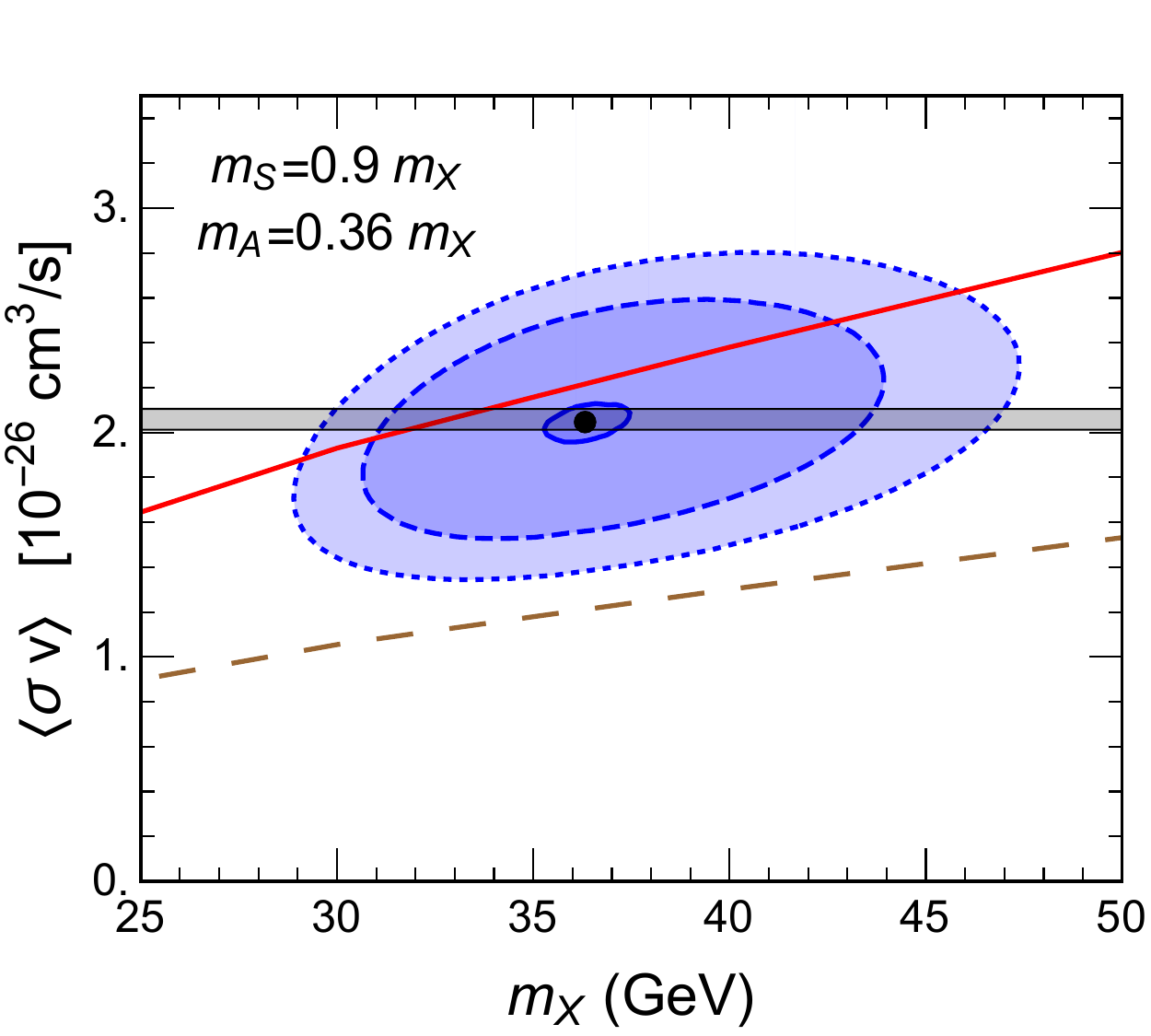}\hskip0.9cm
\includegraphics[width=0.34\textwidth]{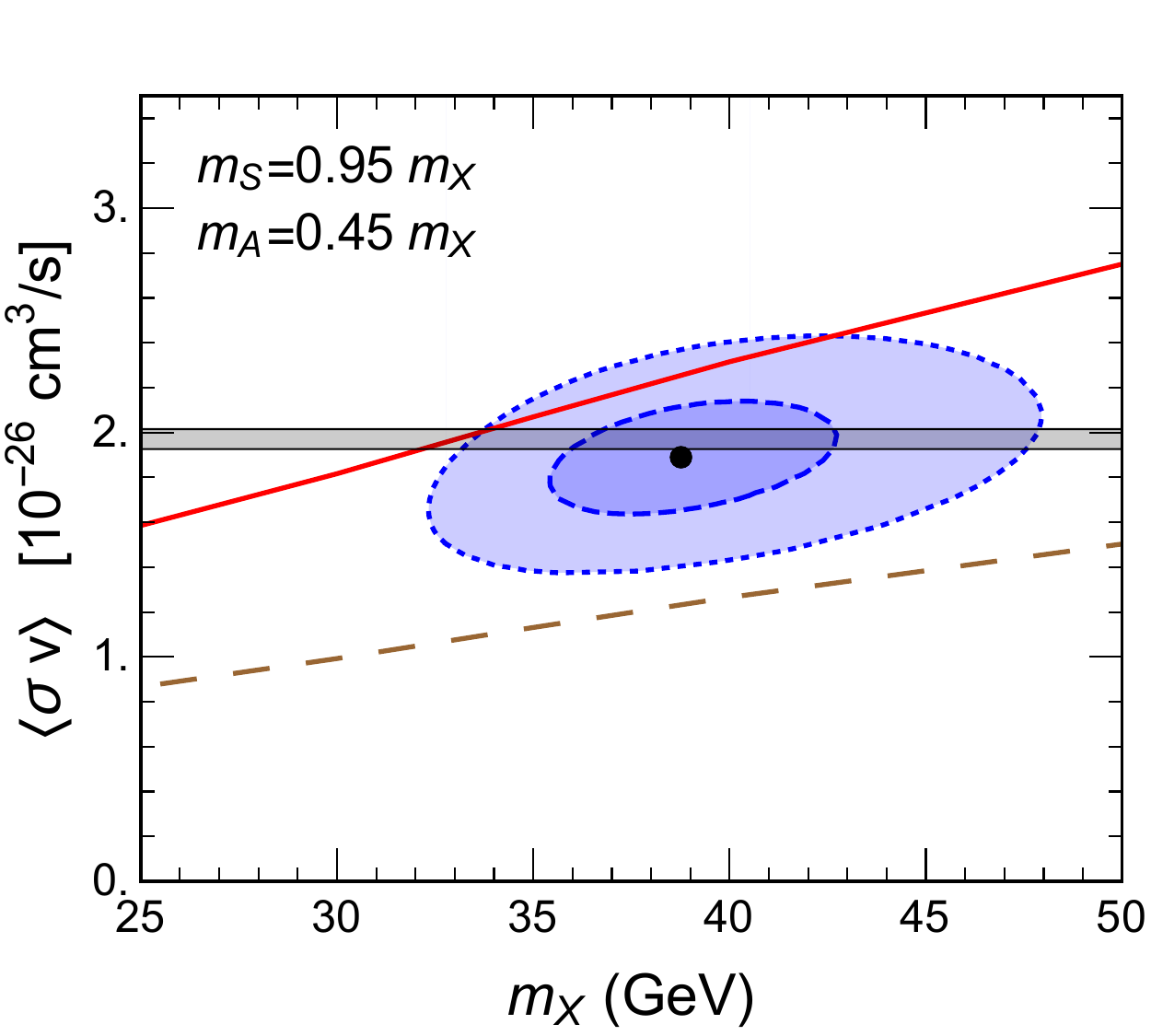} 
\caption{Allowed parameter regions in the dark matter mass $m_X$ and low-velocity annihilation cross section $\langle \sigma v\rangle$ plane.  The GC fitted regions, shown in the blue color with solid, dashed and dotted boundaries, satisfy $p$-value $\geq 0.2, 0.1, 0.05$, respectively.  
The black dot marks the GC best-fit point. In conventional WIMP scenario, the thermal relic density can be accounted for by the narrowed grey region. All GC results refer to $\rho_\odot = \text{0.357 GeV/cm}^3$ and  $\gamma=1.2$.
The 95\% C.L. upper bound and projected limit from  Fermi-LAT observations of dSphs  are denoted as the solid red and long-dashed brown lines, respectively. }
\label{fig:gcmxsv}
\end{center}
\end{figure}

\begin{figure}[t!]
\begin{center}
\includegraphics[width=0.39\textwidth]{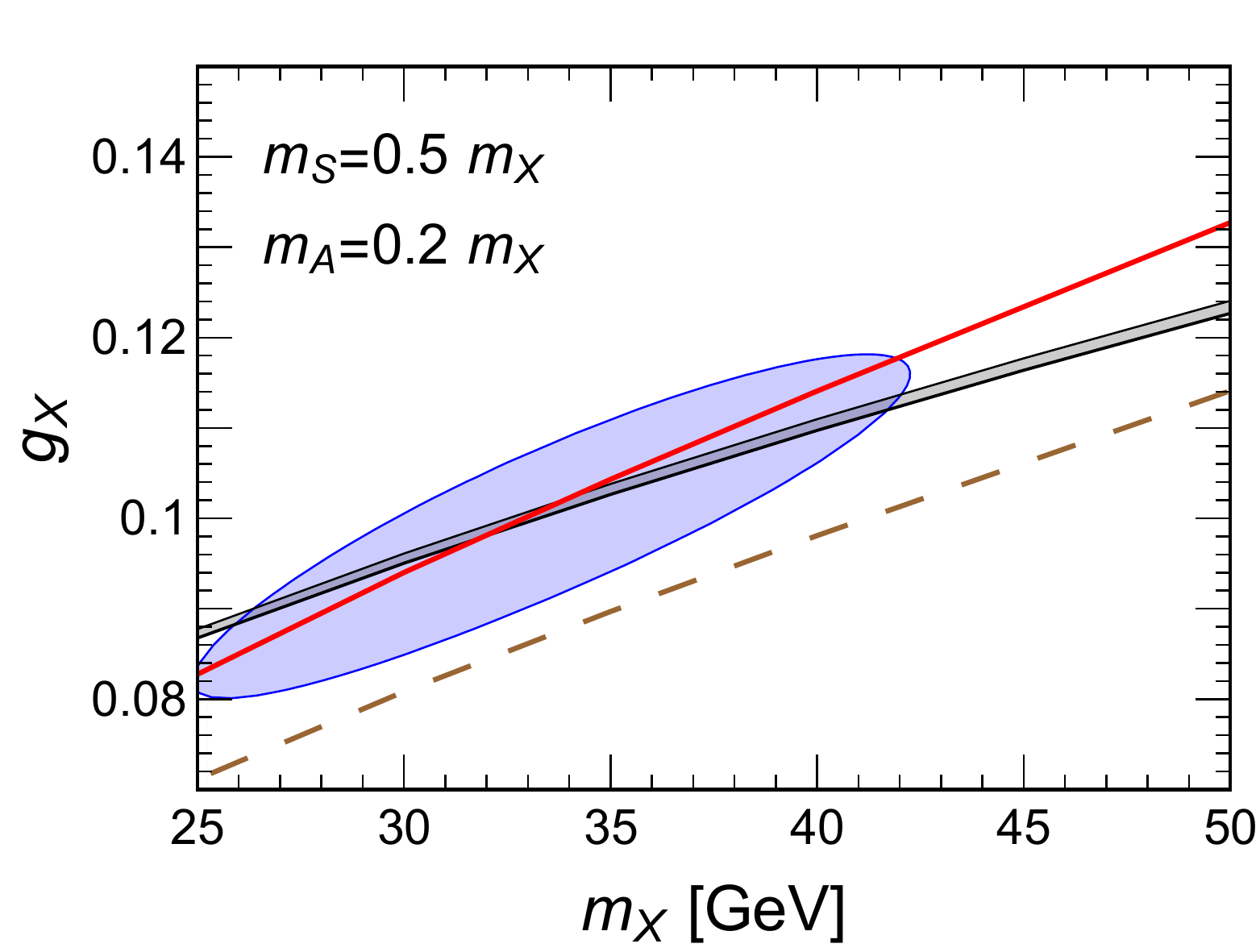}\hskip0.8cm
\includegraphics[width=0.39\textwidth]{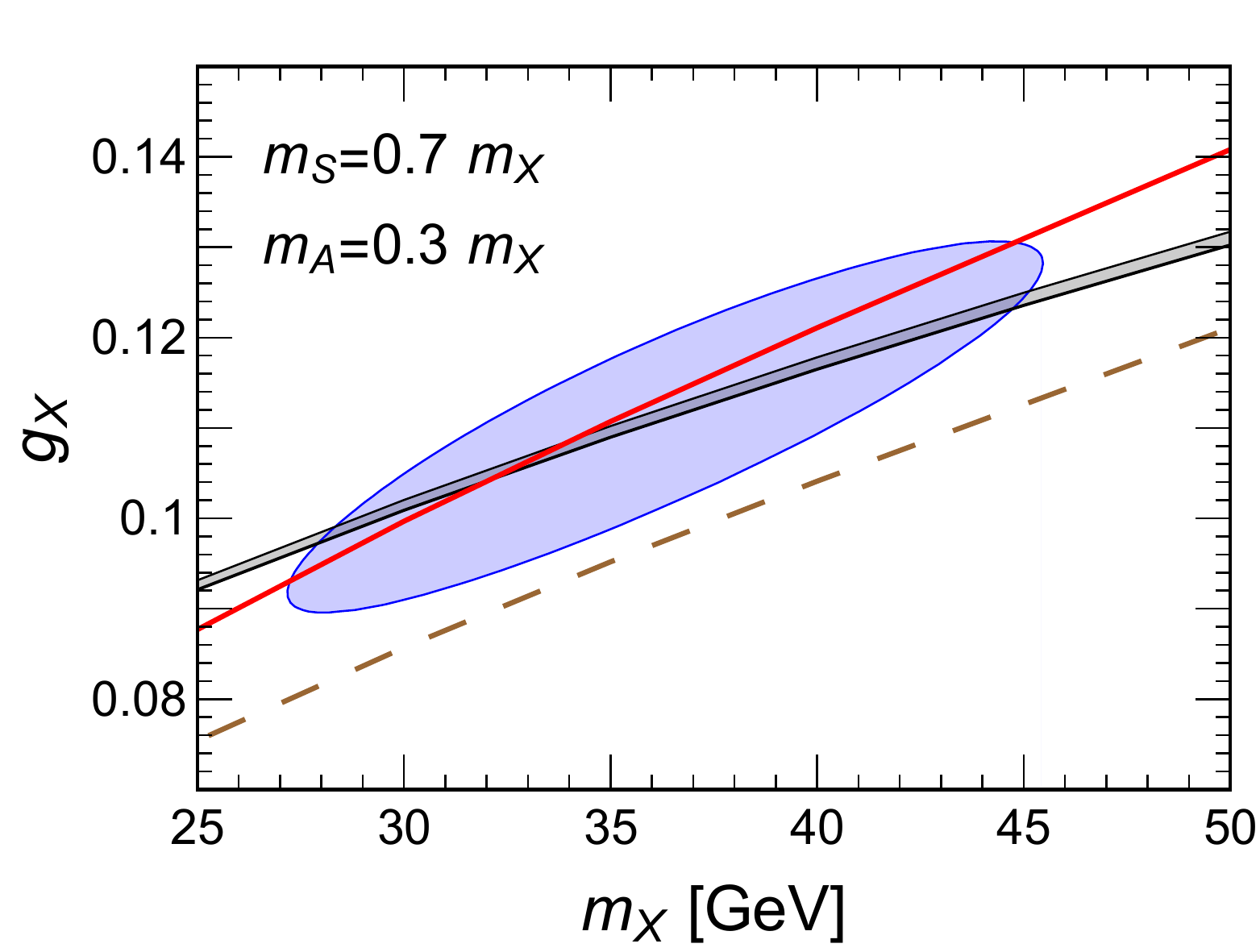} \\
\includegraphics[width=0.39\textwidth]{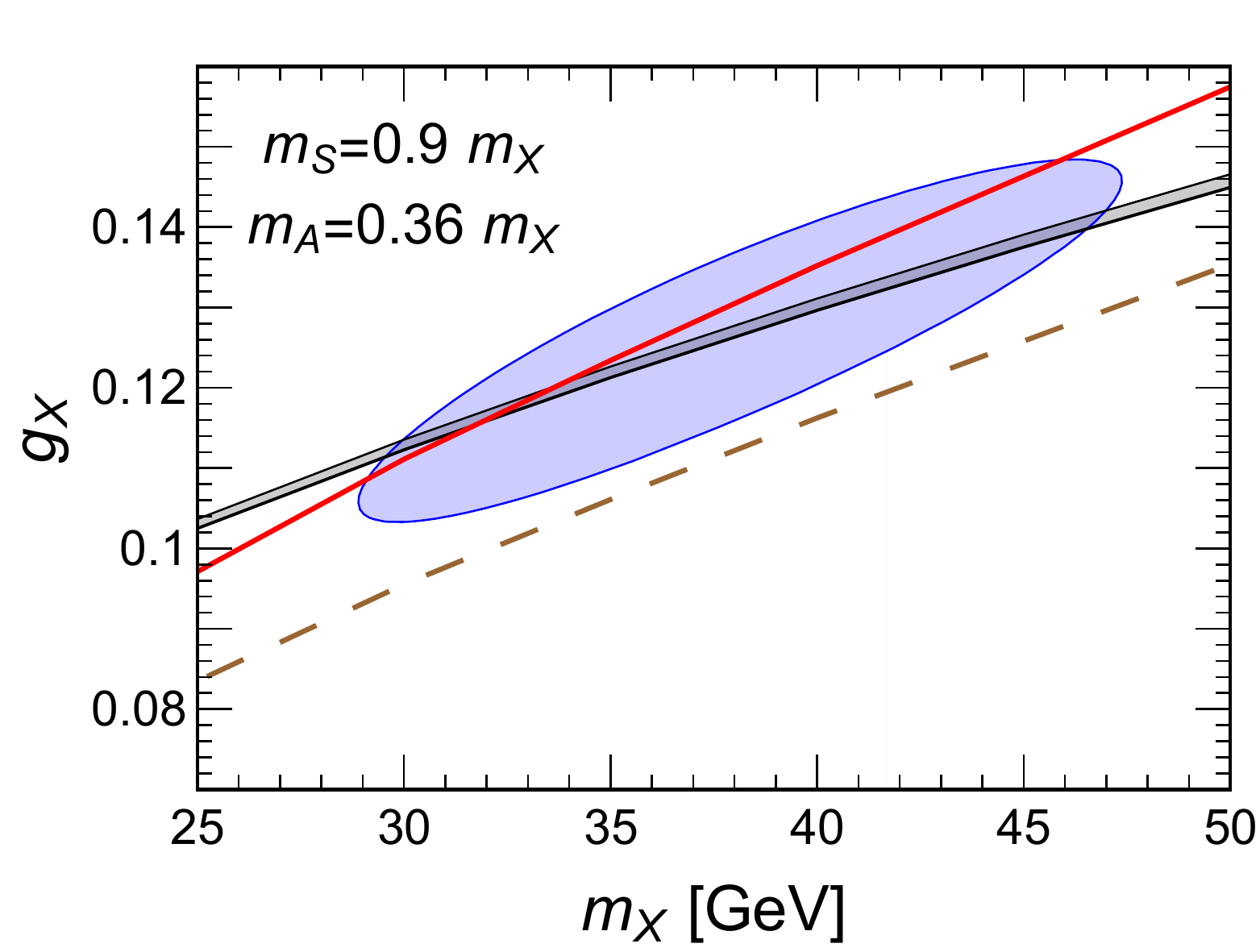}\hskip0.8cm
\includegraphics[width=0.39\textwidth]{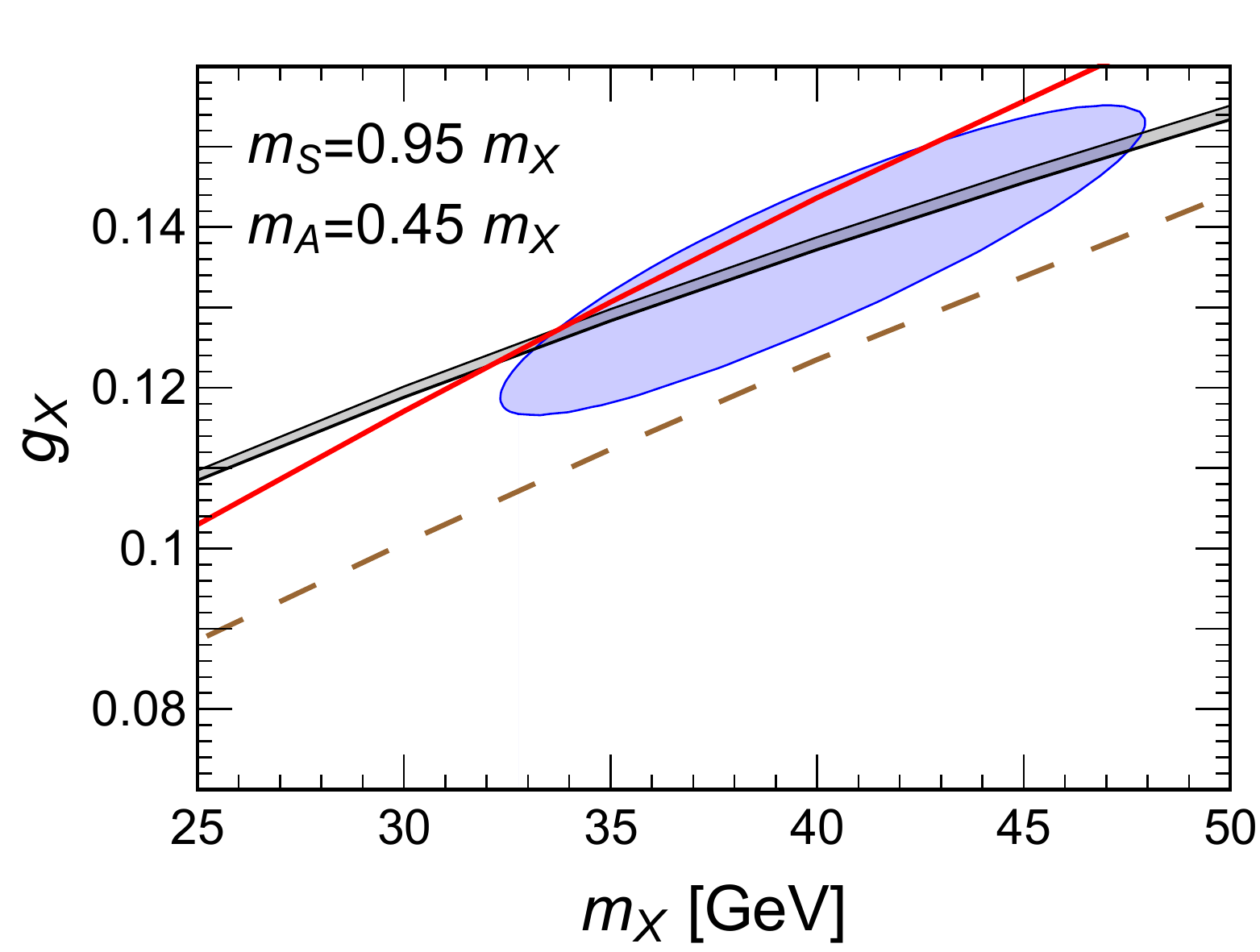} 
\caption{Same as Fig.~\ref{fig:gcmxsv}, but in the $(m_X, g_X)$ plane, where the blue shaded region delineated with the blue line provides a good fit  to the GC gamma-ray data  with $p$-value $\geq 0.05$.}
\label{fig:gxmx}
\end{center}
\end{figure}

\subsection{Results}

In Fig.~\ref{fig:cascade-tau}, we have depicted the two-step cascade DM annihilation process into the final state $\tau$'s, which is the dominant mechanism to explain the GC gamma-ray excess in the present study.
The DM annihilation diagrams, relevant to the indirect search and also to the relic abundance, are shown in Fig.~\ref{fig:vdm_ann};  the resulting cross sections and related discussions are collected in Appendix~\ref{app:annXS}.

The DM two-step cascade annihilation process, increasing the final gamma-ray multiplicity and therefore resulting in a broader gamma-ray spectrum,  provides a better fit to the GC data compared with that obtained from the DM annihilation directly into the tau pair. 
For illustration, in Fig.~\ref{fig:gcfit}, we show the GC gamma-ray energy spectrum \cite{Calore:2014xka} compared our the best-fit model prediction.
 The corresponding GC fitted regions, featuring by the $p$-values, are shown in Fig.~\ref{fig:gcmxsv} on the plane of $m_X$ and low-velocity annihilation cross section for four cases of $(m_S, m_A)=$  $(0.5 m_X, 0.2 m_X)$, $(0.7 m_X, 0.3 m_X)$, $(0.9 m_X, 0.36 m_X)$, and $(0.95 m_X, 0.45 m_X)$, where
$\rho_\odot=0.357~\text{GeV/cm}^3$ and $\gamma=1.2$ have been adopted, so that the best GC fit is consistent with the WIMP relic abundance. 
The gauge coupling constant $g_X$ in this model can thus be determined, and given as a function of $m_X$ in Fig.~\ref{fig:gxmx}.
Note that, because the decay width of $S$  is much less than $m_S$  and, on the other hand, the annihilation $XX\to SS$ via an $s$-channel $h$ or $H$ exchange is highly suppressed, these two effects can be negligible. The detailed discussion will be given in Sec.~\ref{sec:thetadelta-1}.
We find that this model can provide a good fit to the GC gamma-ray excess spectrum for the regions with $m_X\sim 28 - 37 \ (25-50)$~GeV, $m_A \sim 4 - 13 \  (3.6 - 25)$~GeV and $m_X \gtrsim m_S \gtrsim 2m_A$, where $p$-value could be $\gtrsim 0.2 \ (0.05)$. 

Compared with the DM annihilation directly into the tau pair, the two-step cascades increase the final gamma-ray multiplicity by a factor of $\sim 4$, such that if the dark matter mass is still the same, the resulting energy of the GeV photon peak will be reduced by a factor $\sim 4$. Therefore, to fit the observed GeV gamma-ray excess, we need to increase $m_X$, i.e. the initial energy, by a factor $\sim 4$ in magnitude. On other hand, having the resulting changes for the final photon multiplicity and $m_X$, we thus know from Eq.~(\ref{eq:gammaflux}) that the annihilation cross section also needs to be enlarged by a factor of $\sim 4$ to fit the gamma-ray spectrum.

Figs.~\ref{fig:gcmxsv} and \ref{fig:gxmx} show the current 95\% C.L. upper bound and projected limit from the gamma-ray observations of dSphs. The projected limit approximately rescales with the square root of the data size and the square root of the number of targets \cite{Anderson:2015rox}.
Following the estimate given in Ref.~\cite{Charles:2016pgz}, we conservatively assume that  the 15-yr gamma-ray emission data can be successfully collected from the observation of 60 dSphs.
Thus, the  projected sensitivity on the $\langle\sigma v\rangle$  will be further improved by a factor of $\sim 1.83$, and, as shown in Figs.~\ref{fig:gcmxsv} and \ref{fig:gxmx}, the present model is very likely to be probed in the near future. 

In Figs.~\ref{fig:gcmxsv} and \ref{fig:gxmx}, we also show the region which is allowed by the correct DM relic abundance in the conventionally thermal WIMP scenario, for which we have rescaled the thermally averaged annihilation cross section at freeze-out temperature to its corresponding value defined at the low-velocity limit, where the effective number of degrees of freedom (DoF) $g_* \simeq85.5$ corresponding to $T\simeq 1.9$~GeV (and $x\equiv m_X/T \simeq 22$) is adopted \cite{Gondolo:1990dk,Cerdeno:2011tf}.
Note that for a too small coupling constant  $\lambda_{SAA}$, the $S$ particle cannot maintain its chemical equilibrium with the thermal bath, such that 
the dark sector particles to be out of equilibrium with the bath when $T\lesssim m_{X(S)}$. For this case, we will show in Sec.~\ref{sec:thetadelta-2} that the allowed DM annihilation cross section to have the relic abundance could be (much) larger than that in the conventional WIMP scenario.

Three remarks for the gamma-ray fit are in order here. First, for the two-step cascade process, the kinematical condition, $m_X \geq m_S \geq 2 m_A \geq 4 m_\tau$, needs to be satisfied. Second,  considering variation of the local dark matter density from 0.357 to 0.2 (or 0.6) GeV/cm$^3$ and the halo slope $\gamma$  from 1.2 to 1.1 (or 1.3), the low-velocity annihilation cross section and $g_X$ in the GC gamma-ray fit would be further raised (or lowered) by factors of 4.10 and $\sim 4.10^{1/4}$ (or 0.27 and $\sim 0.27^{1/4}$), respectively.
Third,  the uncertainties of the observed $J$ values of dSphs are subject to determination of DM mass profile which is assumed to be spherically symmetric and to have negligible binary motions \cite{Lindholm:2015thesis}.

\section{Determination of mixing angles, $\theta$ and $\delta$}\label{sec:thetadelta}

\subsection{Constraints from invisible Higgs decays and two-step cascade annihilation}\label{sec:thetadelta-1}

\begin{figure}[t!]
\begin{center}
\includegraphics[width=0.39\textwidth]{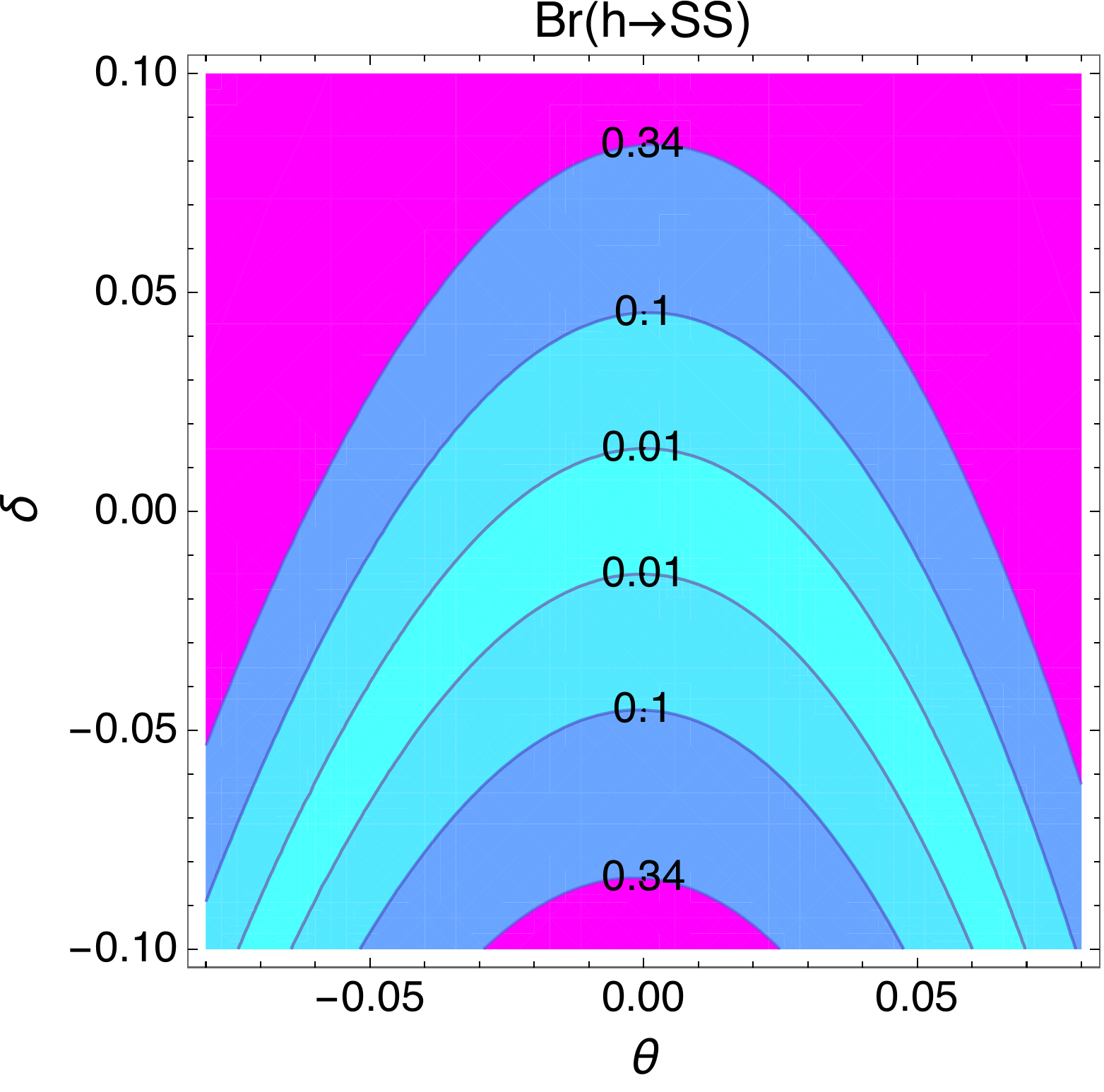}\hskip0.9cm
\includegraphics[width=0.40\textwidth]{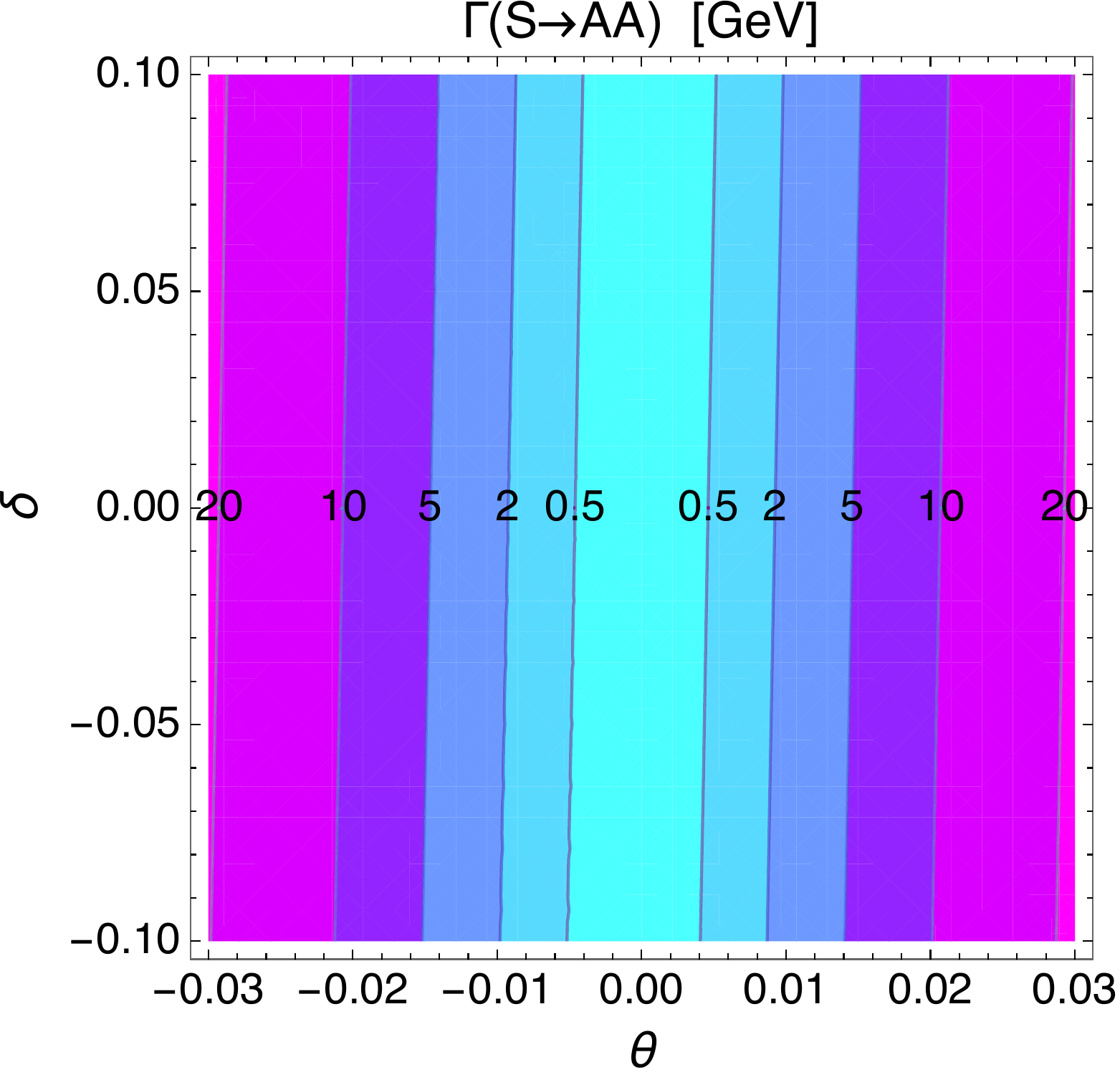}\\
\vskip0.3cm
\includegraphics[width=0.40\textwidth]{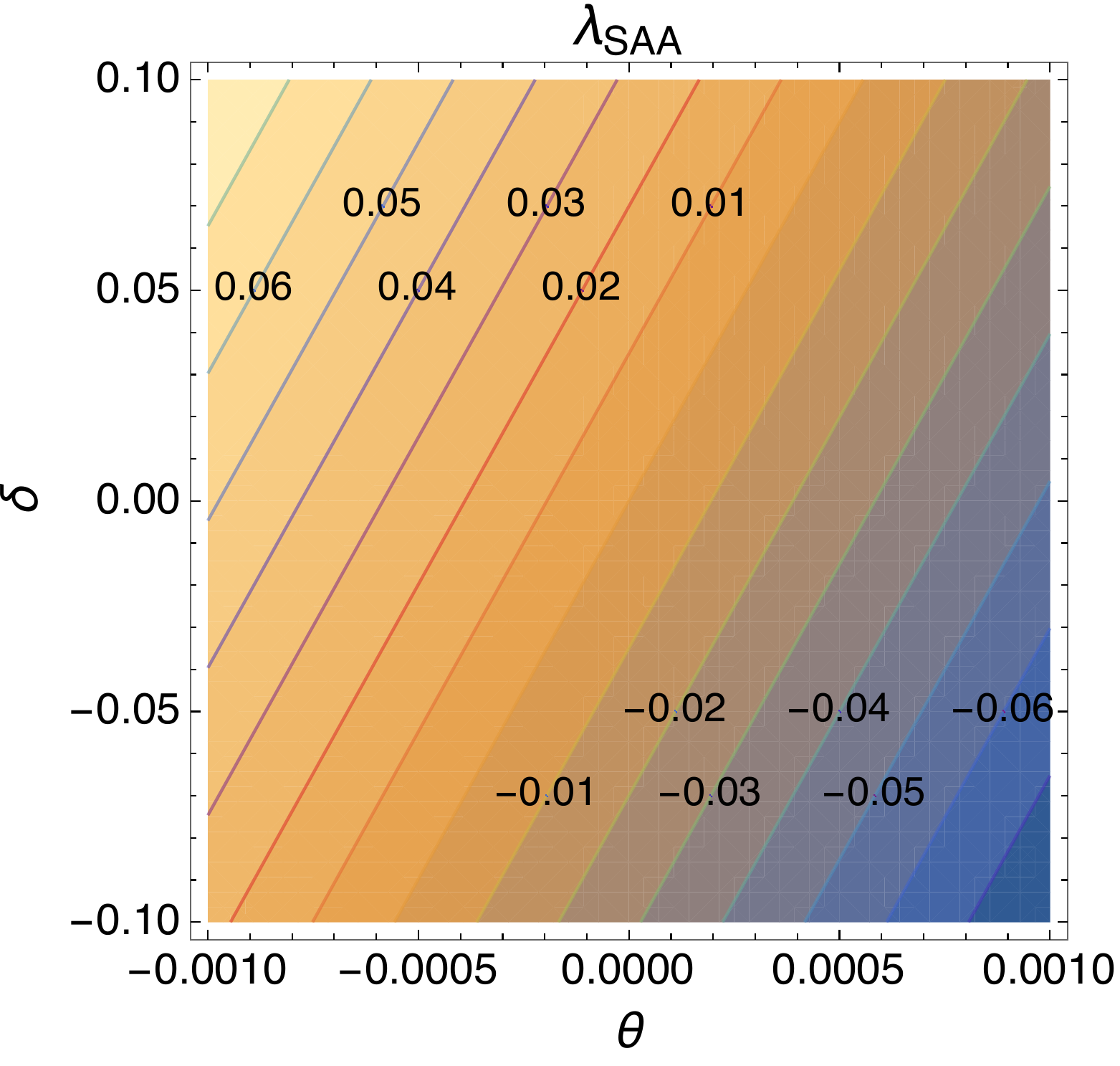} \hskip0.77cm
\includegraphics[width=0.40\textwidth]{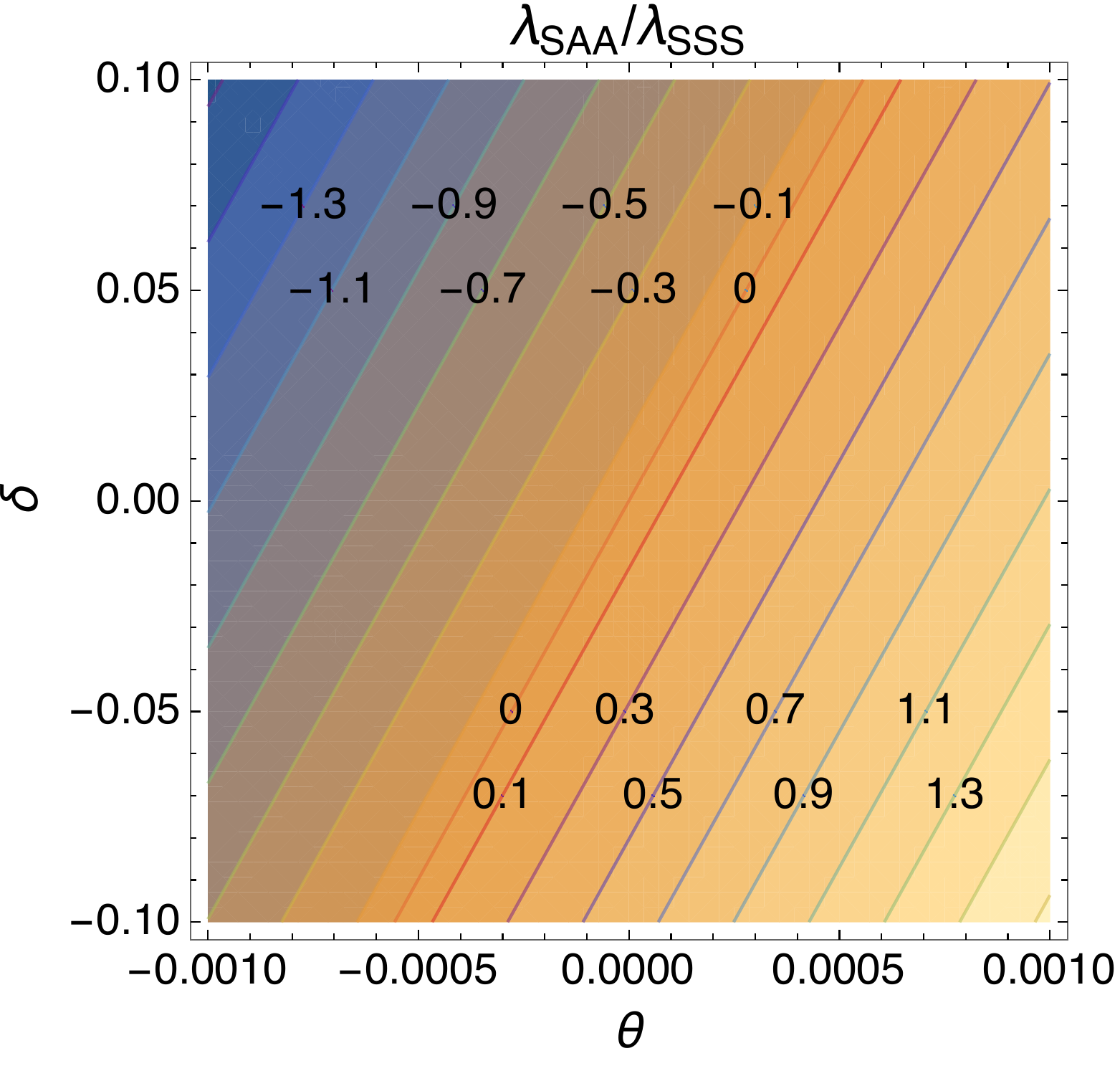}
\vskip0.3cm
\includegraphics[width=0.40\textwidth]{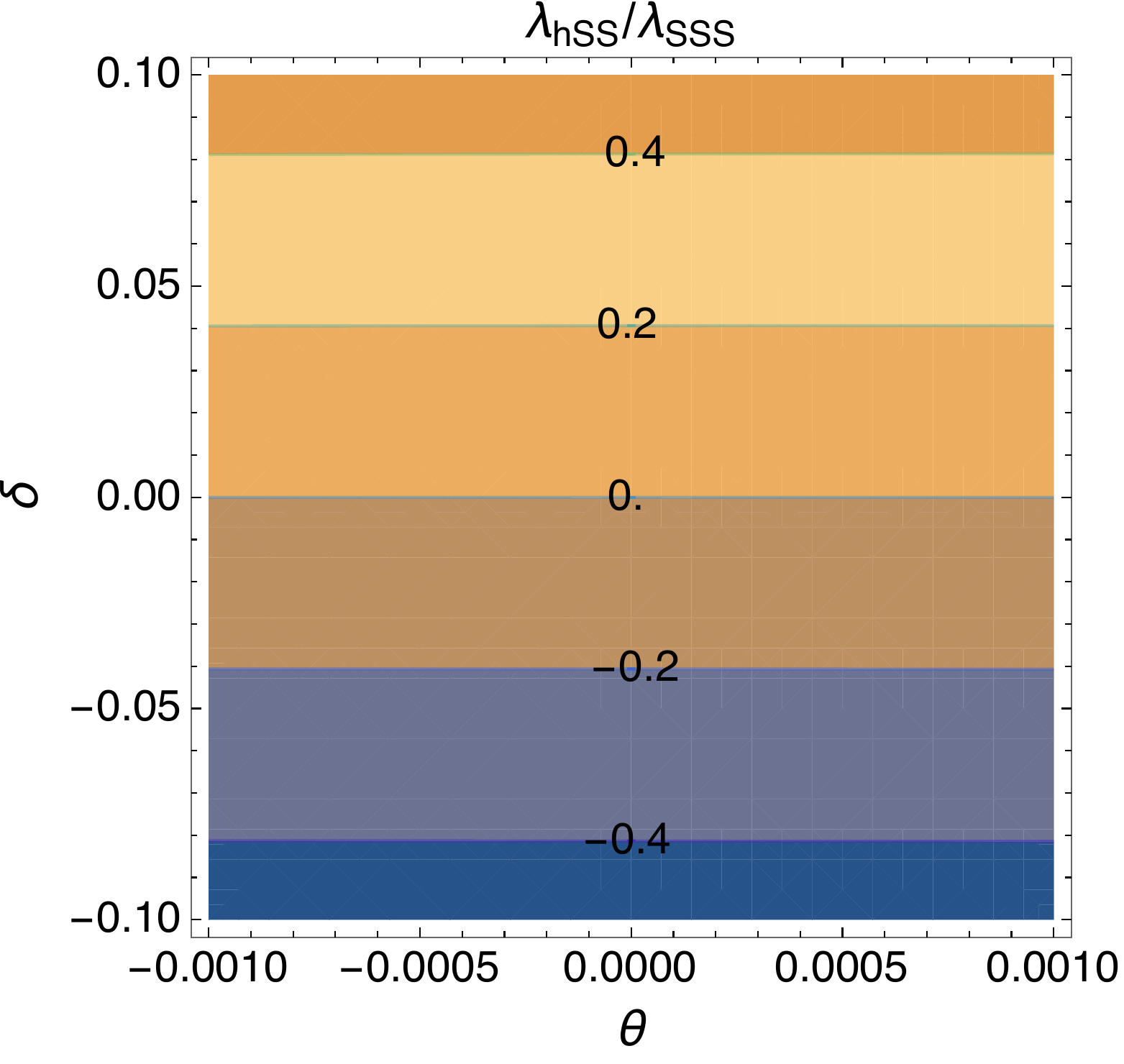} \hskip0.77cm
\includegraphics[width=0.40\textwidth]{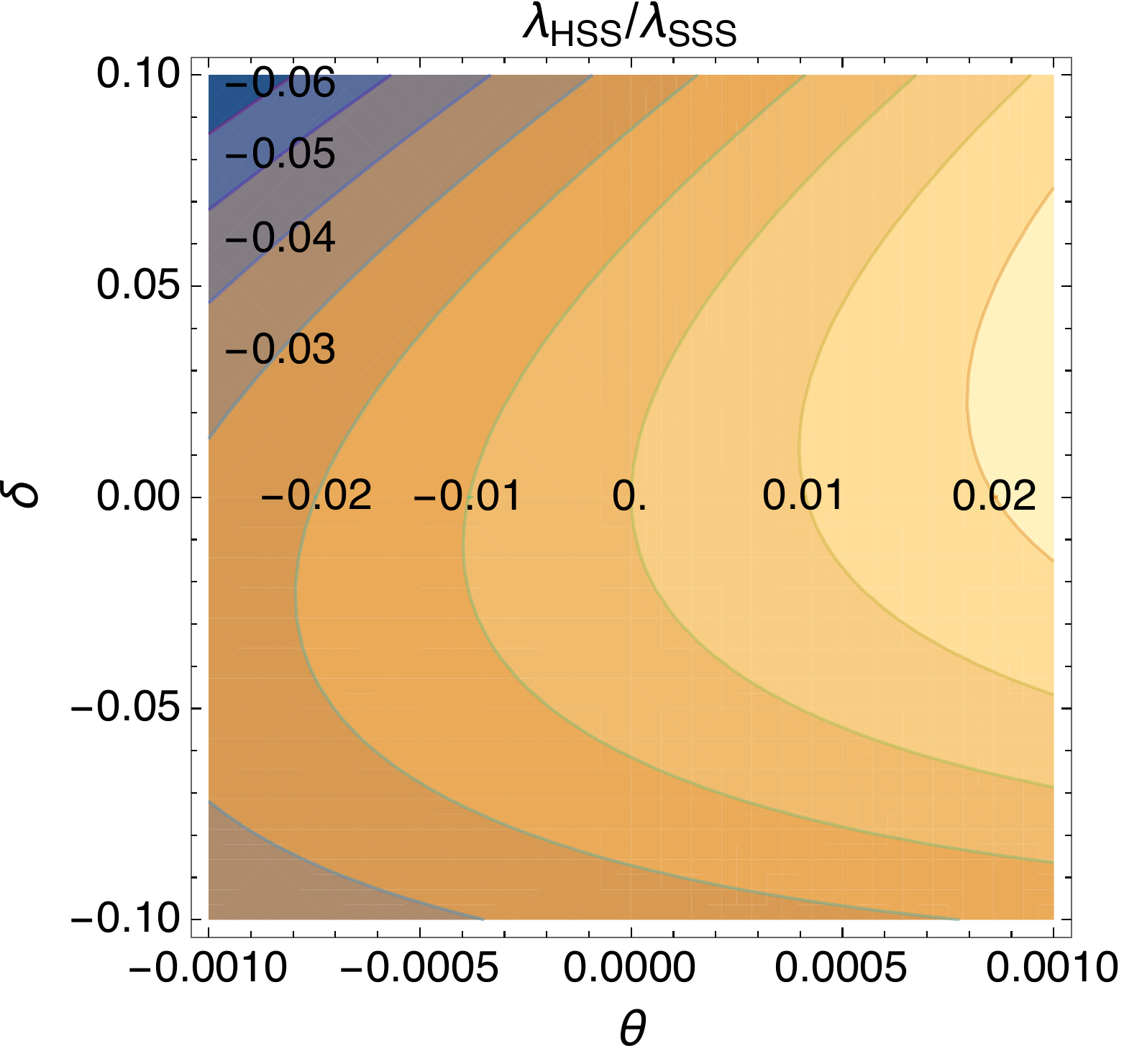}
\caption{ The contour plots on the $(\theta, \delta)$ plane for ${\rm Br}(h \to SS), \Gamma(S\to AA), \lambda_{SAA}$,  $\lambda_{SAA}/\lambda_{SSS}$, $\lambda_{hSS}/\lambda_{SSS}$, and $\lambda_{HSS}/\lambda_{SSS}$, where the parameters given in Eq.~(\ref{eq:parameters-1}) are taken.
}
\label{fig:thetadelta-1}
\end{center}
\end{figure}

Taking into account the GC gamma-ray excess which results from the two-step cascade annihilations of the vector dark matter into the final state $\tau$'s in this model, we have found that the mass of dark matter lies within $25\sim 50$~GeV, along with $m_S\lesssim m_X$ and $m_A\lesssim m_S/2$, as shown in Figs.~\ref{fig:gcmxsv} and \ref{fig:gxmx}. Therefore, because $m_h> 2 m_S $, the decay $h\to SS$ is allowed and is followed by $S\to AA$ and subsequently $A \to \tau^+ \tau^-$. This is relevant for search for the exotic Higgs decay with 8 $\tau$'s in the final state. The exotic Higgs decays with 4 $\tau$'s or other modes in the final state was discussed in Ref.~\cite{Curtin:2013fra}.  Here and in the following sections, we will take parameters,
\begin{align}\label{eq:parameters-1}
& m_X=40~\text{GeV}, m_S=35~\text{GeV}, m_A=15~\text{GeV},  \nonumber\\
& m_H=m_{H^\pm}=M=300~\text{GeV}, g_X=0.123,  \tan\beta=35, \beta-\alpha=0.062909, 
\end{align}
as a benchmark in the discussions, and we have $\lambda_{hAA}\simeq0$, $\text{Br}(h\to AA)\simeq 0$ due to the adopted values of $\tan\beta$ and $\beta-\alpha$. In Fig.~\ref{fig:thetadelta-1}, we show the contour plot for ${\rm Br}(h\to SS)$ on the $(\theta, \delta)$ plane, in comparison with a 95\% C.L. limit: ${\rm Br} (h\to \text{beyond SM})<34\%$, which was fitted with the Higgs produced via SM couplings \cite{pdg}.   

A more stringent constraint can be obtained by requiring that the two-step cascade annihilation to the final state $\tau$'s is dominant over the 
one-step cascade process described by the $s$-channel $XX\to AA$ followed by $A\to \tau^+ \tau^-$. For this requirement, we will therefore restrict $\langle \sigma v\rangle_{XX\to AA} / \langle \sigma v\rangle_{XX\to SS} $ be to less than 0.05. 
Because the $s$-channel contributes  about 15\% to the total $XX\to SS$ cross section, we thus need to have $|\lambda_{SAA}/\lambda_{SSS}| \lesssim 0.48$ and $|\lambda_{SAA}| \lesssim 0.022$ (see also Fig.~\ref{fig:thetadelta-1}), where $\lambda_{SSS} =-3 g_X m_S^2 /(m_X v) \simeq -0.046$.  Imposing the constraints required by ${\rm Br} (h\to \text{beyond SM})<34\%$ and $|\lambda_{SAA}| \lesssim 0.022$, we then obtain
\begin{align}
|\theta| \lesssim 0.001, \quad |\delta|\lesssim 0.088.
\end{align}

There may exist points with $\lambda_{SAA}\simeq 0$ on the $(\theta, \delta)$ plane for $\delta\neq 0$. If we avoid this tiny region around that points, where the other decay modes are also highly suppressed due to the very small values of $\theta$ and $\delta$, we always have ${\rm Br}(S\to AA) \simeq 1$. In Fig.~\ref{fig:thetadelta-1}, the contour plot for the $S\to AA$ decay width is shown on the $(\theta, \delta)$ plane. In this (two-step cascade annihilation dominant) case, we have $\Gamma_S/m_S\lesssim 6.7\times 10^{-4}$ for $|\theta| \lesssim0.001$, where $\Gamma_S$ is the total width of the $S$ boson; the value of the width can thus be negligible in the calculation. On the other hand, compared with $XX\to SS$ via an $s$-channel $S$ exchange, as shown in Fig~\ref{fig:vdm_ann}(b),  if the mediator is replaced by $h$ (or $H$), the resulting cross section is suppressed not only by the propagator of the heavier boson but also by the couplings squared: $(s_\delta \lambda_{hSS}/\lambda_{SSS})^2$  (or  $(s_\theta \lambda_{HSS}/\lambda_{SSS})^2$), where $s_\delta$ (or $s_\theta$) comes from the $X$-$X$-$h$ (or $X$-$X$-$H$) vertex,  and the coupling ratios $ \lambda_{hSS}/\lambda_{SSS}$ and  $\lambda_{HSS}/\lambda_{SSS}$ are shown in Fig.~\ref{fig:thetadelta-1}. Therefore,  the annihilation $XX \to SS$ via a heavier mediator, $h$ or $H$, is negligible in the calculation; the conclusion is also valid for the 0-step cascade annihilation via an $s$-channel $h$ (or $H$) exchange since these cross sections are also suppressed by a factor of $s_\delta^2$ (or $s_\theta^2$) resulted from  the $X$-$X$-$h$  (or $X$-$X$-$H$) vertex.

 In Fig.~\ref{fig:thetadelta-1}, we show the contour results of $\lambda_{SAA}$ and $\lambda_{SAA}/\lambda_{SSS}$ on the $(\theta, \delta)$ plane. As indicated  in the relevant $(\theta, \delta)$ region, the $\lambda_{SAA}$ is much more sensitive to the variation of $\theta$, compared with its dependence on $\delta$. In the following analysis, the $\delta$ is simply set to be zero, and the dependence of the DM relic density on the effective coupling can be related to the variation of $\theta$. If taking $\delta=0$, the constraint from the two-step cascade annihilation gives $|\theta| \lesssim 0.00043$.
 Our conclusion can be easily extended to the case with $\delta \neq 0$.

\subsection{Constraints from dark matter freeze-out and relic abundance}\label{sec:thetadelta-2}

\subsubsection{Coupled Boltzmann equations with interactions: $X X \leftrightarrow S S$, $S S \leftrightarrow AA$, and $S \leftrightarrow AA $}\label{sec:codecay-boltzmann}

The interplay of the DM particles and SM particles mostly results from  the interaction $X X \leftrightarrow S S$ followed by $S\leftrightarrow AA$ and $S S \leftrightarrow AA $ together with $A \leftrightarrow \tau^+ \tau^-$ and $A A \leftrightarrow \tau^+\tau^-$. For $S S \rightarrow AA $ and $A A \rightarrow \tau^+\tau^-$, their annihilation cross sections are summarized in Appendix~\ref{app:eqHiddenVisible}, while for $S\rightarrow AA$ and  $A \rightarrow \tau^+ \tau^-$, their partial decay widths are given in Eqs.~(\ref{eq:S-partial-width-1})-(\ref{eq:A-partial-width-2}).
Note that when the hidden sector particles become nonrelativistic at temperatures $T\lesssim m_{X,S}$, the cannibal annihilations could play important roles; such effects will be separately discussed  in Sec.~\ref{sec:can-boltzmann}. 
The evolutions of the number densities, $n_X$ and $n_S$, for  $X$ and $S$, respectively, are described by the coupled Boltzmann equations, 
\begin{align}
\frac{dn_X}{dt} + 3 H n_X=  & -\langle \sigma v \rangle_{XX\to SS} \bigg(n_X^2 -(n_X^{\text{eq}})^2 \frac{n_S^2}{(n_S^{\text{eq}})^2} \bigg) \,, \label{eq:boltz-1} \\
\frac{dn_S}{dt} + 3 H n_S = & - \langle \Gamma \rangle_{S\to AA} \bigg(n_S -n_S^{\text{eq}}  \frac{n_A^2}{(n_A^{\text{eq}})^2}  \bigg) 
  -\langle \sigma v \rangle_{SS\to AA} \bigg(n_S^2 -(n_S^{\text{eq}})^2  \frac{n_A^2}{(n_A^{\text{eq}})^2} \bigg) \nonumber\\
   & -\langle \sigma v \rangle_{SS\to XX} \bigg(n_S^2 -(n_S^{\text{eq}})^2 \frac{n_X^2}{(n_X^{\text{eq}})^2} \bigg) \,, \label{eq:boltz-2}
\end{align}
where $n_i^{\rm eq}$ is the equilibrium number  density for the particle $``i"$, and $\langle \sigma v \rangle$ and $\langle \Gamma \rangle$ are  respectively the thermally averaged cross section and decay rate,  corresponding to the specific process denoted in the subscript.
For the present vector DM+type-X N2HDM, the interaction between the CP-odd Higgs boson $A$ and lepton $\tau$, via $A \leftrightarrow \tau^+ \tau^- $ and $AA \leftrightarrow \tau^+ \tau^-$, is strong enough to maintain the chemical and thermal equilibrium between $A$ and the SM particles in the early Universe until the DM is completely freeze-out, i.e., $n_A= n_A^{\text{eq}}$; in other words, during the relevant epoch, like other SM particles, the $A$ boson is in thermal equilibrium with the reservoir. 
On the other hand, the interaction strength between $S$ and $A$ depends on the effective coupling $\lambda_{SAA}$, which is a function of $\theta$ and $\delta$ (see also Fig.~\ref{fig:thetadelta-1}).

To solve Eqs.~(\ref{eq:boltz-1}) and (\ref{eq:boltz-2}), we define the normalized yields,
\begin{align}
y_X(x) &=\sqrt{\frac{\pi}{45 G}} m_X g_*^{1/2}   \langle \sigma v \rangle_{XX\to SS}  Y_X (x) \,, \label{eq:yx} \\
y_S(x) &=\sqrt{\frac{\pi}{45 G}} m_X g_*^{1/2}   \langle \sigma v \rangle_{XX\to SS}  Y_S (x) \label{eq:ys}\,,
\end{align}
where $Y_X \equiv  n_X/ s$ and $Y_S \equiv n_S/s$ are respectively the dark matter and mediator  number densities normalized by the total entropy density,  $g_*^{1/2} = h_{\rm eff} [1+(1/3) (d\ln h_{\rm eff} / d\ln T)] /g_{\rm eff}^{1/2}$  is the effectively total number of relativistic DoF, and $x\equiv m_X/T$ is the variable that will be used instead of time $t$. Here $g_{\rm eff}$ and $h_{\rm eff}$ are the effective DoF for the energy density and entropy density, respectively \cite{Gondolo:1990dk}.
 Using the new defined quantities, we can rewrite these two Boltzmann equations into the following forms
\begin{align}
\frac{d y_X}{dx} =& - \frac{1}{x^2} \bigg( y_X^2 - (y_X^{\text{eq}})^2 \frac{y_S^2}{ (y_S^{\text{eq}})^2} \bigg) \,, \label{eq:boltzmann-1}\\
\frac{d y_S}{dx} =&  -x \frac{ \sqrt{90}}{\pi} M_{\rm pl} \frac{g_*^{1/2}}{h_{\text{eff}} } \frac{ \langle \Gamma \rangle_{S\to AA}}{m_X^2}
 (y_S -y_S^{\text{eq}} )  
 -  \frac{1}{x^2}\frac{ \langle \sigma v \rangle_{SS\to AA} }{ \langle \sigma v \rangle_{XX\to SS} }  \Big[ y_S^2 - ( y_S^{\text{eq}} )^2 \Big] \nonumber\\
 & - \frac{1}{x^2}\frac{ \langle \sigma v \rangle_{SS\to XX} }{ \langle \sigma v \rangle_{XX\to SS} }  
 \bigg( y_S^2 - (y_S^{\text{eq}})^2 \frac{y_X^2}{ (y_X^{\text{eq}})^2} \bigg)  \,,  \label{eq:boltzmann-2}
 \end{align}
where $M_{\rm pl}=(8 \pi G)^{-1/2}=2.44\times 10^{18}$~GeV is the reduced Planck mass, and the values of $y_i$  in equilibrium are given by
\begin{equation}
y_i^{\rm eq} (x) =g_i \frac{ \sqrt{90}}{2 \pi^3} M_{\rm pl} \frac{g_*^{1/2}}{h_{\rm eff}} m_X \bigg(x \frac{m_i}{m_X} \bigg)^2 \langle \sigma v \rangle_{XX\to SS}  K_2 \bigg(x \frac{m_i}{m_X} \bigg) \,,
\end{equation}
with the subscript index $``i" \equiv X$ or $S$,  and $g_X=3$ and $g_S=1$ being the internal degrees of freedom of the $X$ (dark matter) and $S$ (mediator) particles, respectively.  Using the relations, which are inferred from the Boltzmann equation \cite{Dodelson:2003}, 
\begin{align}
& (y_X^{\rm eq})^2 \langle \sigma v \rangle_{XX\to SS}  = (y_S^{\rm eq})^2 \langle \sigma v \rangle_{SS\to XX} \,,\\
& \langle \Gamma \rangle_{S\to AA} = \Gamma_{S\to AA}  \frac{K_1(x\cdot m_S/m_X) }{ K_2(x\cdot m_S/m_X)} \,,
\end{align}
 where $\Gamma_{S\to AA}$ is the $S\to AA$ decays width given in the rest frame of $S$, and $K_i$ is the modified Bessel function of second kind, we can further recast Eq.~(\ref{eq:boltzmann-2})  into the form,
\begin{align}
\frac{d y_S}{dx} =& - x \frac{ \sqrt{90}}{\pi} M_{\rm pl} \frac{g_*^{1/2}}{h_{\text{eff}} } \frac{\Gamma_{S\to AA}}{m_X^2}
\frac{K_1(x\cdot m_S/m_X) }{ K_2(x\cdot m_S/m_X)}  (y_S -y_S^{\text{eq}} )  
 -  \frac{1}{x^2}\frac{ \langle \sigma v \rangle_{SS\to AA} }{ \langle \sigma v \rangle_{XX\to SS} }  \Big[ y_S^2 - ( y_S^{\text{eq}} )^2 \Big] \nonumber\\
 & - \frac{1}{x^2} \bigg[ \frac{(y_X^{\text{eq}})^2}{(y_S^{\text{eq}})^2} y_S^2 -  y_X^2 \bigg]  \,.  \label{eq:boltzmann-2.1}
 \end{align}
Because both the annihilation processes, $SS\to AA$ and $XX\to SS$, occur through the s-wave, for simplicity, in the following discussion we approximate $\langle \sigma v \rangle_{SS\to AA} \simeq \langle \sigma v \rangle_{SS\to AA}^{(0)}$ and $ \langle \sigma v \rangle_{XX\to SS}\simeq  \langle \sigma v \rangle_{XX\to SS}^{(0)}$  using their leading values in $v\to 0$ limit.   Note that in the Boltzmann equation, the contribution due to the $SS\leftrightarrow AA$ interaction is usually much smaller than the process $S\leftrightarrow AA$; in other words, in Eq.~(\ref{eq:boltzmann-2.1}), the second term of the right hand side (RHS) is negligible, especially when the co-decay occurs with $|\lambda_{SAA}|\lesssim 1.03\times 10^{-8}$, corresponding to $|\theta| \lesssim  2\times 10^{-10}$ if $\delta=0$.

\begin{figure}[t!]
\begin{center}
\includegraphics[width=0.36\textwidth]{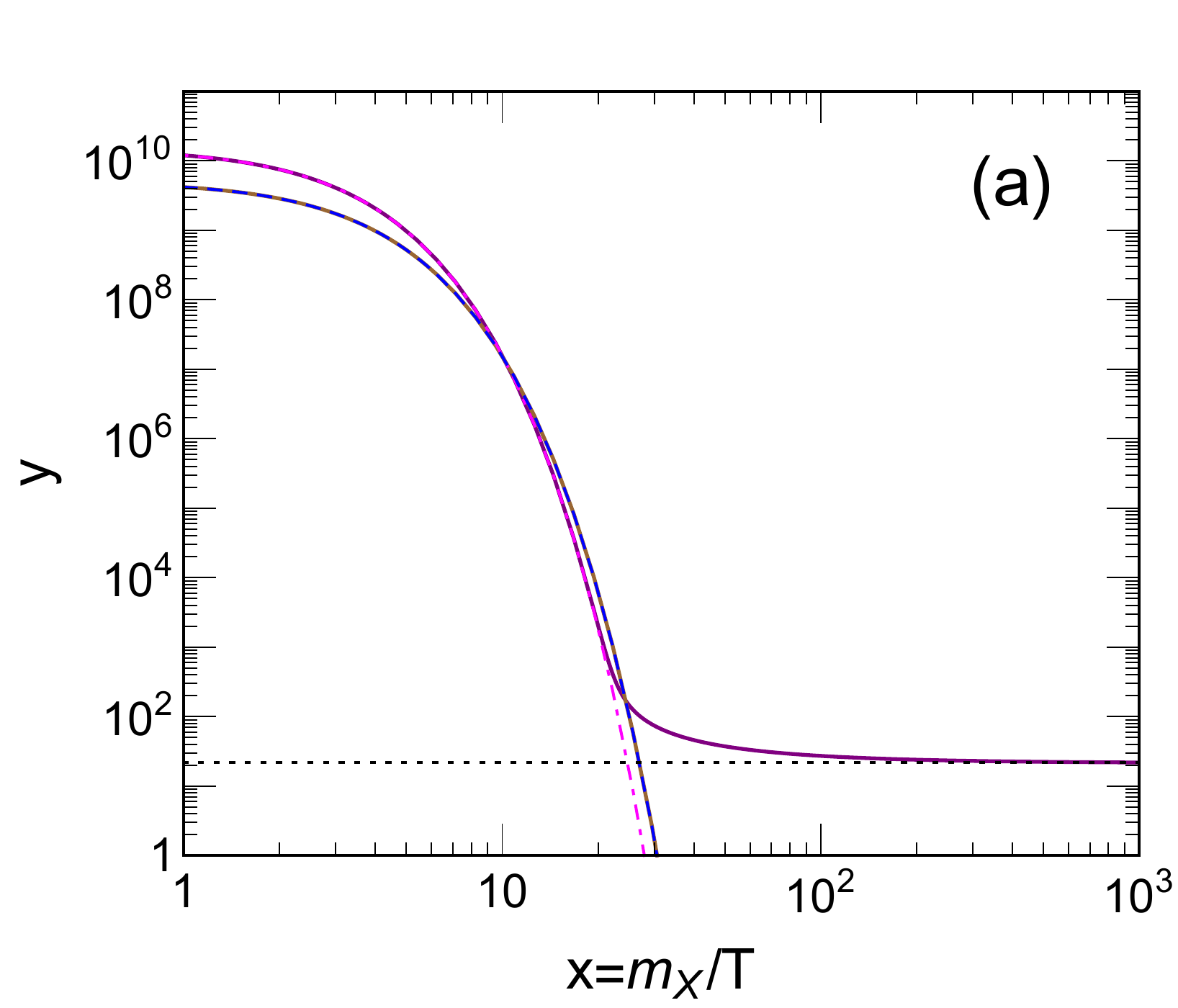}\hskip0.8cm
\includegraphics[width=0.36\textwidth]{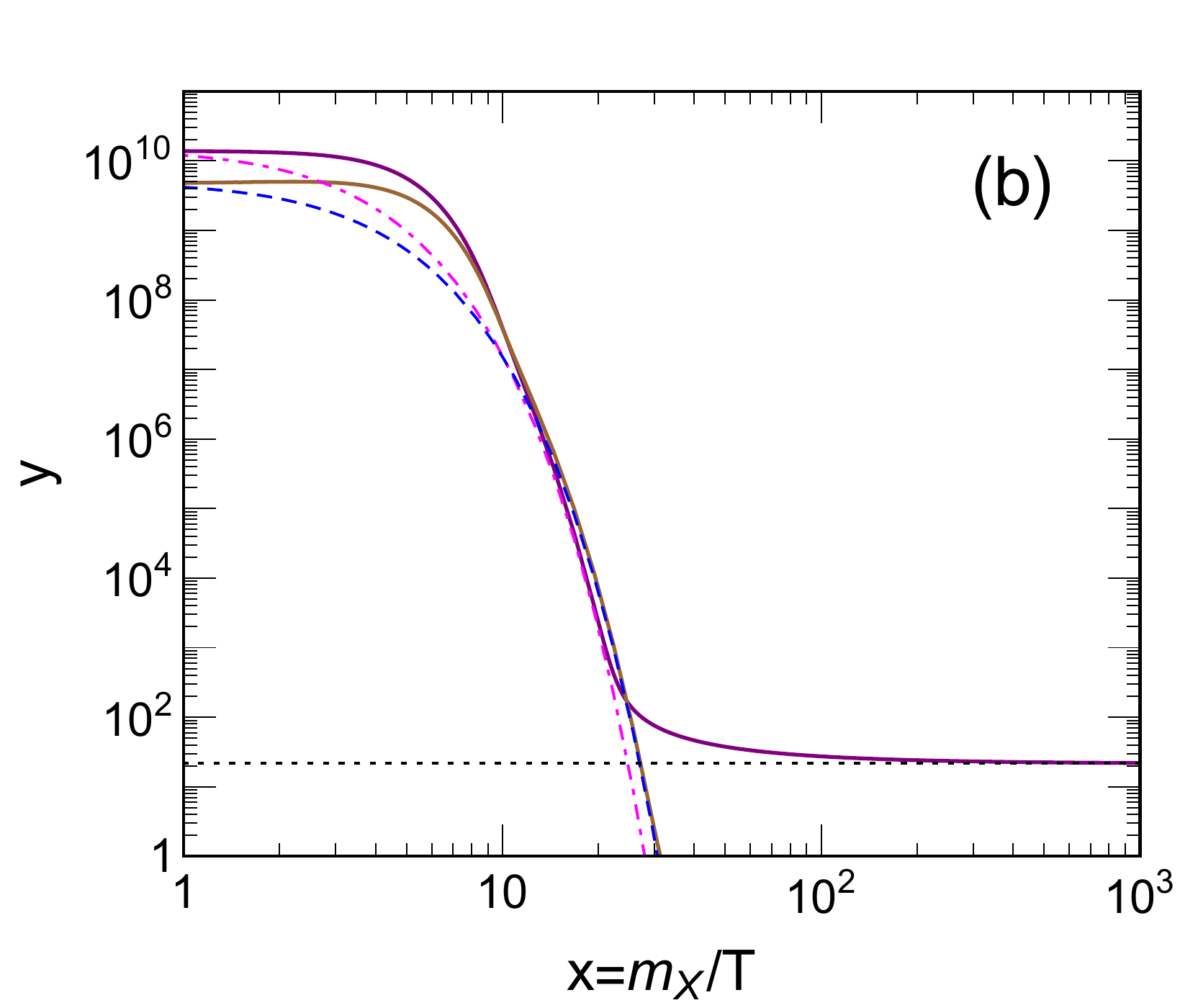}\\
\vskip0.3cm
\includegraphics[width=0.36\textwidth]{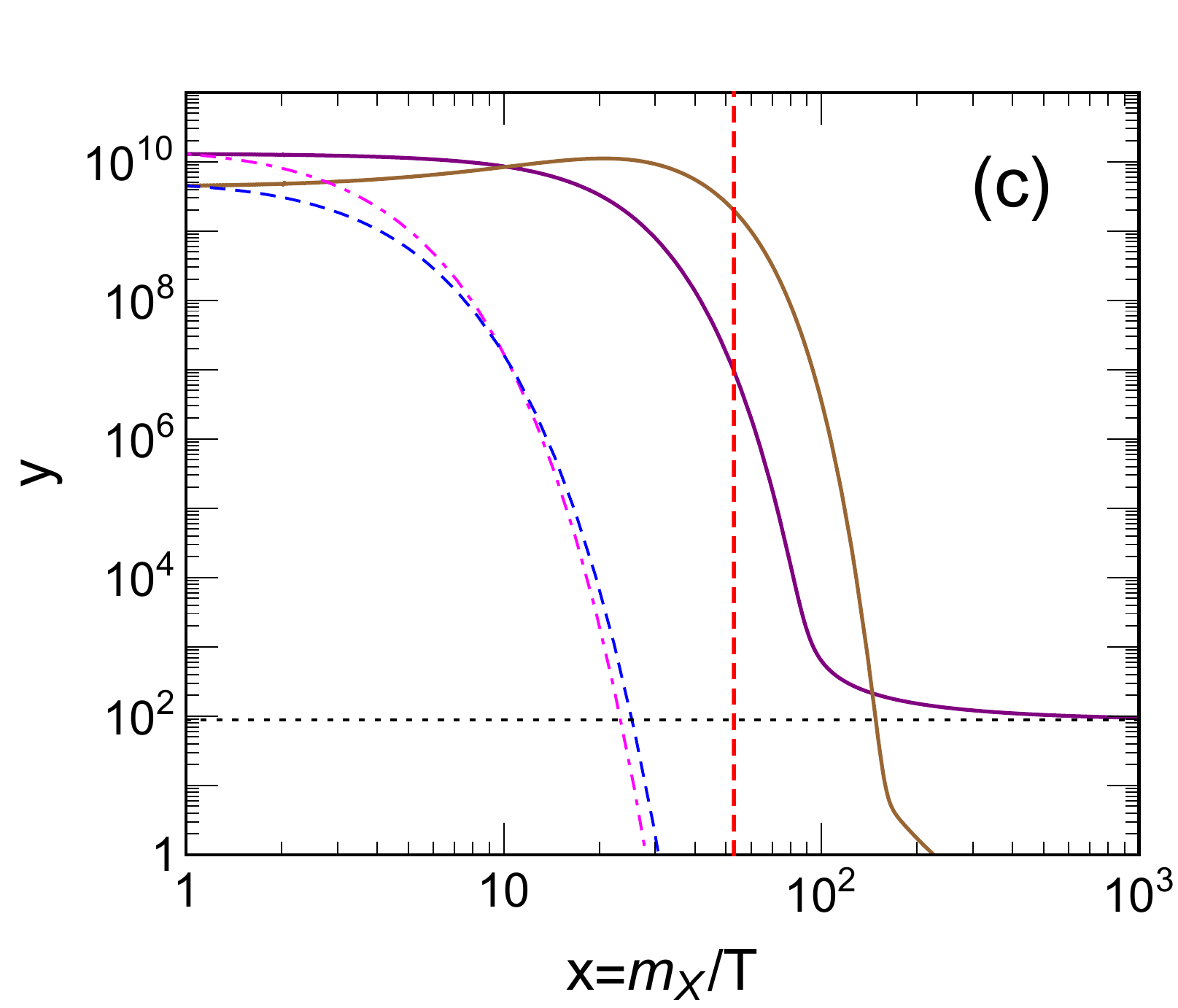} \hskip0.8cm
\includegraphics[width=0.36\textwidth]{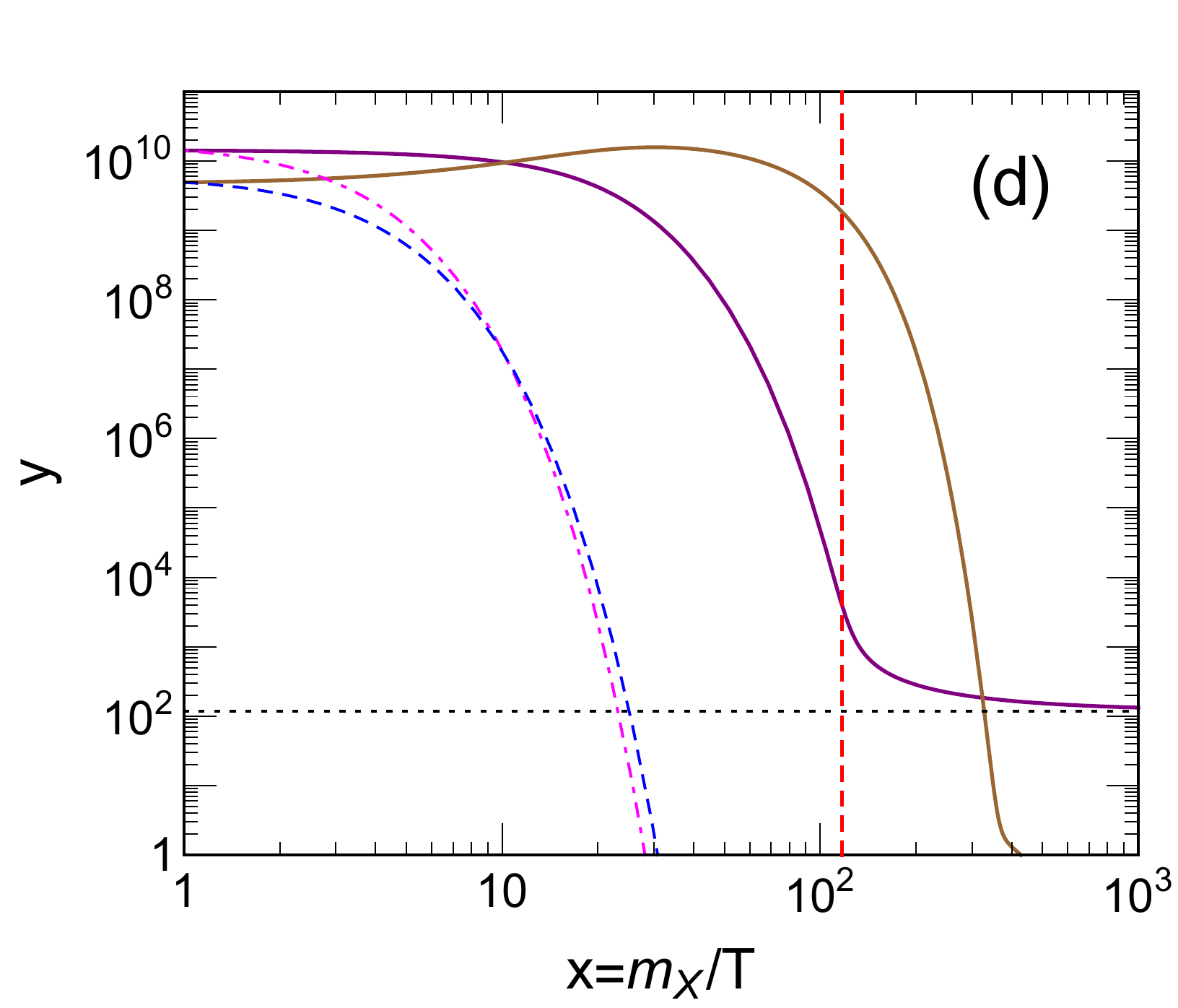}
\caption{ Evolutions of the dark matter yield $y_X$ and mediator yield $y_S$, where the parameters given in Eq.~(\ref{eq:parameters-1}) are used. The purple solid and brown solid curves stand for $y_X$ and $y_S$, respectively, while the magenta dashed-dotted and  blue dashed curves show their corresponding yields under thermal equilibrium. The horizontal line denotes the asymptotic yield of the dark matter, $y_X^\infty =x_f$.  In (a), (b), (c), and (d), we separately use $(\theta,\delta)= (0.00043, 0), (2\times 10^{-10}, 0), (1.2\times 10^{-11}, 0)$, and $(5.1\times 10^{-12}, 0)$ as inputs, for which the resulting $\lambda_{SAA}$ are $-0.022, -1.03\times 10^{-8}, -6.16\times 10^{-10}$ and $-2.62\times 10^{-10}$, and the corresponding $x_f$ are $22, 22, 88$, and $117$, respectively. The vertical dashed (red) lines in (c) and (d) indicate the values of $x_\Gamma$ obtained by Eq.~(\ref{eq:width-sol2}).
}
\label{fig:relic}
\end{center}
\end{figure}

After DM freeze out, so that  $y_X^2 \gg (y_X^{\rm eq}\cdot  y_S/y_S^{\rm eq})^2 $, the Boltzmann equation given in Eq.~(\ref{eq:boltzmann-1}) can reduce to 
\begin{align}
\frac{d y_X}{dx} \approx & - \frac{1}{x^2}  y_X^2\,.
\end{align}
Integrating this approximate equation from the freeze-out epoch $x_f (=m_X/T_f)$ until  very late times $x_\infty (\gg x_f)$, one can obtain
\begin{equation}
\frac{1}{y_\infty} - \frac{1}{y_f} = \frac{1}{x_f} - \frac{1}{x_\infty}    
\quad  \Rightarrow  \quad
y_\infty \simeq x_f \,,
\end{equation} 
where $y_\infty \ll y_f$.
In the following study, along with $\delta=0$, we use the parameters given in Eq.~(\ref{eq:parameters-1}), $m_X=40$~GeV, $m_S=35$~GeV, $m_A=15$~GeV, $g_X=0.123$, $m_H=m_{H^\pm}=M=300$~GeV, $\tan\beta=35$, and $\beta-\alpha=0.062909$ as a benchmark. 
The temperature-dependences of $g_*$ and $h_{\rm eff}$  given in Fig.~1 of Ref.~\cite{Cerdeno:2011tf} are adopted, and their corresponding values, obtained iteratively, at freeze-out temperature are further used in Fig.~\ref{fig:relic}.

 Note that there are three different types of relic results which may occur during the dark matter freezes out:
 
 \vskip0.2cm
 \noindent{\bf (i) The conventional WIMP dark matter}. 
If the $S \leftrightarrow  AA$ and/or  $S S\leftrightarrow AA$ interaction(s) are/is strong enough to maintain the chemical equilibrium of  the mediator $S$ with the CP-odd Higgs $A$ during the epoch of the dark matter freeze-out, i.e. $y_S= y_S^{\rm eq}$, then the solution of Eq.~(\ref{eq:boltzmann-1}) is the same as the conventional WIMP dark matter scenario; the corresponding mixing angle $\theta$ satisfies $7\times 10^{-10} <|\theta| < 0.00043$ if taking $\delta=0$. As an example shown in Fig.~\ref{fig:relic}(a), using $(\theta,\delta)= (0.00043, 0)$, which corresponds to  $\lambda_{SAA} \simeq -0.022$, we obtain $x_f=22$.  For the values of $|\lambda_{SAA}| \lesssim 0.022$, we have found that $\langle \sigma v\rangle_{XX\to AA} / \langle \sigma v\rangle_{XX\to SS} \lesssim 0.05$, so that the GC gamma-ray excess originating from the two-step cascade annihilation to the final state $\tau$'s is still dominant over the 
one-step cascade process described by $XX\to AA$ followed by $A\to \tau^+ \tau^-$. 

\vskip0.2cm
\noindent{\bf (ii) The unconventional WIMP dark matter}.
If $|\theta|$ is less than $7\times 10^{-10}$ but still larger than $2\times 10^{-10}$,  the nonrelativistic dark sector particles start to decouple from the thermal bath as for $x\sim1$. However, for this case,  the dark sector can reach again thermal equilibrium with the reservoir before the DM freeze out, so that the dark matter is still WIMP-like, and has the same freeze-out temperature and thermally averaged annihilation cross section as the WIMP case.   As an example, using $(\theta,\delta)= (2\times 10^{-10}, 0)$, which corresponds to  $\lambda_{SAA} \simeq -1.03\times 10^{-8}$, we show the result in Fig.~\ref{fig:relic}(b). See also Fig.~\ref{fig:xftheta}.

\vskip0.2cm
\noindent{\bf (iii) The co-decaying dark matter}. 
This scenario, for which the corresponding mixing angle is $|\theta| < 2 \times 10^{-10}$ if taking $\delta=0$, is characterized by $  y_S \gg y_S^{\rm eq} $ and  $y_X^2 \gg (y_X^{\rm eq}\cdot  y_S/y_S^{\rm eq})^2 $ at $x=x_f$.
For this type of the scenario, featuring by a small $|\lambda_{SAA}|\lesssim 1.03\times 10^{-8}$, when the dark sector particles become to be nonrelativistic ($x\approx 1$), they start to decouple from the thermal reservoir, and their total yields, $y_X+y_S$, (which is related to the total number density of the dark sector in the co-moving frame) tend to remain constant. 
Then after a time interval, $\Delta t_\Gamma \approx 2 H^{-1}$ (during the radiation dominated epoch) $\sim \Gamma_S^{-1}$, for a degenerate case $m_X\approx m_S$,  the dark matter $X$ as well as the mediator $S$ undergoes an exponential decay 
until freeze-out, of which at the temperature  $x_f$, the Hubble expansion rate $H$ becomes larger than the $XX \to SS$ annihilation rate. This process is described by the  ``co-decaying dark matter" mechanism which was proposed by Dror, Kuflik, and Ng  \cite{Dror:2016rxc} for the degenerate case. 
In the following, we will further discuss a generic case, including the degenerate and non-degenerate ones. We will show that if $\Delta t_\Gamma$ is large enough, the exponential suppression for $y_X$ can occur much earlier than $y_S$ due to a significantly suppressed up-scattering rate for the process $SS\to XX$, such that the DM freeze-out time is earlier than the time that $S$ undergoes an exponential decay. See also Figs.~\ref{fig:relic}(c)-(d).

 We then further estimate the value of $\Delta t_\Gamma$ for the co-decaying DM. There are two different cases: (i) $m_X=m_S$, and (ii) $m_X>m_S$. Considering the case of $m_X=m_S$, we sum Eqs.~(\ref{eq:boltzmann-1}) and (\ref{eq:boltzmann-2.1}), and neglect the second term of RHS of the latter equation,  
\begin{equation}
\frac{d (y_X+y_S)}{dx} \simeq - C x  y_S \,, \label{eq:decaywidth-1}
\end{equation}
with
\begin{equation}
C \equiv \frac{ \sqrt{90}}{\pi} M_{\rm pl} \frac{g_*^{1/2}}{h_{\text{eff}} } \frac{\Gamma_{S\to AA}}{m_X^2} \,.
\end{equation}
Taking the initial conditions: $y_X(1) = y_X^{\rm eq}(1)$, $y_S(1) = y_S^{\rm eq}(1)$, and $y_X^{\rm eq}(1) / y_S^{\rm eq}(1) = g_X/g_S =3$, and approximating $y_X+y_S\simeq 4 y_S$,
we get the solution to Eq.~(\ref{eq:decaywidth-1}),
\begin{equation}
\frac{y_S(x_\Gamma)}{y_S(1)} \simeq e^{-\frac{C}{8} (x_\Gamma^2 -1)} 
\defeq e^{-1} \,, \label{eq:width-sol1-1}
\end{equation}
with $x_\Gamma \simeq \sqrt{1+ 8/C}$.
Because of $\Delta t_\Gamma \simeq (1/2) x_\Gamma^2 C/\Gamma_{S\to AA}$, this solution can thus be rewritten as 
\begin{equation}
\frac{y_S(x_\Gamma)}{y_S(1)} \simeq e^{-\Gamma_{S\to AA}\Delta t_\Gamma/4 }  \,, \label{eq:width-sol1-2}
\end{equation}
for which the lifetime of $S$ is $\Delta t_\Gamma \simeq 4 \, \Gamma_{S\to AA}^{-1}$.
We find that the result shown in Eq.~(\ref{eq:width-sol1-1}) or Eq.~(\ref{eq:width-sol1-2}) can be a good approximation for the case with $x_\Gamma \gtrsim 20$, where the hidden Higgs $S$ can have a sufficient time to satisfy the approximation $y_S -y_S^{\rm eq}\approx y_S$ which has been taken in Eq.~(\ref{eq:decaywidth-1}).

For the case of $m_X>m_S$,  the value of $x_\Gamma$ depends not only on $\theta$ but also on the mass difference of $X$ and $S$.
 If the difference of $m_X$ and $m_S$ is sizable enough, then  the down-scattering rate, $XX\to SS$, could be significantly  larger than the up-scattering rate, $SS\to XX$. Under this condition and after a sufficient time with $x_\Gamma \gtrsim 20$, one could have $y_X^{\rm eq} /y_S^{\rm eq} \ll 1$ and $y_X\ll y_S$,
 so that the second and third terms on the right hand side of Eq.~(\ref{eq:boltzmann-2.1}) are negligible, and this equation can thus be approximated as \begin{equation}
\frac{d y_S}{dx} \simeq - C x  y_S \,. \label{eq:decaywidth-2}
\end{equation}
From this equation, we get
\begin{equation}
\frac{y_S(x_\Gamma)}{y_S^{\rm in}(1)} \simeq e^{-\frac{C}{2} (x_\Gamma^2 -1)} 
\defeq \frac{e^{-1}}{4} \,, \label{eq:width-sol2}
\end{equation}
where the effective initial $S$ yield, $y_S^{\rm in}(1)$, can be approximated as $y_X(1) + y_S(1) \simeq 4 y_S(1)$, because the down scattering is larger than the upper one, and the mostly initial $X$ could scatter into $S$ before the $S$ boson undergoes an exponential decay. 
Therefore, if $m_X - m_S$ is sizable enough, we have $x_\Gamma \simeq \sqrt{1+ 4.8/C}$ and $y_S(x_\Gamma)/y_S(1) \simeq 4 e^{-\Gamma_{S\to AA}\Delta t_\Gamma}$, for which the lifetime of $S$ is $\Delta t_\Gamma \simeq 2.4 \, \Gamma_{S\to AA}^{-1}$.

In general, for a co-decay process with $x_\Gamma \gtrsim 20$,  we  get $ \sqrt{1+ 4.8/C} \lesssim x_\Gamma \lesssim \sqrt{1+ 8/C}$. Taking the same parameters which have been used, but adopting $m_S$ as a free parameter, we find that $x_\Gamma = \sqrt{1+ 4.8/C}$ is a good approximation provided that $2m_A< m_S \lesssim 35$~GeV.
Figs.~\ref{fig:relic}(c) and (d) show the values of $x_\Gamma$ to be about 51 and 117, respectively, where the latter satisfies $x_\Gamma \approx x_f$.   For a process with a much larger $x_\Gamma$ as shown in Figs.~\ref{fig:relic}(c) and (d), the inequality  $y_S>y_X$, i.e. $n_S>n_X$, becomes much more noticeable and, due to a suppressed up-scattering rate, the exponential suppression for $n_X$ occurs much earlier than $n_S$.

\subsubsection{Including the cannibal interactions, $X X S\leftrightarrow S S$,  $S S S \leftrightarrow XX$,  $XXX\leftrightarrow XS$, $S S S\leftrightarrow SS$ and $XSS\leftrightarrow XS$, in the coupled Boltzmann equations: a more complete treatment}\label{sec:can-boltzmann}

It was stressed in Refs. \cite{Farina:2016llk,Pappadopulo:2016pkp} that when the hidden sector is out of the thermal equilibrium with the bath, the hidden particles may undergo so-called cannibalism before DM freeze-out. Here we include the cannibal interactions,   $X X S\leftrightarrow S S$, $S S S \leftrightarrow XX$,  $XXX\leftrightarrow XS$, $S S S\leftrightarrow SS$  and $XSS\leftrightarrow XS$, which are number changing annihialtions, in the Boltzmann equations,
\begin{align}
\frac{dn_X}{dt} + 3 H n_X=  &   \cdots 
  -  \langle \sigma v^2 \rangle_{XXS\to SS} \bigg[n_X^2 n_S -(n_X^{\text{eq}})^2 \frac{n_S^2}{(n_S^{\text{eq}})} \bigg]
   + \frac{1}{3} \langle \sigma v^2 \rangle_{SSS\to XX} \bigg[ n_S^3 -(n_S^{\text{eq}})^3 \frac{n_X^2}{(n_X^{\text{eq}})^2} \bigg] \nonumber\\
  &  -  \frac{1}{3} \langle \sigma v^2 \rangle_{XXX\to XS} \bigg[ n_X^3 -n_X n_S \frac{(n_X^{\text{eq}})^2}{n_S^{\text{eq}}} \bigg]     \,, \label{eq:boltz-cann1} \\
\frac{dn_S}{dt} + 3 H n_S =&  \cdots 
 - \frac{1}{6} \langle \sigma v^2 \rangle_{SSS\to SS} \bigg[ n_S^3 -n_S^2 n_S^{\text{eq}}   \bigg]    
  - \frac{1}{2} \langle \sigma v^2 \rangle_{XSS\to XS} \bigg[ n_X n_S^2 - n_X n_S n_S^{\text{eq}}   \bigg]     \nonumber\\
 & + \frac{1}{2} \langle \sigma v^2 \rangle_{XXS\to SS} \bigg[n_X^2 n_S -(n_X^{\text{eq}})^2 \frac{n_S^2}{(n_S^{\text{eq}})} \bigg] 
   - \frac{1}{2} \langle \sigma v^2 \rangle_{SSS\to XX} \bigg[ n_S^3 -(n_S^{\text{eq}})^3 \frac{n_X^2}{(n_X^{\text{eq}})^2} \bigg] \nonumber\\
     &  + \frac{1}{6} \langle \sigma v^2 \rangle_{XXX\to XS} \bigg[ n_X^3 -n_X n_S \frac{(n_X^{\text{eq}})^2}{n_S^{\text{eq}}} \bigg]  
    \,, \label{eq:boltz-cann2}
\end{align}
where the dots are all terms of the right hand side of Eqs.~(\ref{eq:boltz-1}) and (\ref{eq:boltz-2}), respectively, and $\langle \sigma v^2 \rangle$ is the thermally averaged annihilation cross section for the $3\to 2$ cannibal process.  One can refer to Ref.~\cite{Berlin:2016gtr} for a general from of  $3\to 2$ scattering rates. In each cannibal term of the above equations, the factor (including the relative sign) in front of the cross section equals to $\Delta n_i/N!$, where $1/N!$ is for avoiding the double counting for the initial number density in the reaction with $N$ being the number of the identical particles of the initial states for the relevant cross section, and $\Delta n_i$ is  the number change for the hidden particle with $i\equiv X$ for Eq.~(\ref{eq:boltz-1}) or $\equiv S$ for (\ref{eq:boltz-2}); for instance, for the process $SSS\to XX$, we have $N = 3$, but $\Delta n_X = 2$, $\Delta n_S = -3$, resulting in the factor $1/3$ and $-1/2$ shown in the corresponding terms in Eqs.~(\ref{eq:boltz-cann1}) and (\ref{eq:boltz-cann2}), respectively.

To calculate $3\to 2$ thermally averaged annihilation cross sections for the nonrelativistic dark sector particles with $x\lesssim 1$, we take the low-velocity approximation, $\langle \sigma v^2\rangle \simeq \sigma v^2$, by neglecting the correction of ${\cal O}(T/m_X)$:
\begin{align}
 \langle \sigma v^2 \rangle_{XXS\to SS}  & \simeq \frac{[(2m_X-m_S) (2m_X +3 m_S)]^{1/2}}{128 \pi m_X^4 m_S (2m_X +m_S)} 
g_X^6 \frac{96275}{5184}  
    \Bigg( 1- \frac{4.33\delta}{m_X} + \frac{5.25 \delta^2}{m_X^2} - \frac{0.68\delta^3}{m_X^3}\Bigg)  ,  \\
 \langle \sigma v^2 \rangle_{SSS\to XX} & \simeq \frac{\sqrt{9-4m_X^2/m_S^2}}{384 \pi m_S^3 m_X^2}  g_X^6
 \frac{1377}{16}  
    \Bigg( 1 + \frac{7.76\delta}{m_X} + \frac{18.3 \delta^2}{m_X^2} - \frac{42.3\delta^3}{m_X^3}\Bigg)  , 
\end{align} 
\begin{align}
     \langle \sigma v^2 \rangle_{XXX\to XS}  & \simeq \frac{[(16m_X^2-m_S^2) (4m_X^2 -m_S^2)]^{1/2}}{1152 \pi m_X^7 } 
g_X^6 \frac{1855}{9}  
    \Bigg( 1+ \frac{8.39\delta}{m_X} - \frac{1.04 \delta^2}{m_X^2} + \frac{6.64\delta^3}{m_X^3}\Bigg)  ,  \\
 \langle \sigma v^2 \rangle_{SSS\to SS}  & \simeq \frac{\sqrt{5}}{384 \pi} \frac{18225}{16 m_S^5} \Bigg(\frac{g_X m_S}{m_X}\Bigg)^6  , \\
  \langle \sigma v^2 \rangle_{XSS\to XS}  & \simeq \frac{ [3(2m_X+m_S) (2m_X +3 m_S)]^{1/2}}{128 \pi m_X^3 m_S (m_X +2m_S)^2} 
g_X^6 \frac{21425}{54}  
    \Bigg( 1- \frac{2.89\delta}{m_X} + \frac{3.08 \delta^2}{m_X^2} - \frac{1.56\delta^3}{m_X^3}\Bigg) , 
\end{align}
with $\delta \equiv m_X - m_S$. Here, because the expressions of  $X X S\rightarrow S S$, $S S S \rightarrow XX$,  $XXX \rightarrow XS$, and $XSS\rightarrow XS$ are lengthy, with good approximations we thus expand their amplitudes squared up to ${\cal O} (\delta^3/m_X^3)$.  Again, we further rewrite Eqs.~(\ref{eq:boltz-cann1}) and (\ref{eq:boltz-cann2}) as
\begin{align}
\frac{d y_X}{dx} =  & \cdots  
  + \frac{\pi} {\sqrt{90}} \frac{h_{\text{eff}} }{g_*^{1/2}} \frac{m_X^2}{M_{\rm pl} }
   \frac{1}{x^5}
    \Bigg[
    - \frac{ \langle \sigma v^2 \rangle_{XXS\to SS}}{  (\langle \sigma v\rangle_{XX\to SS})^2 } 
    \bigg( y_X^2 y_S - \frac{y_S^2}{y_S^{\text{eq}}} (y_X^{\text{eq}})^2  \bigg)   \nonumber \\
  & + \frac{1}{3}
    \frac{ \langle \sigma v^2 \rangle_{SSS\to XX}}{  (\langle \sigma v\rangle_{XX\to SS})^2 } 
    \bigg( y_S^3 - \frac{y_X^2}{(y_X^{\text{eq}})^2} (y_S^{\text{eq}})^3  \bigg) 
     - \frac{1}{3}
    \frac{ \langle \sigma v^2 \rangle_{XXX\to XS}}{  (\langle \sigma v\rangle_{XX\to SS})^2 } 
    \bigg( y_X^3 - \frac{y_X y_S (y_X^{\text{eq}})^2}{ y_S^{\text{eq}}} \bigg)     
      \Bigg]
  \,,  \label{eq:boltzmann-cann-1}  \\
  \frac{d y_S}{dx} =  & \cdots  
  +  \frac{\pi} {\sqrt{90}} \frac{h_{\text{eff}} }{g_*^{1/2}} \frac{m_X^2}{M_{\rm pl} }
   \frac{1}{x^5}   \Bigg[
       - \frac{1}{6} 
  \frac{ \langle \sigma v^2 \rangle_{SSS\to SS}}{  (\langle \sigma v\rangle_{XX\to SS})^2 }  
   \Big( y_S^3 - y_S^2  y_S^{\text{eq}}  \Big) \nonumber\\
   &    - \frac{1}{2} 
  \frac{ \langle \sigma v^2 \rangle_{XSS\to XS}}{  (\langle \sigma v\rangle_{XX\to SS})^2 }  
   \Big( y_X y_S^2 - y_X y_S  y_S^{\text{eq}}  \Big) 
   + \frac{1}{2} \frac{ \langle \sigma v^2 \rangle_{XXS\to SS}}{  (\langle \sigma v\rangle_{XX\to SS})^2 } 
    \bigg( y_X^2 y_S - \frac{y_S^2}{y_S^{\text{eq}}} (y_X^{\text{eq}})^2  \bigg)  
 \nonumber\\
  &   
  -  \frac{1}{2}
    \frac{ \langle \sigma v^2 \rangle_{SSS\to XX}}{  (\langle \sigma v\rangle_{XX\to SS})^2 } 
    \bigg( y_S^3 - \frac{y_X^2}{(y_X^{\text{eq}})^2} (y_S^{\text{eq}})^3  \bigg)
      + \frac{1}{6}
    \frac{ \langle \sigma v^2 \rangle_{XXX\to XS}}{  (\langle \sigma v\rangle_{XX\to SS})^2 } 
    \bigg( y_X^3 - \frac{y_X y_S (y_X^{\text{eq}})^2}{ y_S^{\text{eq}}} \bigg)     
   \Bigg]  \,.  
    \label{eq:boltzmann-cann-2}
 \end{align}
 \vskip0.45cm
As shown in Fig.~\ref{fig:relic}(b), (c) and (d), for the case of  $|\theta| < 7\times 10^{-10}$, when $x\lesssim 1$, the dark sector particles become nonrelativistic and are kinetically decoupled from the thermal bath  due to small $S \leftrightarrow AA$ and $SS \leftrightarrow AA$ interaction rates, so that the comoving number density of the total hidden sector particles remain  constant before the time that the hidden Higgs bosons undergo the exponential decay. 
However, in the present case, the cannibal annihilations cannot be neglected. We take Eq.~(\ref{eq:boltz-cann1}) as an example to give a qualitative analysis on cannibalization as follows. 
The left hand side is of order $n_X H$, while the right hand side due to $XXS\to SS$ is of order $ n_X^2 n_S\langle \sigma v^2 \rangle_{XXS\to SS} $. Therefore, if the $XXS\to SS$ reaction rate is much larger than the expansion rate, the only way to maintain the equality of Eq.~(\ref{eq:boltz-cann1}) is to have $n_X =n_X^{\text{eq}}$ and $n_S = n_S^{\text{eq}}$, about which one can obtain the same conclusion using either $SSS \to XX$, $XXX\to XS$, $SSS \to SS$, $XSS \to XS$, or Eq.~(\ref{eq:boltz-cann2}) in doing a similar analysis.  

\begin{figure}[t!]
\begin{center}
\includegraphics[width=0.36\textwidth]{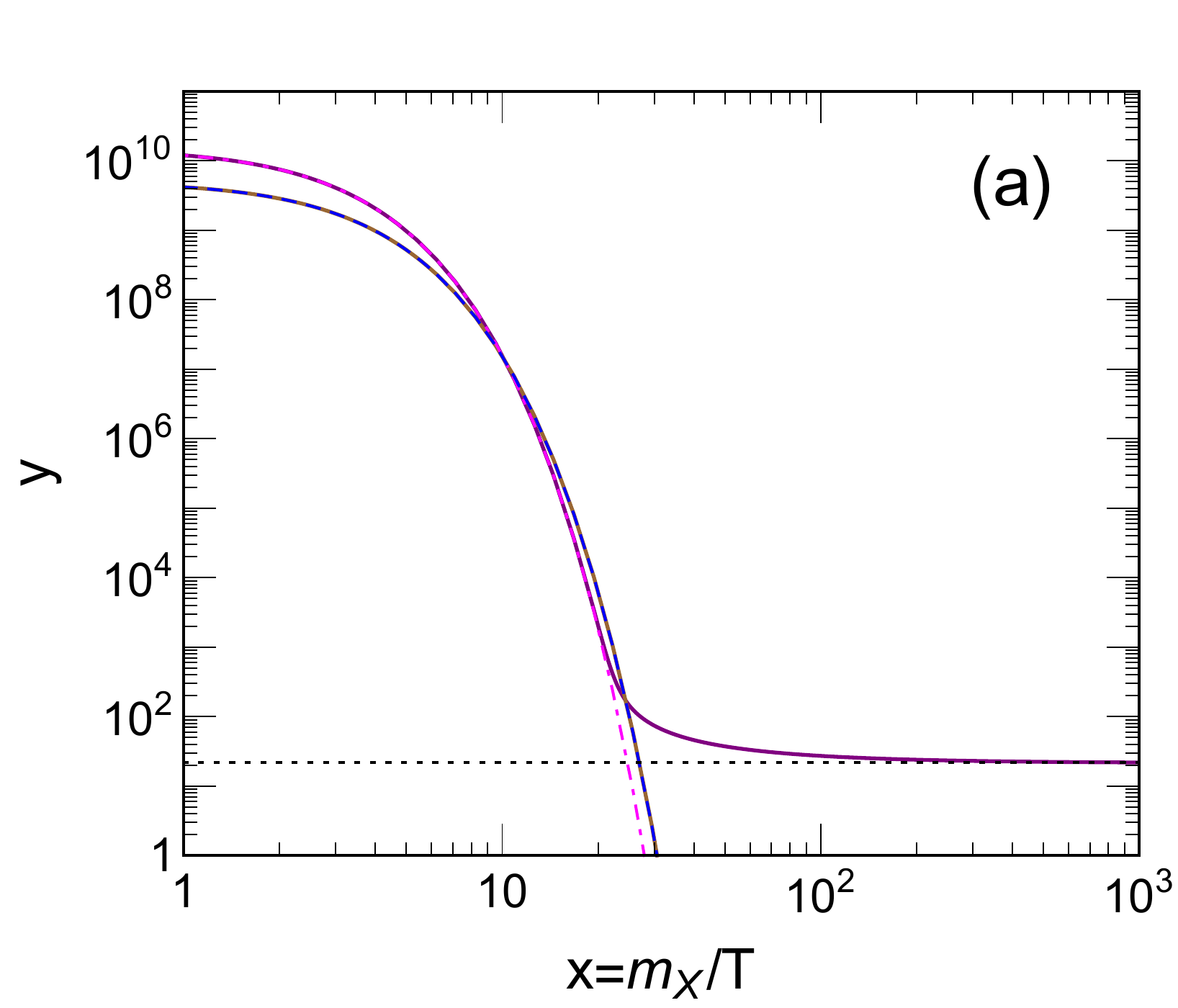}\hskip0.8cm
\includegraphics[width=0.36\textwidth]{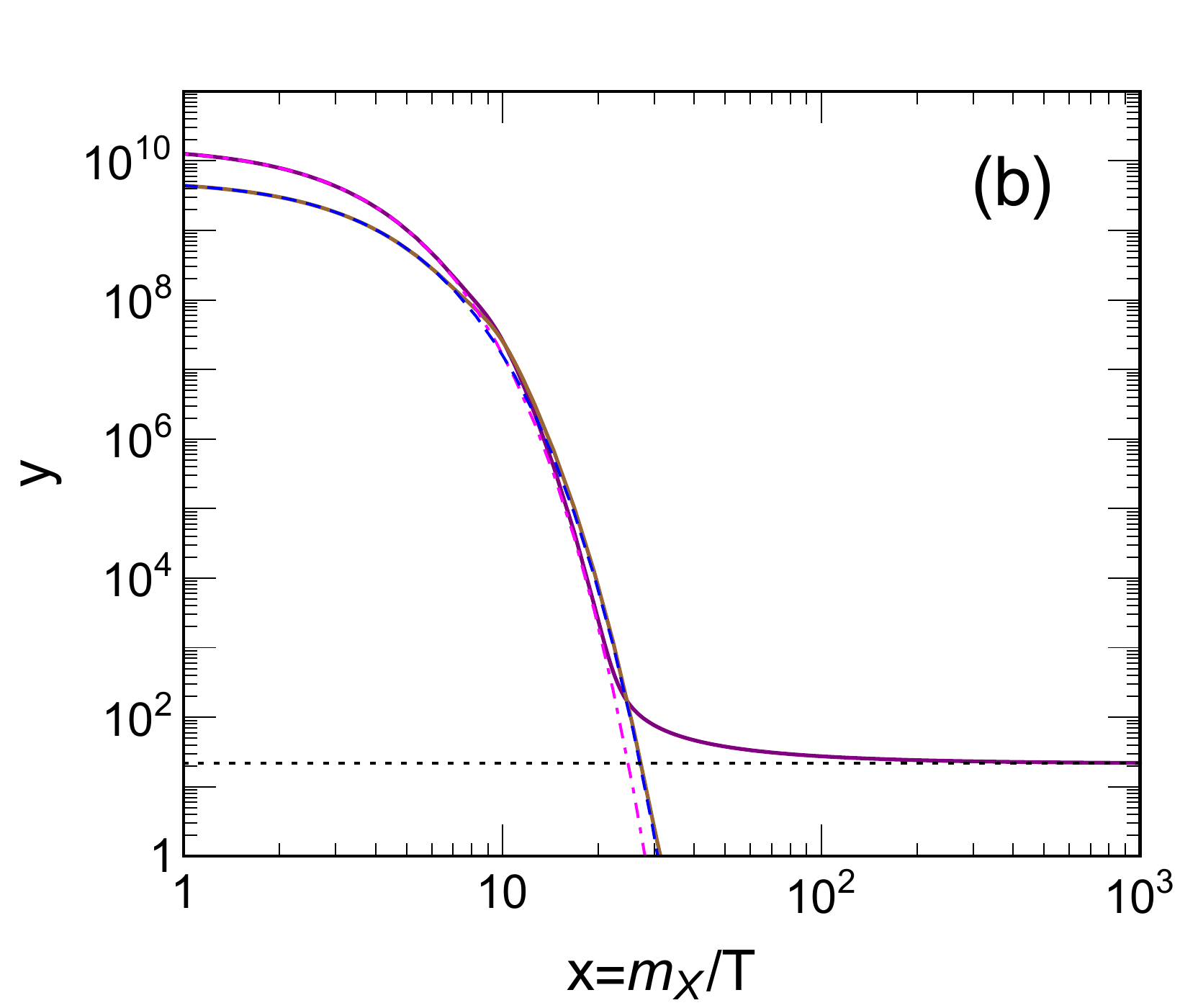}\\
\vskip0.3cm
\includegraphics[width=0.36\textwidth]{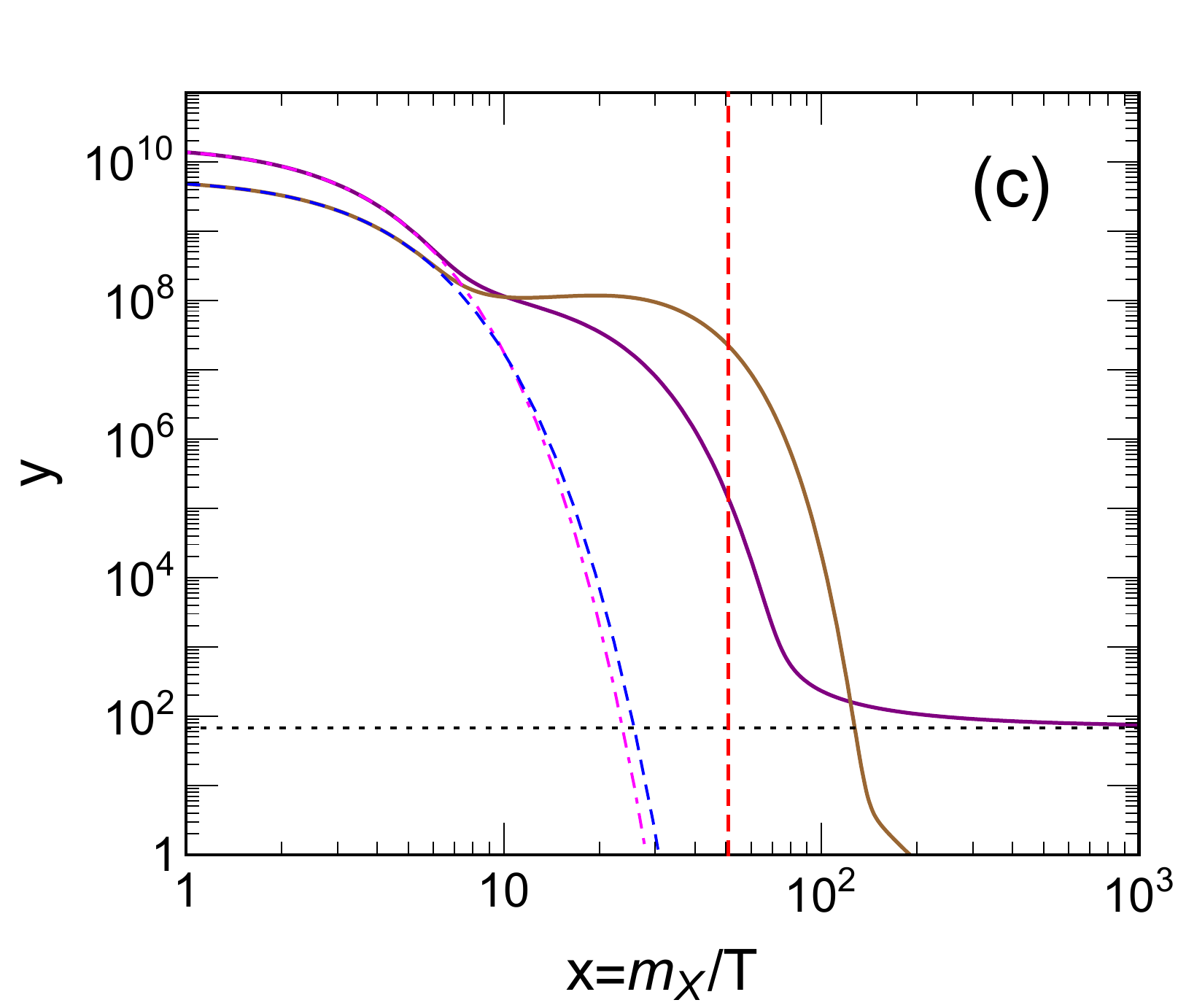} \hskip0.8cm
\includegraphics[width=0.36\textwidth]{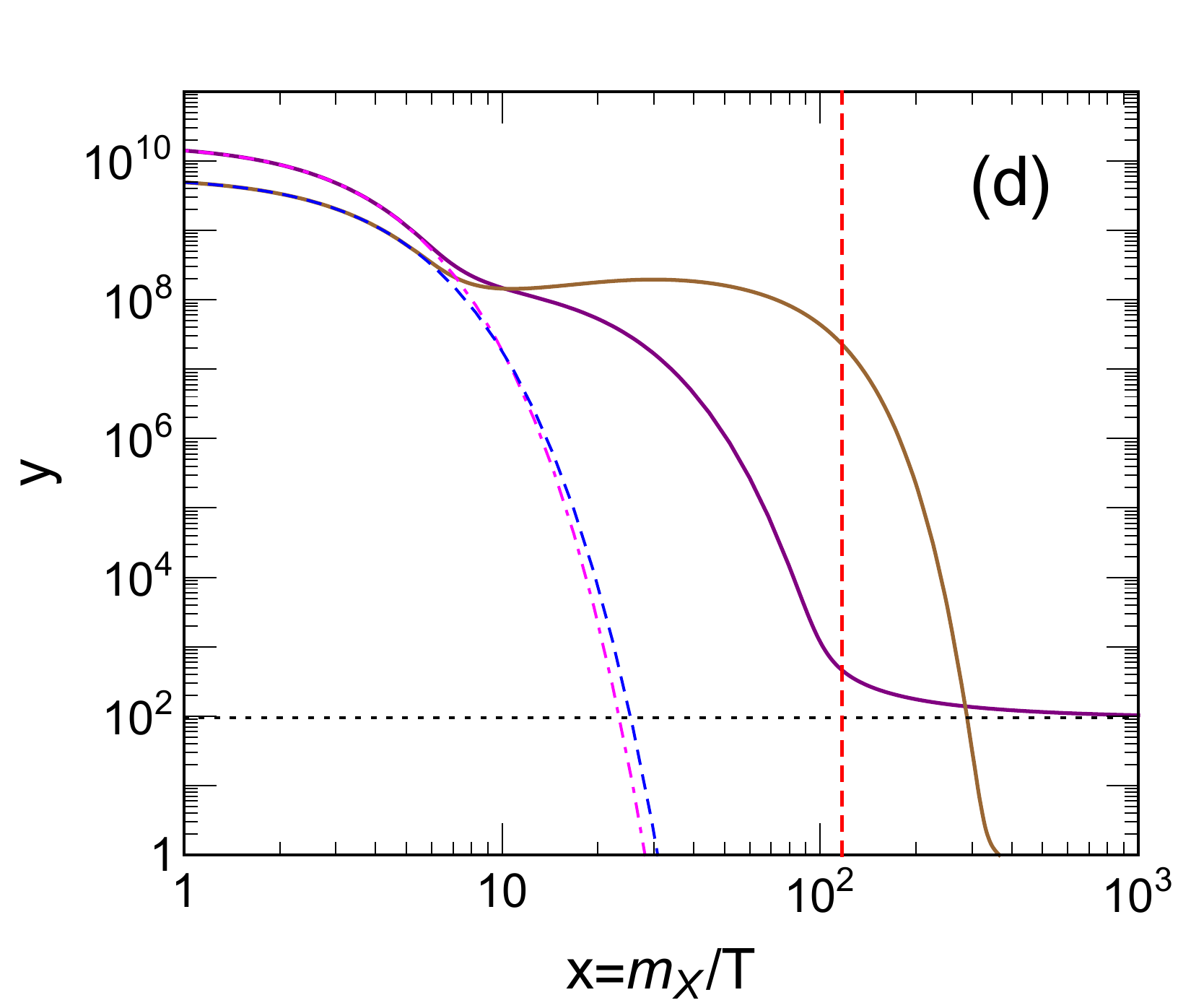}
\caption{ Same as Fig.~\ref{fig:relic} but the $3\leftrightarrow 2$ interactions are included in the Boltzmann equations.  In (a) and (b), the resulting $x_f$ is still to be 22, whereas in (c) and (d), the values of $x_f$ are 70 and 95, respectively. We show the same corresponding $x_\Gamma$ as Fig.~\ref{fig:relic}, for which the value is 51 in (c) and 117 in (d).
}
\label{fig:can-relic}
\end{center}
\end{figure}

Using the same parameters as in Sec.~\ref{sec:codecay-boltzmann}, and further including the $3\leftrightarrow 2$ interactions in the Boltzmann equations,  in Fig.~\ref{fig:can-relic} we show the results of the normalized yields. 
Compared with Fig.~\ref{fig:relic}(c) and (d), at $x_\Gamma$, which has been obtained  in Eq.~(\ref{eq:width-sol2}),  the corresponding value of $y_S$ (and also the number density of $S$) reduces 2 orders of magnitude due to the cannibal effect.
Fig.~\ref{fig:can-relic}(c) and (d) are typical examples about the cannibally co-decaying DM, where, at $x =x_{can} \approx 7$, the cannibal annihilation rate becomes less than the expansion rate of the Universe, so that the number densities of $X$ and $S$ no longer track up the behavior of the Boltzmann suppression.  The resulting $x_f=70$ for Fig.~\ref{fig:can-relic}(c) and 95 for (d) are significantly smaller than that with the cannibal interactions neglected.

In concluding this section, we  would like to discuss the relations and constraints between the parameters and observables. The $XX\to SS$ annihilation cross section and $x_f$ can be related to each other via the DM relic abundance.
The dimensionless density parameter of the present-day DM relic abundance, determined to be  $\Omega_{\rm{DM}}  =(0.1198 \pm 0.0026)/h^2$ from the observations  \cite{pdg,Ade:2013zuv}, is  given by \cite{Gondolo:1990dk}
\begin{equation}
 \Omega_{\rm DM} =\frac{m_X n_X}{\rho_c}= \frac{Y_X^\infty s_0 m_X}{\rho_c} 
 \simeq  \frac{1.04 \times 10^9\  {\rm GeV}^{-1}}{ J \sqrt{8 \pi g_*} M_{\rm pl} h^2}, 
 \label{eq:abundance}
  \end{equation}
where $Y_X^\infty$, related to $y_X^\infty$ via Eq.~(\ref{eq:yx}),  is the post-freeze-out value of $Y_X$, $\rho_c = 3 H_0^2/ (8\pi G)$ is the present critical density, $h\simeq0.678$ is  the present-day Hubble constant $H_0$ in units of  $100~\text{km}\, s^{-1}  \text{Mpc}^{-1}$,
$s_0=2891~\text{cm}^{-3}$ is the present-day entropy, and
\begin{equation}
 J= \int_{x_f}^\infty \frac{\langle \sigma v\rangle_{XX\to SS}}{x^2} dx \approx  \frac{ \langle \sigma v\rangle^{(0)}_{XX\to SS}}{x_f} \,.
\label{eq:j}
\end{equation}

\begin{figure}[t!]
\begin{center}
\includegraphics[width=0.6\textwidth]{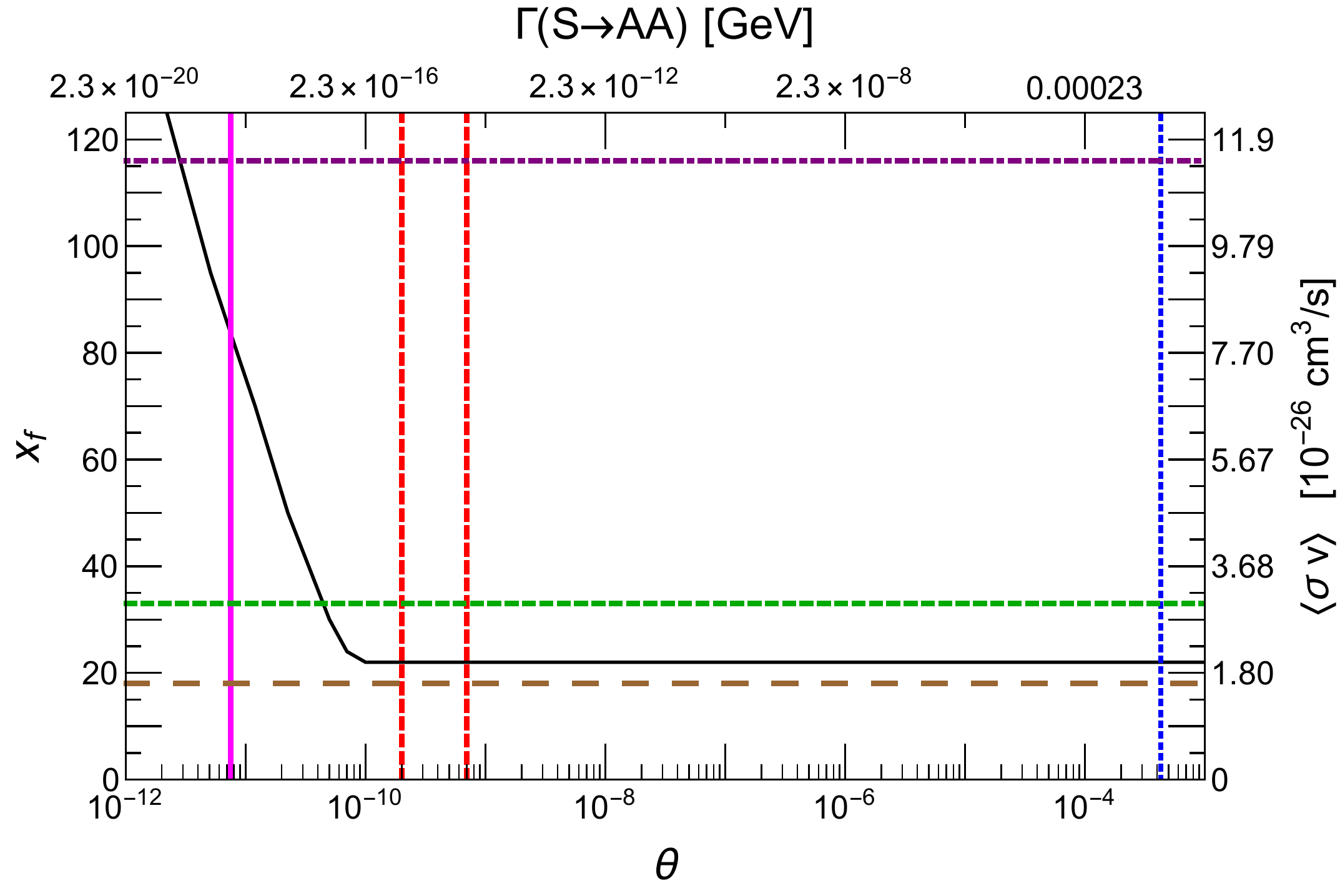}
\caption{ $x_f$ (or the $XX\to SS$ annihilation cross section $\langle \sigma v\rangle$) as a function of $\theta$ (or $\Gamma(S\to AA))$, denoted as the black curve to show the relic result, where we have adopted $\delta=0$ and the parameters given in Eq.~(\ref{eq:parameters-1}). When the $\theta$ is smaller than that denoted by the vertical dotted (blue) line, the GC gamma-ray excess is dominated by the 2-step cascade annihilation. For a $\theta$ smaller than that denoted by right dashed (red) line, the dark sector decouples from the thermal bath when $x \lesssim 1$.   The range of $\theta$ between the two vertical dashed (red) lines denotes that the DM $X$ as well as the mediator $S$ can be again in thermal equilibrium with the thermal bath before freeze out. The horizontal dot-dashed (purple) line shows the maximum value of the annihilation cross section that could still account for the GC gamma-ray excess. 
The 95\% C.L. upper limit and project sensitivity for observation of dSphs are denoted as horizontal dashed (green) and long-dashed (brown) lines, respectively.
For a $\theta$ smaller than that denoted by the vertical solid (magenta) line, we have $x_\Gamma >x_f$.
}
\label{fig:xftheta}
\end{center}
\end{figure}

In Fig.~\ref{fig:xftheta}, we show 
$x_f$ as a function of $\theta$, where the relations between $x_f$ and the $XX\to SS$ annihilation cross section $\langle \sigma v\rangle$, and between $\theta$ and $\Gamma(S\to AA)$ are depicted. For simplicity, we have approximated the thermally averaged DM annihilation cross section by using its leading $s$-wave value in $v\to 0$ limit. Note that the vertical axis for the annihilation cross section has been rescaled by use of the temperature-dependent $g_*$ given in Ref.~\cite{Cerdeno:2011tf}.
Moreover, in this lepton-specific (type-X) N2HDM portal vector dark matter model, the total width of $S$ satisfies $\Gamma_S \simeq \Gamma(S\to AA)$ with taking $\delta=0$.

In Fig.~\ref{fig:xftheta}, if $\theta$ is less than 0.00043, denoted by the vertical dotted (blue) line, the GC gamma-ray annihilation is dominated by the 2-step cascade DM annihilation. For $\theta\lesssim 7\times 10^{-10} $, corresponding to the left hand side of the right dashed (red) line, the dark sector decouples from the thermal reservoir when $x \approx 1$. However, within the range $ 2\times 10^{-10} \lesssim \theta\lesssim 7\times 10^{-10}$, i.e. $-1.03\times10^{-8} \gtrsim \lambda_{SAA} \gtrsim -3.60 \times 10^{-8}$, which is in between the two vertical dashed (red) lines, due to the co-decay and cannibal annihilation of the dark sector, the re-thermalized dark sector can be again in thermal equilibrium with the thermal reservoir before the DM freeze out; see also Fig.~\ref{fig:relic}(b) and Fig.~\ref{fig:can-relic}(b).

For a case with a large value of $x_\Gamma$,  the normalized yield $y_S$ does not tend to decay exponentially until  the time $t \sim t_\Gamma$.
Due to a significantly suppressed up-scattering rate, the exponential suppression for $y_X$ can occur much earlier than that for $y_S$. Moreover, the cannibal annihilations also reduce the value of $x_f$. Therefore, as shown in Fig.~\ref{fig:can-relic}(d),  we can have $x_\Gamma \gtrsim x_f$. 
In Fig.~\ref{fig:xftheta}, for $\theta< 7.5 \times 10^{-12} $ denoted by vertical solid (magenta) line, we have $x_\Gamma > x_f $.    
It should be noted that the approximation for  the temperature dependence of DoF given in Ref.~\cite{Cerdeno:2011tf}  breaks down during the QCD phase transition which may occur in the temperature  range of $150-400$~MeV, i.e. corresponding to $x  \sim 100-267$ for $m_X=40$~GeV.  For simplicity, the freeze-out results shown in Figs.~\ref{fig:relic}(d) and \ref{fig:xftheta} are assumed to occur before the QCD phase transition.

We then further discuss the big bang nucleosynthesis (BBN) and cosmic microwave background (CMB) constraints. The requirement of avoiding the $n/p$ ratio and $^4$He abundance to deviate from the standard BBN \cite{Kawasaki:2000en}
is to have $\Gamma_S^{-1}\lesssim 1$~sec, which imposes a quite relaxed bound of $| \lambda_{SAA}| \gtrsim 2.7\times 10^{-13}$, $|\theta| \gtrsim 5.3\times 10^{-15}$ with taking $\delta=0$. As for the Planck result for the CMB \cite{Ade:2015xua} which provides a probe of the DM annihilation at the time of recombination, $t_{\rm CMB}\sim$ 380,000 yrs, the resulting bound is highly insensitively  dependent on the number of cascade steps (see Fig.~11 of Ref.~\cite{Elor:2015bho}), and gives $\langle \sigma v\rangle \lesssim (7-19)\times 10^{-26}$~${\rm cm^3}/{\rm s}$. The current CMB constraint is modestly weaker than that given by the observations of dSphs.

Note that by varying the local dark matter density from 0.357~GeV/cm$^3$ to 0.2~GeV/cm$^3$ and the halo slope $\gamma$  from 1.2 to 1.1,  the annihilation cross section allowed by the GC data fit, shown in Fig.~\ref{fig:gcmxsv}, would be further extended by a factor of $\sim 4.10$ upward. In Fig.~\ref{fig:xftheta}, the horizontal dot-dashed (purple) line depicts  the 95\% C.L. upper limit on the annihilation cross section ($\langle \sigma v\rangle \simeq 1.15\times 10^{-25}~\text{cm}^3/{\rm s}$), corresponding to $x_f\simeq 116$, to account for the GC gamma-ray excess due to variations of $\rho_\odot$ and $\gamma$.
In Fig.~\ref{fig:xftheta}, we also show the current constraint from the observations of the dSphs, and the projected sensitivity for observations of 60 dSphs with 15-year data collection.   The  dSph projected limit on $\langle\sigma v\rangle$  is likely to be improved by a factor of $\sim 1.83$.

\section{Discussions and Conclusions}\label{sec:conclusion}

\begin{figure}[t!]
\begin{center}
\includegraphics[width=0.40\textwidth]{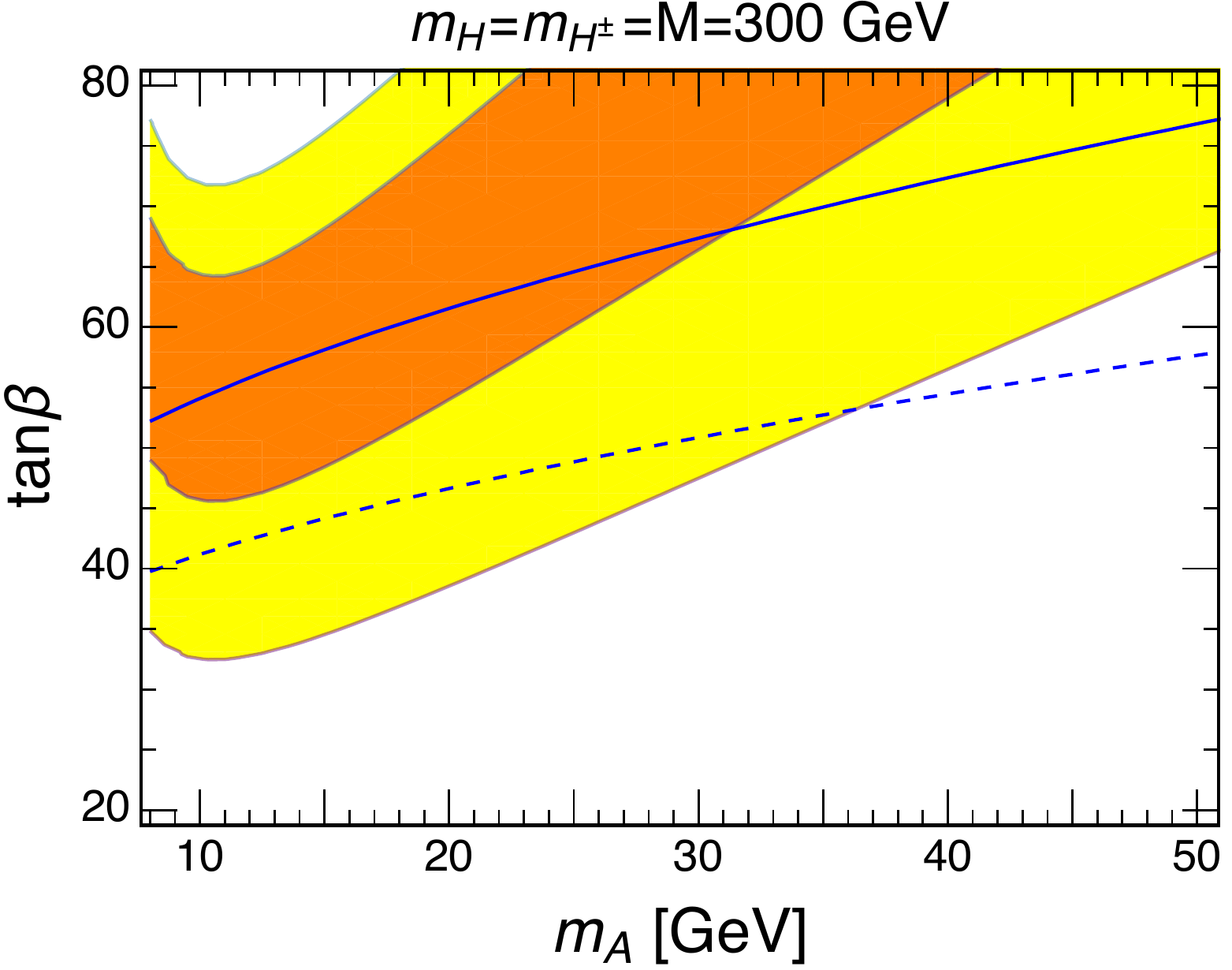}\hskip0.9cm
\includegraphics[width=0.40\textwidth]{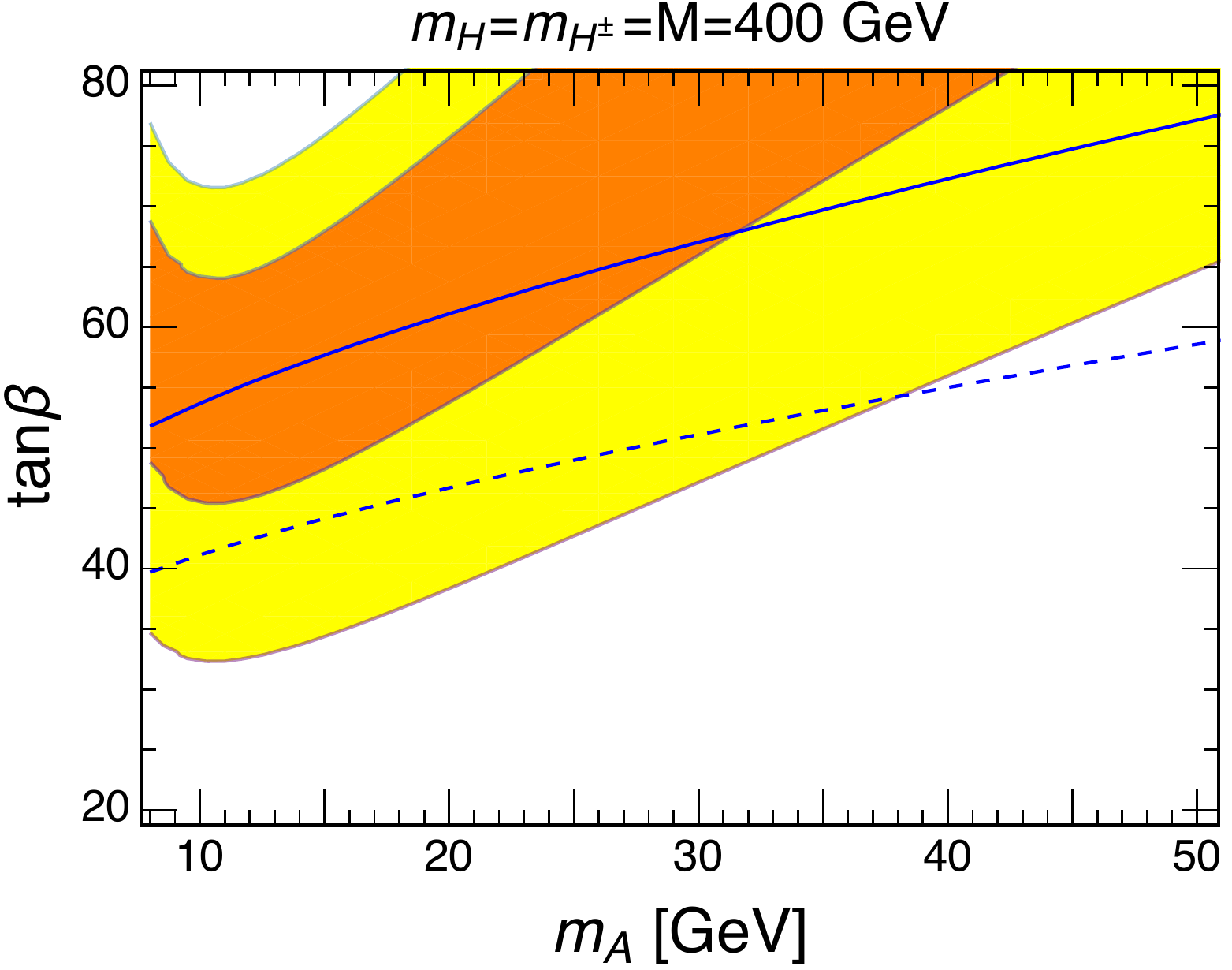}\\
\caption{The $g-2$ allowed regions with $1\sigma$ C.L. (orange) and $2\sigma$ C.L. (yellow) on the $m_A$-$\tan\beta$ plane vs. 95\% C.L.  
and 99\% C.L. upper limits denoted as the dashed and solid blue lines, respectively, from  the lepton universality and $\tau$ decay constraints.
}
\label{fig:g-2}
\end{center}
\end{figure}

Before making conclusion, we would like to discuss the parameter space in favor of the measurement of the muon anomalous magnetic moment $a_\mu \equiv (g-2)
/2$ \cite{pdg,Bennett:2006fi}, which can especially constrain the parameters, $m_A$ and $\tan\beta$. The discrepancy between the experiments and SM prediction is more than $3\sigma$ confidence level \cite{Broggio:2014mna,Davier:2010nc}:
\begin{align}
a_\mu^{\rm exp} - a_\mu^{\rm SM} =(262 \pm 86)\times 10^{-11} \,.
\end{align}
 In this secluded DM model, the mixing angles between the hidden Higgs boson and the visible two-Higgs doublets are very small, such that the dominant contributions to $a_\mu$ are from the visible sectors.
It was shown in Ref.~\cite{Cheung:2001hz} that the two-loop Barr-Zee diagrams can give sizable contributions to $a_\mu$. For our model, in contrast with the one-loop result, the two-loop diagram containing an $A$ propagator with  $m_A$ being ${\cal O}(10)$~GeV gives positive contributions to $a_\mu$ and its magnitude is even larger than that of one-loop. We collect the relevant $a_\mu$ calculations in the lepton-specific N2HDM in Appendix~\ref{app:g-2}.

In Fig.~\ref{fig:g-2}, we show 
the $g-2$ allowed regions at $1\sigma$ C.L. and $2\sigma$ C.L. on the $m_A$-$\tan\beta$ plane vs. 95\% C.L.  
and 99\% C.L. upper limits from  the lepton universality and $\tau$ decays, where we impose $\lambda_{hAA}=0$ to have $\text{Br}(h\to AA)=0$, and then show the results for two different values of $\sqrt{M^2}=m_H=m_{H^\pm}=300$ and 400~GeV. A larger $m_H$ can suppress the negative two-loop correction, whereas the output seems to be insensitive to variation of its value.  We also take $m_S=35$~GeV as input. However, because the result is insensitive to the values of the $\theta$ and $\delta$ due to their smallness, 
the hidden Higgs boson is secluded from the SM related phenomenology, just as the present case; thus the $g-2$ result that we have shown here is basically consistent with that in the type-X 2HDM.
The $\tau$ decays and lepton universality, which was stressed in Ref.~\cite{Abe:2015oca},  can provide a stringent bound on parameter space favored by $g-2$.  The relevant formulas and data are summarized in Appendix~\ref{app:taudecay}. As shown in Fig.~\ref{fig:g-2},  the result, capable of accounting for the muon $g-2$ anomaly at the 2$\sigma$ level but  without violating the constraint from the lepton universality and $\tau$ decays at 95\% C.L., favors the parameter space in which $8\lesssim m_A \lesssim37$~GeV and $30 \lesssim \tan\beta \lesssim 55$ for $m_H\sim$ 300-400~GeV.

It is interesting to note that the $B_s\to \mu^+ \mu^-$ data disfavor $m_A\lesssim 10$~GeV for  $m_H=m_{H^\pm}=300$~GeV,  but  the constraint is less restrictive for larger $m_H$ and $m_{H^\pm}$  \cite{Abe:2015oca}. Moreover, a precise measurement can also provide a stringent constraint on $m_A$ and $M^2$ (see Eq.~(\ref{eq:ghll}) for reference).

We give our conclusion as follows.  We have built a hidden sector dark matter model with dark discrete $Z_2$ symmetry which is the remnant of the spontaneously broken dark $U(1)_\text{dm}$ gauge group.  After symmetry breaking, the hidden sector contains vector dark matter, along with a real hidden Higgs mediating the DM interactions to the visible sector through the mixing among the neutral Higgs bosons in the model, while  the visible sector is the lepton-specific 2HDM.

In this model,  the GC gamma-ray excess is mainly due to the 2-step cascade annihilation describing that the vector dark matter particles annihilate to the pairs of hidden scalars, $S$, which subsequently undergo  the decays $S\to AA$ and then $A\to \tau^- \tau^+$.  
We find the parameter space with $m_X\sim 25-50$~GeV, $m_A \sim 3.6 - 25$~GeV, and $m_X \gtrsim m_S \gtrsim 2m_A$ can provide a good fit to the GC gamma-ray excess spectrum.  The obtained mass of the CP-odd Higgs $A$ in the GC excess fit can explain the muon $g-2$ anomaly at the 2$\sigma$ level without violating the stringent constraints from the lepton universality and $\tau$ decays.
  
Provided that the GC gamma-ray excess is generated by the 2-step cascade annihilation, the interplay  of the hidden sector and visible sector mainly arises from the interaction  $S\leftrightarrow AA$ which is dominant over $SS  \leftrightarrow AA$, where the interaction coupling depends on the mixing angles between the hidden Higgs boson and visible ones.
  We have shown that three different freeze-out types for the DM relic abundance, depending on such mixing angles, may occur, and correspond to the magnitudes of coupling constant $\lambda_{SAA}$ to be  (i) in between $3.60 \times 10^{-8}$ and $0.022$, (ii) in between $1.03\times10^{-8}$  and $3.60 \times 10^{-8}$, and (iii) less than   $1.03\times10^{-8}$.
  
 For the type (i), the hidden Higgs (mediator) can be in chemical and thermal equilibrium with the bath, and the DM particles, consistent the thermal WIMP scenario,  are Boltzmann suppressed until freeze out. For the type (ii), when the temperature of the thermal bath decreases with  $T \lesssim m_X, m_S$, the nonrelativistic particles in the dark sector decouple from the bath. Nevertheless,  the dark sector can be again in thermal equilibrium with the bath before freeze out, so that the dark matter particles behave like WIMPs finally.  See Figs.~\ref{fig:can-relic}(b) and \ref{fig:xftheta} for  example. For the type (iii) which is a typical case of the cannibally co-decaying DM, the dark sector decouples from the thermal background at the temperature below their masses, but undergoes a cannibal phase first. After that,  the total comoving number density  of the dark sector will not exponential suppressed  until the time that the mediator decaying to the CP-odd Higgs boson pair occurs. For this type, the exponential suppression for the comoving dark matter number density  could occur much earlier than that for the mediator due to a significantly kinematical suppression on the up-scattering  ($SS\to XX$) rate. See Sec.~\ref{sec:thetadelta-2} for details.

The vector dark matter in the hidden sector does not directly couple to the visible sector\footnote{Precisely speaking,  the dark matter can couple to the visible sector via the (visible) Higgs bosons, but, due to the negligibly small coupling constants,  such contributions are highly suppressed.}, but instead annihilates into the``short-lived" hidden Higgs bosons which decay through a small coupling into the CP-odd Higgs bosons. Considering the BBN bound, the lifetime of the short-lived hidden Higgs is required to be less than 1 sec, which imposes a quite relaxed constraint on the coupling $|\lambda_{SAA}| \gtrsim 2.7\times10^{-13}$.
Therefore, due to the very small coupling constant, $|\lambda_{SAA}| <0.022$,  the DM in the hidden sector is secluded from detections in the direct searches or colliders even though the resultant DM annihilation cross section in the case of type (iii) could be much larger than that in the thermal WIMP scenario, as shown in Fig.~\ref{fig:xftheta}. 

However, the DM annihilation signals are not suppressed in a general hidden sector model. We have shown the  constraints from the observations of dSphs and from the 15-year Fermi-LAT projected sensitivity of 60 dSphs, where  the  projected limit on $\langle\sigma v\rangle$  might be improved by a factor of $\sim 1.83$. The observations of dSphs  thus provide a promising way to test this hidden dark matter model in the near future.

\acknowledgments \vspace*{-1ex}
I would like thank Joshua T. Ruderman for helpful correspondence.
 This work was supported in part by the Ministry of Science and Technology, Taiwan, under Grant No. 105-2112-M-033-005.

\appendix

\section{Theoretical constraints}\label{app:th-constraints}

The parameters in the Type-X N2HDM scalar potential are subjected to the following theoretical constraints.

\noindent{\underline{\bf Perturbativity}}

To make sure the validity of the perturbative expansion, we impose the couplings to satisfy
\begin{equation}
g_X^2< 4  \pi, \  |\lambda_i| < 4 \pi, \  \text{for}\   i=1,\ \dots, 8 \,.
\end{equation}

\noindent{\underline{\bf Vacuum stability}}

To have a potential bounded from below, the allowed parameter regions satisfy the conditions \cite{JWittbrodt2016,Muhlleitner:2016mzt},
\begin{equation}
 \Bigg( 
 \lambda_1, 
 \lambda_2, 
 \lambda_6 > 0, 
 \sqrt{\lambda_1 \lambda_6} + \lambda_7 > 0,
 \sqrt{\lambda_2 \lambda_6} + \lambda_8 > 0, 
 \nonumber
 \sqrt{\lambda_1 \lambda_2} + \lambda_3 + D > 0,
 \lambda_7 + \sqrt{\frac{\lambda_1}{\lambda_2}} \lambda_8 \ge 0 
 \Bigg) \label{eq:omega1}
\end{equation}
or 
\begin{align}
& \Bigg( 
\lambda_1, 
\lambda_2, \lambda_6 > 0,
 \sqrt{\lambda_2 \lambda_6} \ge \lambda_8 > -\sqrt{\lambda_2 \lambda_6},
 \sqrt{\lambda_1 \lambda_6} > - \lambda_7 \ge \sqrt{\frac{\lambda_1}{\lambda_2}} \lambda_8,
  \nonumber
\\
&
\sqrt{(\lambda_7^2 - \lambda_1 \lambda_6)(\lambda_8^2 -\lambda_2
  \lambda_6)} > \lambda_7 \lambda_8 - (D+\lambda_3) \lambda_6 \Bigg)
\;,
\label{eq:omega2}
\end{align}
with
\begin{equation}
D = \left\{ \begin{array}{lll} \lambda_4 - \lambda_5 & \mbox{for} &
    \lambda_4 > \lambda_5 \\ 0 & \mbox{for} & \lambda_4 \le \lambda_5 \end{array}\right.
\;.
\end{equation}

\noindent{\underline{\bf Tree-level perturbative unitarity}}

The tree-level perturbative unitarity is obtained by requiring that all absolute eigenvalues of the two-body scalar scattering matrix are less than $8\pi$.
The constraints are given by \cite{JWittbrodt2016,Muhlleitner:2016mzt,Krause:2017mal}
\begin{align}
|\lambda_3 - \lambda_4| &< 8 \pi \label{eq:ev1} \,, \\
|\lambda_3 + 2 \lambda_4 \pm 3 \lambda_5| &< 8 \pi \,, \\
\left| \frac{1}{2} \left( \lambda_1 + \lambda_2 + \sqrt{(\lambda_1 -
     \lambda_2)^2 + 4 \lambda_4^2}\right) \right| &< 8\pi \,, \\
\left| \frac{1}{2} \left( \lambda_1 + \lambda_2 + \sqrt{(\lambda_1 -
     \lambda_2)^2 + 4 \lambda_5^2}\right) \right| &< 8\pi \,, \\
|a_{1,2,3}| &< 8\pi \,, \label{app:pu-constraints}
\end{align}
where $a_{1,2,3}$ are the real roots of the following equation,
\begin{eqnarray}
&&4\left(-27 \lambda_1 \lambda_2 \lambda_6 + 12 \lambda_3^2 \lambda_6 + 12
\lambda_3 \lambda_4 \lambda_6 + 3 \lambda_4^2 \lambda_6 + 6 \lambda_2
\lambda_7^2 - 8 \lambda_3 \lambda_7 \lambda_8 - 4 \lambda_4 \lambda_7
\lambda_8 + 6 \lambda_1 \lambda_8^2 \right) \nonumber\\
&&+ x \left(36 \lambda_1 \lambda_2 -
16\lambda_3^2 - 16\lambda_3 \lambda_4 - 4 \lambda_4^2 + 18 \lambda_1
\lambda_6 + 18 \lambda_2 \lambda_6 - 4\lambda_7^2 - 4\lambda_8^2\right)
\nonumber\\
&& +x^2 
\left(-6 (\lambda_1 + \lambda_2) -3 \lambda_6\right) + x^3
=0 \;. \label{eq:rootsl}
\end{eqnarray}

\section{The quartic couplings in the Higgs potential of the next-to-minimal two-Higgs doublet portal vector DM model}\label{app:quartic}

Keeping terms linear in $\sin\theta$ and $\sin\delta$, the quartic couplings $\lambda_i$ can be expressed in terms of $M$ and the physical Higgs masses as 
\begin{align}
\lambda_1 & \simeq \frac{1} {v^2 c_\beta^2} \left( -M^2 s_\beta^2 + m_h^2 s_\alpha^2 + m_H^2 c_\alpha^2  \right), 
\label{app:lambda1} \\
\lambda_2 & \simeq   \frac{1} {v^2 s_\beta^2} \left( -M^2 c_\beta^2 + m_h^2 c_\alpha^2 + m_H^2 s_\alpha^2 \right),
\label{app:lambda2} \\
\lambda_3 & \simeq  \frac{1} {v^2}  \left[ -M^2  +2m_{H_\pm}^2 
+ \frac{1}{s_\beta^2 c_\beta^2} \big( -m_h^2 s_\alpha c_\alpha + m_H^2 s_\alpha c_\alpha    \big) \right], 
\label{app:lambda3} \\
\lambda_4 & \simeq  \frac{1}{v^2} (M^2 -m_A^2 -2 m_{H^\pm}^2 ) ,
\label{app:lambda4} \\
\lambda_5 & \simeq \frac{1}{v^2} (M^2 - m_A^2) , 
\label{app:lambda5} \\
\lambda_6 & \simeq  \frac{m_S^2 } { v_S^2}   , 
\label{app:lambda6} \\
\lambda_7 & \simeq  \frac{1} {v v_S c_\beta} \left(   m_H^2 c_\alpha s_\theta - m_h^2  s_\alpha  s_\delta  - m_S^2 ( c_\alpha 
s_\theta - s_\alpha s_\delta) \right), 
\label{app:lambda7} \\
\lambda_8  & \simeq  \frac{1} {v v_S s_\beta} \left(   m_H^2 s_\alpha s_\theta + m_h^2  c_\alpha  s_\delta - m_S^2 ( s_\alpha 
s_\theta + c_\alpha s_\delta) \right) \,.   \label{app:lambda8}
\end{align}

\section{Oblique parameters}\label{ew:oblique}

The new physics effects can contribute to the gauge boson vacuum polarization amplitudes, and can be described by the three oblique parameters, $S, T$ and $U$ at the one-loop level.  We adopt the definition of these parameters, which were originally introduced by Peskin and Takeuchi (PT) and expanded to the linear order in $q^2$ \cite{Peskin:1990zt,Peskin:1991sw}. Considering the expansion beyond linear order, the new physics effects, introduced by Maksymyk, Burgess and London (MBL) \cite{Maksymyk:1993zm}, were defined as 6 parameters, $S, T, U, V, W$ and $X$. The PT parameters can be related to the MBL ones as $S_{\rm PT}= S_{\rm MBL} + 4(c_w^2 -s_w^2)X_{\rm MBL}$, $T_{\rm PT}= T_{\rm MBL} $, and $U_{\rm PT}= U_{\rm MBL} + 8s_w^2 X_{\rm MBL}$ \cite{Kundu:1996ah}, where $s_w \equiv \sin\theta_W=0.23129$ and $c_w\equiv \cos\theta_W$ at the scale $\mu=m_Z$.  Taking the limit $s_{\beta-\alpha} \to1$, and keeping terms linear in $\sin\theta$ and $\sin\delta$, we obtain the three PT parameters in the present model from a general multi-Higgs-doublet study in the MBL formulas given in Ref.~\cite{Grimus:2008nb},  and express the approximate result as
\begin{align} 
T & \simeq  \frac{1}{16 \pi^2 \alpha_{\rm em} v^2} 
\left(  F( m_{H^\pm}^2, m_H^2) + F( m_{H^\pm}^2, m_A^2)  - F( m_H^2, m_A^2) \right)   , \\
S & \simeq \frac{1}{24 \pi} \left(
G(m_H^2, m_A^2, m_Z^2) - (2s_w^2 -1)^2 G(m_{H^\pm}^2, m_{H^\pm}^2, m_Z^2)  + \ln \Big( \frac{m_H^2 m_A^2}{m_{H^\pm}^4} \Big)
 \right),  \\
U & \simeq \frac{1}{24 \pi} \left(
G(m_{H^\pm}^2, m_H^2, m_W^2) + G(m_{H^\pm}^2, m_A^2, m_W^2) - G(m_H^2, m_A^2, m_Z^2)  \right. \nonumber\\
 & \left. \qquad \qquad  + (4s_w^4 -1) G(m_{H^\pm}^2, m_{H^\pm}^2, m_Z^2)  
 \right) ,
\end{align}
where $\alpha_{\rm em}=1/127.950$ at the scale $m_Z$, the function $F$ is given by
\begin{align}
  F(m_1^2, m_2^2) = \left\{
      \begin{array}{ll}
&  \frac{m_1^2 + m_2^2}{2} -\frac{m_1^2 m_2^2}{m_1^2 -m_2^2} \ln\frac{m_1^2}{m_2^2}, \quad \: \text{ for } m_1^2 \not=m_2^2,\\
 & 0, \quad \hskip3.7cm  \text{for } m_1^2 =m_2^2 ,\\
      \end{array}
    \right.
\end{align}
and the function $G$, which is in a more complicated form, can be referred to Ref.~\cite{Grimus:2008nb}. In the limit $|m_{H^\pm} -m_H| \ll m_H$, $m_A \ll m_H$, the oblique parameters can be further approximated as 
\begin{align}
T &\approx \frac{1}{32 \pi^2 \alpha_{\rm em} v^2}  m_H (m_{H^\pm} -m_H) ,\nonumber\\
S &\approx -\frac{1}{24 \pi}  \left( \frac{5}{3} +\frac{1}{2} \frac{m_Z^2}{m_H^2}
 -\frac{(2s_w^2-1)^2}{5} \frac{m_Z^2}{m_{H^\pm}^2}+\frac{4 (m_{H^\pm} -m_H) }{m_H } \right),\nonumber\\
U &\approx \frac{1}{12 \pi}  
\left[ 
  -\frac{1}{10} \frac{m_W^2}{m_{H^\pm}^2}\Bigg( 1+ \frac{4s_w^4-1}{c_w^2} \Bigg)
  + \frac{1}{4} m_Z^2 \frac{m_{H^\pm}^2 -c_w^2 m_H^2}{m_H^2 m_{H^\pm}^2} 
  +  \frac{m_{H^\pm} -m_H }{m_H } \right].
\end{align}

\section{Annihilation cross sections}\label{app:annXS}

\subsection{The annihilation process  $XX\to SS$}

The DM annihilation processes for $XX\to SS$, relevant to the calculations of the thermal relic abundance and indirect measurements, are shown in Fig.~\ref{fig:vdm_ann}, and their cross sections can be expressed as
\begin{align}
\sigma v_{\text{lab}}
&= (\sigma v_{\text{lab}})_{\text{4v,s}}  + (\sigma v_{\text{lab}})_{\text{t,u}}  + (\sigma v_{\text{lab}})_{\text{int}}  \,, 
\end{align}
where $ (\sigma v_{\text{lab}})_{\text{4v,s}}$ is the result for the 4-vertex and $s$-channel diagrams,   $(\sigma v_{\text{lab}})_{\text{t,u}}$ is for the $t$- and $u$-channels, and $(\sigma v_{\text{lab}})_{\text{int}}$  is the interference between (4-vertex, $s$) and ($t$, $u$) channels, given by
\begin{align}
(\sigma v_{\text{lab}})_{\text{4v,s}} 
&=
\left(3+ \frac{s ( s-4 m_X^2 )}{4 m_X^4} \right) 
\frac{g_X^2 \sqrt{s-4 m_S^2} }{72 \pi  \left( (s-m_S^2 )^2 +  \Gamma_S^2 m_S^2 \right) \sqrt{s} (s-2 m_X^2 )}
\nonumber\\
& \quad \times \left[
\left(g_X (s- m_S^2) - g_{SSS} m_X \right)^2 + \Gamma_S^2 g_X^2 m_S^2 \right] \,, 
\label{eq:4vs}
\\
 (\sigma v_{\text{lab}})_{\text{t,u}} 
 &=
\frac{g_X^4 \sqrt{s-4 m_S^2} }{288 \pi m_X^4  \sqrt{s}  (s-2 m_X^2)}  \nonumber\\
&
\times \Bigg[4 m_S^4+4 s m_S^2+s^2  
+\frac{2 (m_S^8-8 m_X^2 m_S^6+24 m_X^4 m_S^4-32 m_X^6 m_S^2+48 m_X^8) }{m_S^4-4 m_X^2 m_S^2+m_X^2 s}
\nonumber\\
& \quad 
- \frac{4 \big(3 m_S^8-8 m_X^2 m_S^6  + (4 m_X^2 m_S^2- m_S^4) (8 m_X^4+s^2) -2 m_X^2 (24 m_X^6-2 s^2 m_X^2+s^3) \big) }{( s-2 m_S^2) \sqrt{s-4 m_S^2} \sqrt{s-4 m_X^2}} \nonumber\\
& \qquad
\times \ln \left(\frac{s-2 m_S^2 + \sqrt{s-4 m_S^2} \sqrt{s-4 m_X^2}}{s-2 m_S^2 - \sqrt{s-4 m_S^2} \sqrt{s-4 m_X^2}}\right)
\Bigg] \,,  
\label{eq:tu}
\\
(\sigma v_{\text{lab}})_{\text{int}}
&=g_X^3 
\frac{
 g_X \left( (s-m_S^2 )^2 +  \Gamma_S^2 m_S^2 \right) -  g_{SSS} m_X (s-m_S^2) }{144 \pi m_X^4 \left( \left(s- m_S^2 \right)^2+ \Gamma_S^2 m_S^2\right) \left(s- 2 m_X^2\right) \sqrt{s} \sqrt{s-4 m_X^2}}
\nonumber\\
&
 \times \Bigg[
\sqrt{s-4 m_S^2} \sqrt{s-4 m_X^2}  \big(  s (6 m_X^2 -s) - 2 (2 m_X^2+s ) m_S^2 \big) \nonumber\\
& \quad -
2 \left( (2 m_X^2+s) m_S^4-4 m_X^2  (2 m_X^2+s ) m_S^2+2 m_X^2  (12 m_X^4-2 s m_X^2+s^2 ) \right) \nonumber\\
& \qquad
\times \ln \left(\frac{s-2 m_S^2 + \sqrt{s-4 m_S^2} \sqrt{s-4 m_X^2}}{s-2 m_S^2 - \sqrt{s-4 m_S^2} \sqrt{s-4 m_X^2}}\right)
\Bigg] \,,
\label{eq:int}
 \end{align}
with $s$ being the invariant mass of the DM pair, $v_{\text{lab}}$ being the dark matter relative velocity in the rest frame of one of the incoming particles, and the effective coupling $g_{SSS}\equiv  v \lambda_{SSS} \simeq - 3 g_X m_S^2/m_X$. The thermally averaged annihilation cross section $\langle \sigma v_{\text{M\o l}} \rangle$ is equivalent to $\langle \sigma v_{\rm lab}\rangle$ \cite{Gondolo:1990dk}, and  can be approximated as \cite{Yang:2017zor}
\begin{equation}
\langle \sigma v_{\text{M\o l}} \rangle
\simeq
\frac{2 x^{3/2}}{\sqrt{\pi}}
 \int_{0}^{\infty} \sigma v_{\text{lab}}  \frac{(1+ 2\epsilon) \epsilon^{1/2} }{(1+\epsilon)^{1/4}}
 \left( 1-\frac{15}{4x} + \frac{3}{16x (1+\epsilon)^{1/2}}   \right)
 e^{-\frac{x\epsilon}{(1+\sqrt{1+\epsilon})/2}} d\epsilon
 \,,
 \label{eq:simpliedthermalXS}
 \end{equation}
provided that  $x (\equiv m_X /T) \gg 1$,  where $\epsilon=(s-4m_X^2)/(4m_X^2)$. For the indirect searches, assuming that the dark matter particles are locally in thermal equilibrium, we have  $x^{-1}=v_p^2/(2c^2)$, with $v_p$ the most probable speed of the DM distribution. The value of $v_p/c$ is about ${\cal O}(10^{-5})$ in the dwarf spheroidal satellite galaxies \cite{Simon:2007dq,Walker:2007ju,Martin:2008wj,Geha:2008zr,Walker:2009zp} and ${\cal O}(10^{-3})$ in our Galactic center \cite{Battaglia:2005rj,Dehnen:2006cm}.  In the low-velocity limit, we can have $\langle \sigma v_{\text{M\o l}}\rangle \cong  (\sigma v_{\text{lab}})_{\text{LV}}$ by taking $s\to 4 m_X^2$,
\begin{align}
(\sigma v_{\text{lab}})_{\text{LV}}=&\frac{g_X^2 \sqrt{m_X^2-m_S^2}} {144 \pi m_X^3 \left(2 m_X^2 -m_S^2 \right)^2 \left((4 m_X^2-m_S^2)^2+ \Gamma_S^2 m_S^2\right)  }\nonumber\\
& \times
\Big[g_X^2 \left((4 m_X^2-m_S^2)^2+ \Gamma_S^2 m_S^2\right) (11 m_S^4-28 m_X^2 m_S^2+44 m_X^4 )  
\nonumber\\
& +3 g_{SSS}^2 m_X^2 \left(2 m_X^2-m_S^2\right)^2  
-2 g_{SSS} g_X m_X \left(m_S^6-16 m_X^2 m_S^4+68 m_X^4 m_S^2-80 m_X^6\right) \Big] \,.
\end{align}

\subsection{The annihilation processes, $SS\to AA$ and $AA\to \tau^+ \tau^-$}\label{app:eqHiddenVisible}

In the derivation of the coupled Boltzmann equations, given in Eqs.~(\ref{eq:boltz-1}) and (\ref{eq:boltz-2}), one needs to take into account the chemical equilibrium among the relevant particles. 
The chemical equilibrium between the CP-even scalar $S$ and CP-odd scalar $A$ particles depends on the magnitudes of the decay width for $S\to AA$ and annihilation cross section for $SS\to AA$. The former  is described in Eqs.~(\ref{eq:S-partial-width-1})-(\ref{eq:S-partial-width-f}), and the latter is given by
\begin{align}
  (\sigma v_{\text{lab}})_{SS\to AA} 
  & = \frac{\sqrt{s-4 m_A^2}}{32 \pi   \sqrt{s} (s-2 m_S^2) \left((s-m_S^2)^2 + \Gamma_S^2 m_S^2 \right)} \nonumber\\
 \times & \Bigg[g_{SAA}^2 \, g_{SSS}^2 - 2 g_{SSAA }\, g_{SAA} \, g_{SSS} (s-m_S^2) 
 + \lambda_{SSAA}^2 \left((s-m_S^2)^2 + \Gamma_S^2 m_S^2 \right)  
 \nonumber\\
& 
+\frac{ 2 g_{SAA}^4\left((s-m_S^2)^2 + \Gamma_S^2 m_S^2 \right) }{m_S^4+m_A^2 \left(s-4 m_S^2\right)} 
 \nonumber\\
&  -
\frac{ 8 g_{SAA}^2 \coth ^{-1}\left(\frac{2 m_S^2-s}{\sqrt{s-4 m_A^2} \sqrt{s-4 m_S^2}}\right)  }{\sqrt{s-4 m_A^2} \sqrt{s-4 m_S^2} \left(s-2 m_S^2\right)}
 \bigg(
\Gamma_S^2 m_S^2  \big(g_{SAA}^2+ \lambda_{SSAA} (s-2 m_S^2 ) \big)
\nonumber\\
& -(s- m_S^2) \Big( g_{SAA} \, g_{SSS}  (s-2 m_S^2 ) - g_{SAA}^2 (s- m_S^2) - \lambda_{SSAA}  (2 m_S^4-3 s m_S^2+s^2 ) \Big) \bigg) \Bigg] \,,
\end{align}
with $g_{SAA} \equiv v \lambda_{SAA}$ and $g_{SSS} \equiv v \lambda_{SSS}$.

The chemical equilibrium between $A$ and $\tau$ depends on the magnitudes of the $A\to \tau^+ \tau^-$ decay width and $A A\to \tau^+ \tau^-$ annihilation cross section, for which the former is described in Eqs.~(\ref{eq:A-partial-width-1}) and (\ref{eq:A-partial-width-2}), and the latter is given by 
\begin{align}
  (\sigma v_{\text{lab}})_{AA\to \tau\tau} 
  & = 
\frac{\sqrt{s-4 m_\tau^2}}{8 \pi  \sqrt{s} (s-2 m_A^2)} \nonumber\\
 \times 
 & \Bigg[\frac{2  \bar{g}_{A\tau\tau}^4 \left(2 m_A^4-4 (2 m_\tau^2+s ) m_A^2+s  (4 m_\tau^2+s )\right) }
 { (s - 2 m_A^2) \sqrt{s-4 m_A^2} \sqrt{s-4 m_\tau^2}}
 \ln \Bigg(\frac{ s- 2 m_A^2 + \sqrt{s-4 m_A^2} \sqrt{s-4 m_\tau^2}}{ s -2 m_A^2 - \sqrt{s-4 m_A^2} \sqrt{s-4 m_\tau^2}} \Bigg)
 \nonumber\\
 &
 -\frac{2 \bar{g}_{A\tau\tau}^4 (m_A^4-4 m_\tau^2 m_A^2+8 m_\tau^4 )}{m_A^4-4 m_\tau^2 m_A^2+m_\tau^2 s}
 + \frac{ g_{SAA}^2 \bar{g}_{S\tau\tau}^2 s}{ (s-m_S^2)^2 + \Gamma_S^2 m_S^2}   \Bigg]  \,,
 \end{align}
where $\bar{g}_{A\tau\tau}\equiv (m_\tau /v) g_{A\tau\tau}$ and $\bar{g}_{S\tau\tau}\equiv (m_\tau /v) g_{S\tau\tau}$ with $g_{A\tau\tau} = t_\beta $ and $g_{S\tau\tau} = (-c_\alpha s_\theta + s_\alpha s_\delta)/c_\beta$, as shown in Table~\ref{tab:effyukln2hdm}.

\section{The muon $g-2$ and constraints from $\tau$ decays and lepton universality}\label{app:g2tau}

\subsection{The muon $g-2$}\label{app:g-2}

The contributions to  $a_\mu$ in the present type-X N2HDM are approximately by
the following one- and two-loop results. The one-loop contributions are given by \cite{Dedes:2001nx}
\begin{align}
\Delta a_\mu^{(1)} 
=
\frac{m_\mu^2}{8 \pi^2 v^2} \sum_{j = S, h, H, A, H^{\pm}} g_{j}^2 \, \frac{m_\mu^2}{m_j^2} \, f_j \bigg(\frac{m_\mu^2}{m_j^2} \bigg) \,,
\end{align}
with $g_{H^\pm} \equiv \tan\beta$,  $g_j \equiv g_{j\mu\mu}$ for $S, H$ and $A$ being the normalized Yukawa couplings given in Table~\ref{tab:effyukln2hdm}, and
\begin{align}
f_S (r) =f_h (r) =f_H (r) \simeq -\ln r -7/6 \,, \quad 
f_A (r) \simeq \ln r +11/6 \,, \quad
f_{H^\pm} (r)  \simeq -1/6 \,. \nonumber
\end{align}
The two-loop contributions  $\Delta a_\mu^{(2)}=\Delta a_\mu^{(2-1)}+\Delta a_\mu^{(2-2)}$ are from the Barr-Zee diagrams \cite{Cheung:2001hz,Ilisie:2015tra} with a fermion in the loop, given by
\begin{align}
\Delta a_\mu^{(2-1)} 
=
\frac{m_\mu^2}{8 \pi^2 v^2} \frac{\alpha_{em}}{\pi}
\displaystyle \mathop{\sum_{j = S, h, H,A}}_{f\equiv \text{quark, lepton}} 
N_c^f \,  Q_f^2 \, g_{j\mu\mu} \, g_{j ff} \, \frac{m_f^2}{m_j^2} \, g_j \bigg(\frac{m_f^2}{m_j^2} \bigg) \,,
\end{align}
 and with the charged Higgs in the loop, given by
\begin{align}
\Delta a_\mu^{(2-2)} 
=
\frac{m_\mu^2}{16 \pi^2 } \frac{\alpha_{em}}{\pi}
\sum_{j = S, h, H} 
 g_{j\mu\mu} \, \lambda_{j H^+ H^-} \, \frac{1}{m_j^2} \, g_j \bigg(\frac{m_{H^\pm}^2}{m_j^2} \bigg) \,,
\end{align}
where $N_c^f \equiv \text{3 (1)}$ for quarks (leptons),  $Q_f$ the electric charge of the fermion in the loop, and
\begin{align}
g_i(r)  = \int_0^1 \frac{N_i (x)}{x(1-x) -r} \ln \frac{x (1-x)}{r} \,,
\end{align}
with
$N_h(x) =N_H (x) \equiv 2x(1-x) -1$, $N_A (x)  \equiv 1$, and $N_{H^\pm} \equiv x(1-x)$. Here the triple Higgs couplings $\lambda_{j H^+ H^-}$ equal to  $\lambda_{j AA}$ with $m_A$ replaced by $m_{H^\pm}$; see Eqs.~(\ref{eq:triple-1-hAA}), (\ref{eq:triple-2-HAA}), and (\ref{eq:triple-3-SAA}) for references.

\subsection{Constraints from $\tau$ decays and lepton universality}\label{app:taudecay}

The ratio of the pure leptonic processes can be parametrized as $\Gamma(\ell_1 \to  \nu_1 \ell_2 \bar\nu_{\ell_2}) / \Gamma(\ell_3 \to  \nu_3 \ell_4 \bar\nu_{\ell_4}) \equiv (g_{\ell_1} g_{\ell_2})^2 / (g_{\ell_3} g_{\ell_4})^2$.
In the present  model, three parameters for the $\delta_{\ell \ell'} ( \equiv (g_\ell /g_{\ell^\prime}) -1)$, which describe the deviations from the SM results, can be obtained from the data, $\tau \to \mu \nu \bar\nu, \tau \to e \nu \bar\nu$ and $\mu \to e \nu \bar\nu$ \cite{Amhis:2014hma},  and are given by
\begin{equation} \label{eq:delta}
\delta_{\tau\mu} = \delta_{\rm loop},\quad
\delta_{\tau e} = \delta_{\rm tree} + \delta_{\rm loop},\quad
\delta_{\mu e} = \delta_{\rm tree}, 
\end{equation}
where the theoretical formulas for $\delta_{\rm tree}$ and $\delta_{\rm loop}$ are \cite{Krawczyk:2004na}
\begin{align} \label{eq:delta_tree_loop}
\delta_{\rm tree} &= {m_\tau^2 m_\mu^2 \over 8 m^4_{H^\pm}} t_\beta^4
- {m_\mu^2 \over m^2_{H^\pm}} t^2_\beta {g(m_\mu^2/m^2_\tau) \over f(m_\mu^2/m_\tau^2)}, \nonumber\\
\delta_{\rm loop} &= {G_F m_\tau^2 \over 8 \sqrt{2} \pi^2} t^2_\beta 
\left[1 + {1\over4} \left( H(x_A) + s^2_{\beta-\alpha} H(x_H) + c^2_{\beta-\alpha} H(x_h)
+(s_{\beta-\alpha} s_\theta + c_{\beta-\alpha} s_\delta )^2 H(x_S) \right)
\right], 
\end{align}
with $g(x)\equiv 1+9x-9x^2-x^3+6x(1+x)\ln x$,  $f(x)\equiv 1-8x+8x^3-x^4-12x^2 \ln x$, 
$H(x) \equiv (\ln x) (1+x)/(1-x)$, and $x_\phi=m_\phi^2/m_{H^{\pm}}^2$.
Further considering the data of the semi-hadronic processes $\pi/K \to \mu\nu$ that can also give the value of $\delta_{\tau\mu}$ \cite{Amhis:2014hma}, Chun, Kang, Takeuchi, and Tsai \cite{Chun:2015hsa} obtained the following three independent constraints,
\begin{eqnarray} \label{LUconstraints}
 {1\over\sqrt{2}} \delta_{\rm tree} +\sqrt{2} \delta_{\rm loop} &=& 0.0028 \pm 0.0019,\quad \nonumber\\ 
{\sqrt{3\over2} } \delta_{\rm tree} &=& 0.0022\pm 0.0017,\quad \\
\delta_{\rm loop} &=&0.0001 \pm 0.0014 \nonumber.
\end{eqnarray}
We use the above results to show the constraint on the $m_A$-$\tan\beta$ plane in Fig.~\ref{fig:g-2}.

\end{document}